%% file: revision1.tex
\newif\ifarxiv
\arxivtrue

\ifarxiv
\documentclass[onefignum,onetabnum,pagebackref]{siamart190516}
\else
\documentclass[review,onefignum,onetabnum,pagebackref]{siamart190516}
\fi

\input{ex_shared}
\usepackage{amsmath}

\ifpdf
\hypersetup{
  pdftitle={A Theory of Quantum Subspace Diagonalization},
  pdfauthor={E. N. Epperly, L. Lin, and Y. Nakatsukasa}
}
\fi

\ifarxiv
\else
\fi

\begin{document}

\maketitle

\begin{abstract}
  Quantum subspace diagonalization methods are an exciting new class of algorithms for solving large\rev{-}scale eigenvalue problems using quantum computers.
  Unfortunately, these methods require the solution of an ill-conditioned generalized eigenvalue problem, with a matrix pair corrupted by a non-negligible amount of noise that is far above the machine precision.
  Despite pessimistic predictions from classical \rev{worst-case} perturbation theories, these methods can perform reliably well if the generalized eigenvalue problem is solved using a standard truncation strategy.
  By leveraging and advancing classical results in matrix perturbation theory, we provide a theoretical analysis of this surprising phenomenon, proving that under certain natural conditions, a quantum subspace diagonalization algorithm can accurately compute the smallest eigenvalue of a large Hermitian matrix.
  We give numerical experiments demonstrating the effectiveness of the theory and providing practical guidance for the choice of truncation level.
  Our new results can also be of independent interest to solving  eigenvalue problems outside the context of quantum computation.
\end{abstract}



\section{Introduction}

Quantum computing is a fundamentally new computational paradigm, which has the potential to have a transformative impact on \rev{certain areas of} computational science \rev{\cite{Preskill2018,Preskill2021}.}
One particularly compelling use case for quantum computers is to solve eigenvalue problems related to quantum many-body systems, for which the dimension of the discretized matrix grows exponentially with respect to the number of particles.

Quantum subspace diagonalization (QSD) methods \cite{CG21,HLB+20,KMC+21,MKCd17,MST+20,PM19,SekiYunoki2021,SHE20}, also known as quantum Krylov methods, are an exciting class of quantum algorithms for solving large\rev{-}scale Hermitian eigenvalue problems.
One common key step of these algorithms is to solve a nearly singular generalized eigenvalue \rev{problem}, where each entry of the associated matrix pair can be corrupted by Monte Carlo errors many orders of magnitude larger than the round-off error typically seen in classical computation.
For such noisy generalized eigenvalue problems, \rev{classical perturbation theory fails to explain why such problems could be solved accurately}.
Despite \rev{this}, QSD methods appear to work in practice, at least on some examples and with some procedure to compensate for the measurement error.
This article is addressed squarely at explaining why, in theory, QSD algorithms perform well and how, in practice, errors in the problem data can be effectively dealt with.

While our analysis is centered around the \rev{context of QSD}, the underlying problem of solving an ill-conditioned generalized eigenvalue problem with noisy data is a fundamental linear algebra problem, which emerges in multiple application areas such as electronic structure theory \cite{CaiBaiPaskEtAl2013,JungenKaufmann1992}, control theory \cite{EV82}\rev{, and variational Monte Carlo optimization \cite{SMS20}}.
\rev{Our results may be of particular interest in filter diagonalization methods \cite{MT97,WN95} which were a classical antecedent to QSD methods \cite{PM19} and also require the solution of an ill-conditioned generalized eigenvalue problem.} 

\paragraph{Notation}

Vectors and matrices will be denoted by boldface lower- and upper-case letters respectively.
We denote the conjugate transpose by ${}^{\rev{*}}$.
All matrices and vectors \rev{are} assumed to be over the complex numbers.
The unembellished norm $\norm{\cdot}$ shall refer to the Euclidean norm of a vector $\norm{\vec{x}} = \sqrt{\vec{x}^{\rev{*}}\vec{x}}$ or the spectral norm (largest singular value) of a matrix.
At times\rev{,} we shall also make use of the Frobenius norm $\norm{\mat{B}}_{\rm F} := \sqrt{\tr(\mat{B}^{\rev{*}}\mat{B})}$.
The absolute values of the generalized eigenvalues of a pair $(\mat{H},\mat{S})$ is denoted $|\Lambda(\mat{H},\mat{S})|$.
Relations $\approx$, $\lessapprox$, $\ll$, etc.\ are informal, with no precise mathematical relation being implied.

\subsection{Quantum subspace diagonalization and its numerical challenges}
\label{sec:quant-subsp-diag}

\rev{
  We begin by describing a simple QSD algorithm developed in parallel by Parrish and McMahon \cite{PM19} and Stair, Huang, and Evangelista \cite{SHE20} with a focus on its linear algebraic (Krylov) structure.
  For readers unfamiliar with quantum computation and why it might be advantageous over classical computing for certain problems, we recommend the classic textbook~\cite{NielsenChuang2000} as well as a recent tutorial aimed at a mathematical audience~\cite{Nannicini2020}.
}

\rev{
  Suppose we are interested in the ground-state energy\footnote{\rev{The ground-state energy, by itself, is a useful quantity for applications in electronic structure, as it determines the energy landscape for dynamical simulation.}} (least eigenvalue) $E_0$ of a Hamiltonian operator (Hermitian matrix) $\oper{H}$.
  Our goal shall be to compute an approximation $\tilde{E}_0$ for which the (forward) error $|\tilde{E}_0 - E_0|$ is small.
  For the  problem sizes for which QSD is appropriate, vectors with the dimension of the operator $\oper{H}$ are too large to store classically, so we would like to represent them as states in a quantum computer.
  The QSD method rests on two assumptions:
  \begin{enumerate}
  \item We can efficiently prepare a state $\vec{\varphi}_0$ on the quantum device that has a nontrivial overlap $|\vec{\varphi}_0^*\vec{\psi}_0^{\vphantom{*}}| \gg 0$ to the true ground-state eigenvector $\vec{\psi}_0$.
  \item The time evolution $\vec{\varphi} \mapsto \e^{\iu t \oper{H}} \vec{\varphi}$ can be efficiently simulated on the quantum device.
  \end{enumerate}
  To improve on our initial guess $\vec{\varphi}_0$, we enlarge to a subspace spanned by vectors
  \begin{equation} \label{eq:quantum_subspace}
    \vec{\varphi}_j = \e^{\iu t_j \oper{H}} \quad \textnormal{for } j =0,\ldots,n-1
  \end{equation}
  where $t_j = j\Delta t$ are a time sequence with step size $\Delta t > 0$.
  The subspace $\spn \{ \vec{\varphi}_j \}$ forms a ``unitary Krylov space'' and plays a role analogous to the Krylov subspace in the Lanczos method.
  Eigenvalue estimates for the operator $\oper{H}$ can be computed from this unitary Krylov subspace by applying the Rayleigh--Ritz method \cite[\S11]{Par98}.
  
  In a classical Krylov method, one would usually orthogonalize the basis vectors \eqref{eq:quantum_subspace}.
  Unfortunately, this orthogonalization operation can be inherently difficult to perform on a quantum computer, so we instead work in the basis \eqref{eq:quantum_subspace} as computed.
  In a non-orthogonal basis, the Rayleigh--Ritz eigenvalue estimates are obtained by solving a generalized eigenvalue problem
}
\begin{equation} \label{eq:gep}
  \mat{H} \vec{c} = E \, \mat{S} \vec{c}.
\end{equation}
Here the \emph{projected Hamiltonian} and \emph{overlap} matrices $\mat{H}$ and $\mat{S}$ are Hermitian--Toeplitz matrices defined as
\begin{equation} \label{eq:HS}
  \mat{H}_{jk} = \rev{\vec{\varphi}_j^*{\oper{H}}\vec{\varphi}_k}, \quad \mat{S}_{jk} = \rev{\vec{\varphi}_j^*\vec{\varphi}_k},
\end{equation}
$\vec{c}$ is the reduced Ritz vector, and $E$ is the Ritz value.
Each matrix entry $\mat{H}_{jk},\mat{S}_{jk}$ \rev{can be} estimated via a Monte Carlo sampling procedure \rev{such as the Hadamard test on a quantum computer (see e.g., \cite[Chapter 5]{NielsenChuang2000}, \cite[\rev{App.~D}]{CG21}).}
Once a sufficiently good estimate to $\mat{H},\mat{S}$ is obtained, the generalized eigenvalue problem \eqref{eq:gep} is solved on a classical computer.

\begin{remark}[Other QSD methods]
  \rev{A number of QSD methods have been proposed which differ in how the basis states \eqref{eq:quantum_subspace} are generated and how the eigenvalue estimates are obtained.}
  \rev{Alternative methods for basis vector generation \eqref{eq:quantum_subspace} from one or more initial guesses include}: multiplication by creation and annihilation operators \cite{CRD+18,MKCd17} \rev{and} imaginary-time evolution $\rev{\vec{\varphi}_j := \e^{\iu (\iu t_j)\oper{H}} \rev{\vec{\varphi}_0}}$ \cite{MST+20} (with $t_j\ge 0$).
  \rev{Methods also differ in whether the Rayleigh--Ritz procedure is applied to $\oper{H}$ itself \cite{PM19,SHE20} or the time evolution operator $\e^{\iu\,\Delta t \,\oper{H}}$ \cite{CG21,KMC+21}.}
  For concreteness, we shall focus in this article on \rev{the QSD method as discussed}, though our analysis should have insights for understanding the broader class of QSD algorithms.
\end{remark}

\rev{There are two main questions in the analysis of the (forward) error of the QSD method:
  \begin{enumerate}[label=(\Alph*)]
  \item How to analyze the Rayleigh--Ritz error due to the use of a finite-dimensional unitary Krylov subspace? \label{item:question_1}
  \item How to analyze the error of the generalized eigenvalue problem \eqref{eq:gep} in the presence of the Monte Carlo noise for estimating the matrix entries in $\mat{H},\mat{S}$? \label{item:question_2}
  \end{enumerate}
  Below we first discuss issues related to question\ \ref{item:question_2}, which are particularly challenging from the perspective of numerical linear algebra.
}

\subsection{Numerical Issues with QSD}
\label{sec:numer-issu-with}

\rev{Numerical results indicate that the size of $n$ needed to obtain desired accuracy can be very modest (e.g.\ $10 \le n\le 100$)}, so we are free to use any algorithm~\cite{GolubVanLoan2013,Par98} to solve \rev{the} dense generalized eigenvalue problem \rev{\eqref{eq:gep}}.
However, it is frequently observed that the states $\rev{\vec{\varphi}_0},\ldots,\rev{\vec{\varphi}_{n-1}}$ are very close to being linearly dependent, leading to the matrices $\mat{H}$ and $\mat{S}$ being nearly rank-deficient and the generalized eigenvalue problem \eqref{eq:gep} nearly singular. 
This ill-conditioning is an intrinsic feature of this method since the problem necessarily becomes ill-conditioned if the initial guess $\rev{\vec{\varphi}_0}$ possesses the desirable property of approximately lying in a low-dimensional invariant subspace.

The near-singularity of the problem \eqref{eq:gep} becomes particularly alarming when taken in conjunction with the fact that the matrix elements \eqref{eq:HS} will be corrupted by several types of error when measured from a quantum computer.
Some forms of error, such as discretization error in evaluating the time evolution $\e^{\iu t_j \oper{H}} \rev{\vec{\varphi}_0}$ by a Trotter formula \cite{CST+21} and gate errors, can in principle be systematically controlled on a fault-tolerant quantum device.
However, even on a flawless quantum device, the matrix elements \eqref{eq:HS} still need to be computed \rev{via sampling, which incurs}
Monte Carlo-type $\approx \delta^{-2}$ samples to compute each entry to $\delta$-accuracy.
We shall refer to all of these errors collectively as ``noise''.

Classical perturbation theory \cite[{\S}VI.3]{SS90} as well as modern improvements \cite{ML04} \rev{do not apply} when the perturbation is large enough to make the problem \eqref{eq:gep} singular, which is almost always the case for the QSD algorithm because of sampling error and ill-conditioning.
Indeed, the following variant of the classical example \cite[Eq.~(4.10)]{Wil79} shows that a perturbation just a touch larger than the distance to singularity can send the eigenvalues (originally, $1$ and $2$) to any pair of numbers$\alpha,\beta\in\complex$:
\begin{equation} \label{eq:wilkinson}
  \mat{H} = \twobytwo{2}{0}{0}{\epsilon}, \: \mat{S} = \twobytwo{1}{0}{0}{\epsilon} \xrightarrow{\order(\epsilon) \mbox{ error}}\mat{\tilde{H}} = \twobytwo{2}{\alpha \epsilon}{\beta\epsilon}{0},\: \mat{\tilde{S}} = \twobytwo{1}{\epsilon}{\epsilon}{0}.
\end{equation}
Even the well-conditioned eigenvalue $2$ can be perturbed arbitrarily far if the noise is large enough to make the problem singular!

There is evidence that adversarially chosen perturbations such as \eqref{eq:wilkinson} are pathologically unlikely to occur, with the well-conditioned eigenvalues of a pair $(\mat{H},\mat{S})$ changing only modestly after perturbation.
Indeed, Wilkinson showed that ``most'' $\order(\epsilon)$ perturbations to \eqref{eq:wilkinson} have an eigenvalue near the well-conditioned eigenvalue of 2 \cite{Wil79}, and recent analysis by Lotz and Noferini \cite{LN20} show that some eigenvalues of genuinely singular generalized eigenvalue problems can, in effect, be locally well-conditioned with high probability.

\rev{Even if a good approximation to the ground-state energy is among the eigenvalues of $(\mat{\tilde{H}},\mat{\tilde{S}})$, identifying it can be difficult.
When the pair $(\mat{H},\mat{S})$ is nearly singular, the perturbed problem is almost assured to possess spurious eigenvalues.
To see why this is the case, observe that if $(\mat{H},\mat{S})$ is nearly singular, there exists an eigenpair $(\vec{c},E)$ such that $\mat{H}\vec{c}, \mat{S}\vec{c} \approx \vec{0}$.
Perturbations in $\mat{H}$ and $\mat{S}$ create large changes in both the numerator and the denominator of the Rayleigh quotient $E = \vec{c}^*\mat{H}\vec{c} / \vec{c}^*\mat{S}\vec{c}$; since eigenvalues of a pair with positive definite $\mat{S}$ extremize the Rayleigh quotient, noise can easily introduce fake eigenvalues much smaller than the genuine least eigenvalue of $(\mat{H},\mat{S})$.}
Reliably \rev{distinguishing genuine eigenvalues from such fake eigenvalues} is challenging; we tried many heuristics and all of them failed for a nontrivial fraction of random initializations of the measurement error (see section~\ref{sec:failure-heuristics}).

\rev{To address these issues, we shall solve the eigenvalue problem \eqref{eq:gep} using the following truncation scheme}:
First, compute an eigendecomposition of the matrix $\mat{S}$ and discard all eigenvalues smaller than or equal to a threshold $\epsilon > 0$.
Then, letting $\mat{V}_{>\epsilon}$ denote a matrix whose columns are the non-discarded eigenvectors and $\mat{\Lambda}_{>\epsilon} := \mat{V}_{>\epsilon}^{{\rev{*}}}\mat{S}\mat{V}_{>\epsilon}^{\vphantom{*}}$, we solve the reduced generalized eigenvalue problem
\begin{equation*}
  \mat{V}_{>\epsilon}^{{\rev{*}}}\mat{H}\mat{V}_{>\epsilon}^{\vphantom{*}} \, \vec{c} = \tilde{E} \, \mat{V}_{>\epsilon}^{{\rev{*}}}\mat{S}\mat{V}_{>\epsilon}^{\vphantom{*}} \vec{c},
\end{equation*}
or equivalently find the eigenvalues of $\mat{\Lambda}_{>\epsilon}^{-1/2} \mat{V}^{{\rev{*}}}_{>\epsilon} \mat{S} \mat{V}_{>\epsilon}^{\vphantom{*}}\mat{\Lambda}_{>\epsilon}^{-1/2}$.
This procedure appears to have been first discovered in quantum physics by L\"owdin in 1967 \cite{Low67} (also rediscovered in \cite{GN86}), where it is associated with the \rev{name} \emph{canonical orthogonalization}.
We shall call this procedure \emph{thresholding} and present it in Algorithm~\ref{alg:thresholding} for convenient reference in the rest of the document.
A more careful variant of thresholding from the numerical analysis community was proposed by Fix and Heiberger in 1972 \cite{FH72}, though its authors expressly advise against using it in precisely the setting of the QSD algorithm where the pair $(\mat{H},\mat{S})$ is nearly singular.

\begin{algorithm}[t]
  \caption{Thresholding procedure for solving a noise-corrupted or nearly singular generalized eigenvalue problem $\mat{H}\vec{c} = E\, \mat{H}\vec{c}$.} \label{alg:thresholding}
  \begin{algorithmic}
    \Procedure{Thresholding}{$\mat{H}$, $\mat{S}$, $\epsilon$}
    \State $(\mat{V},\mat{D}) \leftarrow \mathtt{eig}(\mat{S})$
    \State $I \leftarrow \{ i : \mat{D}_{ii} > \epsilon \}$
    \State $\mat{V} \leftarrow \mat{V}(:,I)$
    \State \Return smallest eigenvalue of $(\mat{V}^{\rev{*}}\mat{H}\mat{V},\mat{V}^{\rev{*}}\mat{S}\mat{V})$
    \EndProcedure
  \end{algorithmic}
\end{algorithm}

Despite appearing quite natural, \rev{there are examples where thresholding fails to work and is thus not appropriate for arbitrary Hermitian definite generalized eigenvalue problems}.
For instance, if one applies thresholding with parameter $\epsilon$ to the pair 
\begin{equation}\label{eq:thresholding_bad}
  \mat{H} = \twobytwo{1}{\epsilon}{\epsilon}{\epsilon^2}, \quad \mat{S} = \twobytwo{1}{0}{0}{\epsilon^2},
\end{equation}
one recovers an eigenvalue of $1$, which is far from the genuine eigenvalues $0$ and $2$ of the pair.
Even with the existence of bad examples like \eqref{eq:thresholding_bad}, thresholding appears to be quite reliable at filtering out noise and dealing with the ill-conditioning of the overlap matrix \rev{for QSD-derived pairs $(\mat{H},\mat{S})$} in our experiments, with similar results being observed for an SVD-based truncation strategy in \cite[{\S}II.F]{KMC+21}.
Despite truncation strategies such as canonical orthogonalization and Fix--Heiberger having a fifty year history, we are unaware of any general theory of why these methods work.

\subsection{Overview and main results}
\label{sec:overv-main-results}

We aim to elucidate why the QSD algorithm works when combined with the thresholding procedure, undeterred by the presence of negative examples such as \eqref{eq:wilkinson} and \eqref{eq:thresholding_bad}.
Our main three results are
\begin{enumerate}[label=(\roman*)]
\item In the absence of noise and with an appropriate choice of the time sequence, the QSD procedure with thresholding is accurate.
  \rev{This provides a positive answer to the question\ \ref{item:question_1} in the error analysis of the QSD method due to the  Krylov subspace approximation in the presence of thresholding.}  
 \label{item:1}
\item The thresholded problem is stable under noise \rev{in the sense} that the thresholded problem we solve from the noise-corrupted \rev{pair $(\mat{\tilde{H}},\mat{\tilde{S}})$} is close to the un-perturbed \rev{pair $(\mat{H},\mat{S})$}.
  \label{item:2}
\item If the thresholded problem and its noisy perturbation are sufficiently close (as we establish with the previous result), the well-conditioned and well-separated eigenvalues of the thresholded problem are accurately computed in the presence of noise.
  \rev{Results \ref{item:2} and \ref{item:3} provide a positive answer to the question\ \ref{item:question_2} in the error analysis of the QSD method due to the classical solution of the noisy generalized eigenvalue problem.} \label{item:3}
\end{enumerate}
Together, these results paint a reasonably complete picture of why the QSD algorithm works in the presence of noise when thresholding is used.
To our knowledge, this is the first work providing rigorous analysis of the theoretical efficacy of QSD-type algorithms, both in the noise-free and noisy settings.

Our \rev{first main result} can be summarized informally as follows (a formal statement is presented as Theorem~\ref{thm:main_theorem_formal}):
\begin{inftheorem} \label{thm:main_theorem}
  Suppose the thresholding procedure (Algorithm~\ref{alg:thresholding}) is applied to the perturbed pair $(\mat{\tilde{H}},\mat{\tilde{S}}) = (\mat{H}+\mat{\Delta}_{\mat{H}},\mat{S}+\mat{\Delta}_{\mat{S}})$, which are Hermitian matrices of size $n$.
  Consider the thresholded matrix pair $(\mat{A},\mat{B}) := (\mat{V}_{>\epsilon}^{\rev{*}}\mat{H}\mat{V}_{>\epsilon},\mat{V}_{>\epsilon}^{\rev{*}}\mat{S}\mat{V}_{>\epsilon})$ and let $E_0$ be its least eigenvalue.
  Assume $E_0$ is sufficiently well-separated from other eigenvalues of $(\mat{A},\mat{B})$, let $d_0^{-1}$ denote the condition number of the \emph{eigenangle} $\tan^{-1} E_0$, and suppose the perturbations $\mat{\Delta}_{\mat{H}}$ and $\mat{\Delta}_{\mat{S}}$ have spectral norm not exceeding $\eta$.
  There exists a constant $0 \le \alpha \le 1/2$ such that the recovered eigenvalue $\tilde{E}_0$ from the noise-perturbed pair $(\mat{\tilde{H}},\mat{\tilde{S}})$ using threshold parameter $\epsilon = \Theta\mleft(\eta^{\tfrac{1}{1+\alpha}}\mright)$ satisfies the bound
  \begin{equation} \label{eq:main_eig_informal}
      \mleft| \tan^{-1} \tilde{E}_0^{\rm th} - \tan^{-1} E_0^{\rm th} \mright| \le \order\mleft( d_0^{-1} \eta^{\tfrac{1}{1+\alpha}} \mright).
  \end{equation}
  The implicit constant in the $\order$-notation depends on the eigenvalues of $(\mat{H},\mat{S})$, the spectrum of $\mat{S}$, and $n$.
\end{inftheorem}
Our result shows that, \rev{using the empirically observed value $\alpha = 1/4$ (see section~\ref{sec:value-alpha})}, we are able to recover the smallest eigenvalue \rev{of $(\mat{H},\mat{S})$} (or more specifically its arctangent) with error proportional to $\eta^{4/5}$ times its condition number.
Given examples \eqref{eq:wilkinson} and \eqref{eq:thresholding_bad} showing arbitrarily large errors for small perturbations of nearly singular generalized eigenvalue problems and thresholding, the fact that we are able to obtain any nontrivial error bounds for the QSD algorithm with thresholding may be regarded as surprising.

\rev{Our second main result provides an} end-to-end bound \rev{for the QSD} method.

\begin{inftheorem} \label{thm:main_theorem_big}
  Let $\Delta E_j$ denote the difference between the $j$th smallest and the smallest eigenvalue $E_0$ of $\oper{H}$ and let $\gamma_0$ denote the inner product between the initial state ${\rev{\vec{\varphi}_0}}$ and the true ground-state eigenvector of $\oper{H}$.
  Let $(\mat{H},\mat{S})$ denote the noise-free output of the QSD algorithm for a particular choice of time step, and instate the notation and assumptions of Informal Theorem~\ref{thm:main_theorem}.
  Then
  \begin{equation*}
      \mleft| \tan^{-1} \tilde{E}_0^{\rm th} - \tan^{-1} E_0 \mright| \le \order\mleft( \frac{1-|\gamma_0|^2}{|\gamma_0|^2} \e^{-n\order\mleft(\tfrac{\Delta E_1}{\Delta E_{N-1}}\mright)} + \mleft[ \frac{\Delta E_{N-1}}{|\gamma_0|^2} + d_0^{-1}\mright] \eta^{\tfrac{1}{1+\alpha}} \mright).
  \end{equation*}
\end{inftheorem}

The remainder of this paper is organized as follows.
For expository reasons, we present our main results in the reverse order outlined here.
\rev{Section~\ref{sec:pert-analys-thresh} discusses perturbation analysis for the thresholded problem, leading to} a formalization Theorem~\ref{thm:main_theorem_formal} of Informal Theorem~\ref{thm:main_theorem} in section~\ref{sec:putting_pieces_together}.
\rev{Section~\ref{sec:analys-quant-subsp} discusses Rayleigh--Ritz errors due to approximation by the finite-dimensional unitary Krylov subspace and thresholding procedure.}
We then present additional results in section~\ref{sec:addit-resul} which are independent of the rest of the presentation.
We draw particular attention to Theorem~\ref{thm:thresholding}, which shows that thresholding applied to a general pair $(\mat{H},\mat{S})$ recovers the least eigenvalue accurately if it is well-conditioned.
This does not contradict the bad example \eqref{eq:thresholding_bad} since both its eigenvalues are ill-conditioned with condition numbers $\Theta(\epsilon^{-1})$.
We conclude with numerical experiments (section~\ref{sec:numer-exper}) and conclusions (section~\ref{sec:conclusions}).

\section{Perturbation Analysis for the Thresholded Problem}
\label{sec:pert-analys-thresh}

\rev{In this section, we analyze the effects of noise on the solution of the generalized eigenvalue problem \eqref{eq:gep} using thresholding Algorithm~\ref{alg:thresholding}.}
\rev{The main result of this section is} that well-separated, well-conditioned eigenvalues of the thresholded problem \rev{can be} recovered  accurately in the presence of noise.
Together with section~\ref{sec:analys-quant-subsp} which analyzes both the Rayleigh--Ritz and thresholding errors, this comprises a fairly complete explanation for the success of the QSD algorithm when implemented with thresholding.

Let $\mat{H}$ and $\mat{S}$ denote the exact outputs of the QSD algorithm \eqref{eq:HS} and $\mat{V}$ the eigenvectors of $\mat{S}$ with eigenvalues \rev{greater than} $\epsilon$.
Dependence of $\mat{V} \mathbin{\rev{:=}} \mat{V}_{>\epsilon}$ on $\epsilon$, as in the introduction, has been suppressed for conciseness.
The thresholded problem is described by the pair $(\mat{A},\mat{B}) := (\mat{V}^{\rev{*}}\mat{H}\mat{V},\mat{V}^{\rev{*}}\mat{S}\mat{V})$.\footnote{\rev{To make the output of the thresholding procedure unambiguous, we assume eigenvectors are arranged left-to-right in decreasing order of the corresponding eigenvalues. The ordering convention does not effect the outputs of the thresholding procedure Algorithm~\ref{alg:thresholding}.}}
When implemented on a quantum computer, $\mat{H}$ and $\mat{S}$ are corrupted by noise as $\mat{\tilde{H}} := \mat{H} + \mat{\Delta}_{\mat{H}}$ and $\mat{\tilde{S}} := \mat{S} + \mat{\Delta}_{\mat{S}}$.
As a simple measure of the size of the perturbation, we introduce
\begin{equation} \label{eq:eta}
  \eta := \sqrt{\eta_{\mat{H}}^2 + \eta_{\mat{S}}^2} := \sqrt{ \norm*{\mat{\Delta}_{\mat{H}}}^2 + \norm*{\mat{\Delta}_{\mat{S}}}^2 },
\end{equation}
which represents the noise level.
In principle, one could undertake a careful analysis of the different sources of error (e.g., discretization, gate, and sampling) to obtain probabilistic bounds on $\eta$.
\rev{(We provide such a bound for the sampling error alone in section~\ref{sec:toeplitz}.)}
For \rev{now}, we shall just assume $\eta$ or a good bound for it is known, and the threshold level $\epsilon$ is chosen to be (at least) larger than $\eta$.
With the perturbations $\mat{\tilde{H}}$ and $\mat{\tilde{S}}$ in hand, the \rev{practitioner} computes the large-eigenvalue eigenvectors $\mat{\tilde{V}}$ of the perturbation $\mat{\tilde{S}}$, and constructs the perturbed thresholded problem $(\mat{\tilde{A}},\mat{\tilde{B}}) := (\mat{\tilde{V}}^{\rev{*}}\mat{\tilde{H}}\mat{\tilde{V}},\mat{\tilde{V}}^{\rev{*}}\mat{\tilde{S}}\mat{\tilde{V}})$.
We denote the dimension of \rev{$\mat{\tilde{A}}$} and \rev{$\mat{\tilde{B}}$} as $q$.

We hope to show that the smallest eigenvalue of the pair $(\mat{\tilde{A}},\mat{\tilde{B}})$---i.e., our computed approximation to the ground state energy---is close to the smallest eigenvalue of $(\mat{A},\mat{B})$.
Unfortunately, there are a number of reasons to worry this might not be the case.
First, even if $(\mat{\tilde{H}},\mat{\tilde{S}})$ is close to $(\mat{H},\mat{S})$, it is still possible that $(\mat{\tilde{A}},\mat{\tilde{B}})$ is not close to $(\mat{A},\mat{B})$.
A small perturbation in just $\mat{S}$ can lead to a large perturbation of $\mat{A}$: \rev{For a small parameter $\eta > 0$,}
\begin{align}
  \mat{H} &= \twobytwo{20}{0}{0}{1},\: \mat{S} = \twobytwo{1}{0}{0}{1-\tfrac{\eta}{2}}& &\xrightarrow{\mbox{$\eta$ error}}& \mat{\tilde{H}} &= \twobytwo{20}{0}{0}{1},\: \mat{\tilde{S}} = \twobytwo{1}{0}{0}{1+\tfrac{\eta}{2}} \label{eq:perturbation_threshold} \\
  \mat{A} &= \twobytwo{20}{0}{0}{1},\: \mat{B} = \twobytwo{1}{0}{0}{1-\tfrac{\eta}{2}}& &\xrightarrow{\phantom{\mbox{$\eta$ error}}}& \mat{\tilde{A}} &= \twobytwo{1}{0}{0}{20},\: \mat{\tilde{B}} = \twobytwo{\rev{1+\tfrac{\eta}{2}}}{0}{0}{\rev{1}}. \nonumber
\end{align}
Additionally, $\mat{\tilde{A}}$ and $\mat{A}$ can even be of different sizes if the perturbation causes the number of eigenvalues \rev{larger than} $\epsilon$ to change.
Fortunately, \eqref{eq:perturbation_threshold} suggests that the potential for small errors in $\mat{S}$ to magnify into large errors in $\mat{A}$ \rev{might} have a benign source, with the error in this example caused simply by a reordering of the eigenvectors.
The eigenvalues of the pair $(\mat{\tilde{A}},\mat{\tilde{B}})$ are indifferent to a symmetric reordering of its rows and columns or\rev{,} more generally\rev{,} a \rev{$*$-}conjugation $(\mat{\tilde{A}},\mat{\tilde{B}}) \mapsto (\mat{W}^{\rev{*}}\mat{\tilde{A}}\mat{W},\mat{W}^{\rev{*}}\mat{\tilde{B}}\mat{W})$.
\rev{Thus, it is} sufficient for purposes of analysis to show that $(\mat{\tilde{A}},\mat{\tilde{B}})$ are close to $(\mat{A},\mat{B})$ after $(\mat{\tilde{A}},\mat{\tilde{B}})$ is replaced by an appropriate \rev{$*$-}conjugation.

Assume this issue is addressed, and we obtain a \rev{$*$-}conjugation of $(\mat{\tilde{A}},\mat{\tilde{B}})$ that is close to $(\mat{A},\mat{B})$.
Classical \rev{worst-case} perturbation theory still paints a grim portrait on the sensitivities of the eigenvalues of the thresholded pair $(\mat{A},\mat{B})$.
After possibly replacing $(\mat{\tilde{A}},\mat{\tilde{B}})$ by a \rev{$*$-}conjugation, let $(\mat{\tilde{A}},\mat{\tilde{B}}) = (\mat{A}+\mat{\Delta}_{\mat{A}},\mat{B}+\mat{\Delta}_{\mat{B}})$.
A measure of the difference between $(\mat{A},\mat{B})$ and $(\mat{\tilde{A}},\mat{\tilde{B}})$ is given by 
\begin{equation} \label{eq:chi}
  \chi := \sqrt{ \norm*{\mat{\Delta}_{\mat{A}}}^2 + \norm*{\mat{\Delta}_{\mat{B}}}^2}.
\end{equation}
Let $E$ be the least eigenvalue of $(\mat{A},\mat{B})$, and $\tilde{E}$ be the least eigenvalue of $(\mat{\tilde{A}},\mat{\tilde{B}})$.
The classical perturbation theorem of Stewart \cite[Thm.~3.2]{Ste79} can only show that
\begin{equation} \label{eq:stewart}
  \mleft| \tan^{-1} \tilde{E} - \tan^{-1} E \mright| \le \sin^{-1} \frac{\chi}{c(\mat{A},\mat{B})} \le \sin^{-1} \frac{\chi}{\epsilon}.
\end{equation}
Here,
\begin{equation} \label{eq:crawford}
  c(\mat{A},\mat{B}) := \min_{\norm{\vec{x}} = 1} \sqrt{ \mleft(\vec{x}^{\rev{*}}\mat{A}\vec{x}\mright)^2 + \mleft(\vec{x}^{\rev{*}}\mat{B}\vec{x}\mright)^2 }
\end{equation}
is the Crawford number, \rev{which} is only guaranteed to be \rev{larger than} $\epsilon$ under the standing assumptions.
This leads to pessimistic error bound\rev{s} on the order of $\chi/\epsilon$.
We would much prefer a bound which scales like $\chi$ times the condition number of $\tan^{-1} E$, without an explicit $\epsilon$ dependence.

In the rest of this section, we address these challenges.
The analysis has two parts.
In the first part, we use the Davis--Kahan $\sin\mat{\Theta}$ theorem to show that, after replacing by a \rev{$*$-}conjugation, $(\mat{\tilde{A}},\mat{\tilde{B}})$ is $\chi \le \order(\eta/\epsilon^\alpha)$-close to $(\mat{A},\mat{B})$, where $0\le \alpha \le 1/2$ is a constant.
In particular, $\alpha$ is assured to be no more than $1/2$, with $\alpha = 1/4$ appearing to be more representative in numerical experiments.
For the second part, we use the perturbation theory of Mathias and Li \cite{ML04} to improve on Stewart's bound \eqref{eq:stewart} in the case when the eigenvalue of interest is well-conditioned, obtaining the desired dependence on $\chi$ rather than $\chi/\epsilon$.
We present these pieces in reverse order.

\subsection{Eigenvalue Perturbation Bounds}
\label{sec:eigenv-pert-bounds}

Suppose that, potentially after a redefining $(\mat{\tilde{A}},\mat{\tilde{B}})$ to a \rev{$*$-}conjugation, $(\mat{\tilde{A}},\mat{\tilde{B}})$ and $(\mat{A},\mat{B})$ are separated by a small distance $\chi$, as defined in \eqref{eq:chi}.
We shall show that 
the perturbation theory developed by Mathias and Li \cite{ML04} allows us to significantly improve on the bound \eqref{eq:stewart} furnished by Stewart's theory.
The beauty of the Mathias--Li theory is that the Crawford number in \eqref{eq:stewart} can be replaced by a quantity related to the conditioning of $E$, provided a spectral gap condition is satisfied.
We begin with a mildly specialized version of \cite[Thm.~3.3]{ML04}:

\begin{fact} \label{fact:mathias_li}
  Let $(\mat{A},\mat{B})$ be a pair of $q\times q$ Hermitian matrices with all eigenvalues of $\mat{B}$ larger than $\epsilon$.
  Consider perturbations $(\mat{\tilde{A}},\mat{\tilde{B}}) := (\mat{A} + \mat{\Delta}_{\mat{A}},\mat{B}+\mat{\Delta}_{\mat{B}})$ where $\mat{\Delta}_{\mat{A}}$ and $\mat{\Delta}_{\mat{B}}$ are Hermitian matrices and $\chi$ is defined in \eqref{eq:chi} \rev{and satisfies} $q\chi \le \epsilon$.
  Let $E_0,\ldots,E_{q-1}$ be the eigenvalues of $(\mat{A},\mat{B})$ with unit-norm eigenvectors $\vec{x}_0,\ldots,\vec{x}_{q-1}$.
  Define
  \begin{equation} \label{eq:lower_upper}
    \ell_j = \tan^{-1} E_j - \sin^{-1} \frac{q\chi}{d_j},\quad u_j = \tan^{-1} E_j + \sin^{-1} \frac{q\chi}{d_j}.
  \end{equation}
  where
  \begin{equation} \label{eq:eigenangle_condition}
    d_j := |\vec{x}_j^{{\rev{*}}}(\mat{A}+\iu \mat{B})\vec{x}_j|.
  \end{equation}  
  Let $\big\{\ell^\uparrow_j\big\}_{j=0}^{q-1}$ and $\big\{u^\uparrow_j\big\}_{j=0}^{q-1}$ denote the increasing rearrangements of the bounds $\mleft\{\ell_j\mright\}_{j=0}^{q-1}$ and $\mleft\{u_j\mright\}_{j=0}^{q-1}$.
  Then with $\tilde{E}_0\le \tilde{E}_1\le \cdots \le \tilde{E}_{q-1}$ the eigenvalues of $(\mat{\tilde{A}},\mat{\tilde{B}})$, the following bound holds:
  \begin{equation*}
    \ell^\uparrow_j \le \tan^{-1} \tilde{E}_j \le u^\uparrow_j.
  \end{equation*}
\end{fact}

\begin{figure}[t]
  \centering
    \begin{subfigure}[b]{0.47\textwidth}
    \centering
    \includegraphics[width=\textwidth]{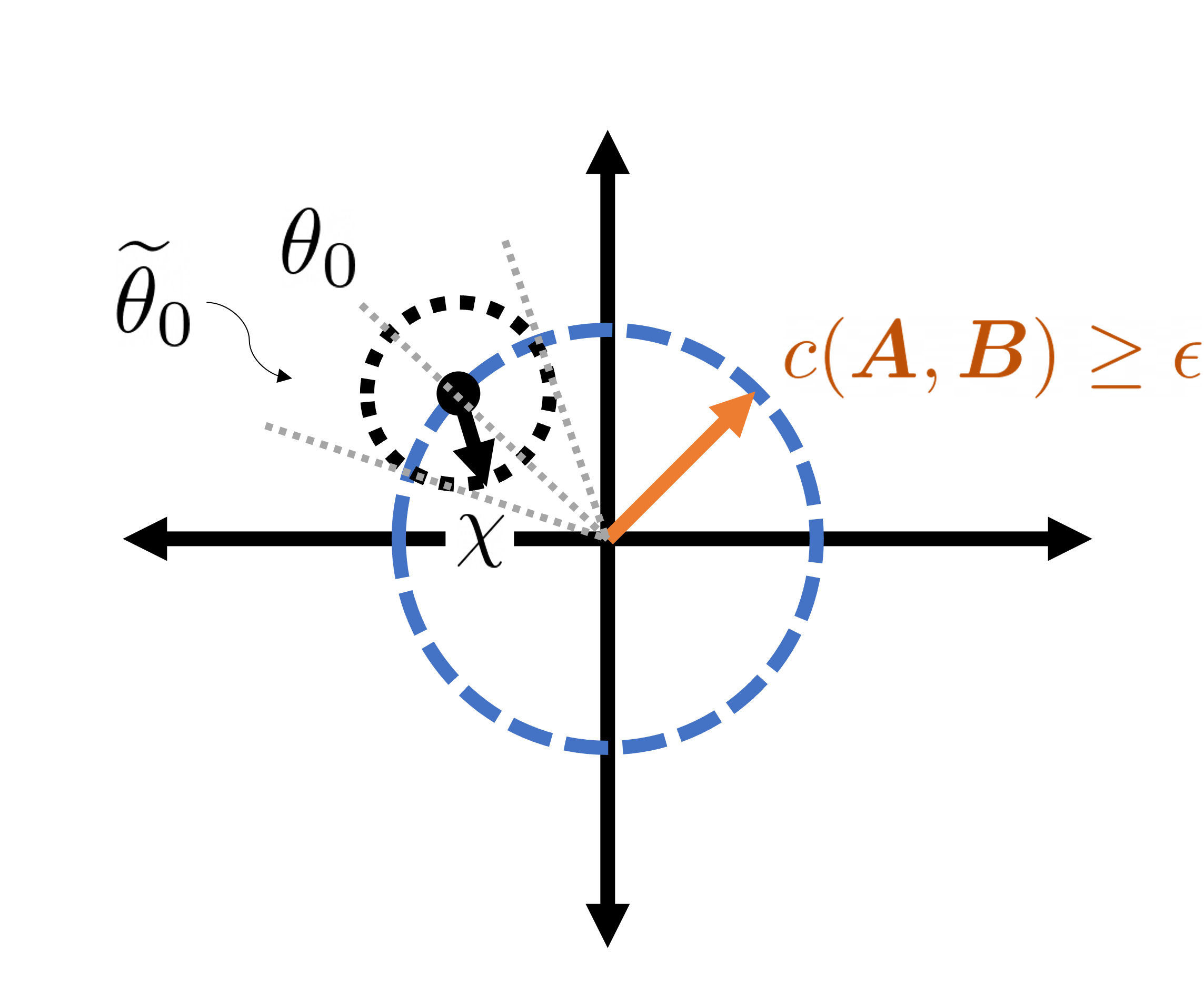}
    \caption{Stewart}\label{fig:stewart}
  \end{subfigure}
  ~
  \begin{subfigure}[b]{0.47\textwidth}
    \centering
    \includegraphics[width=\textwidth]{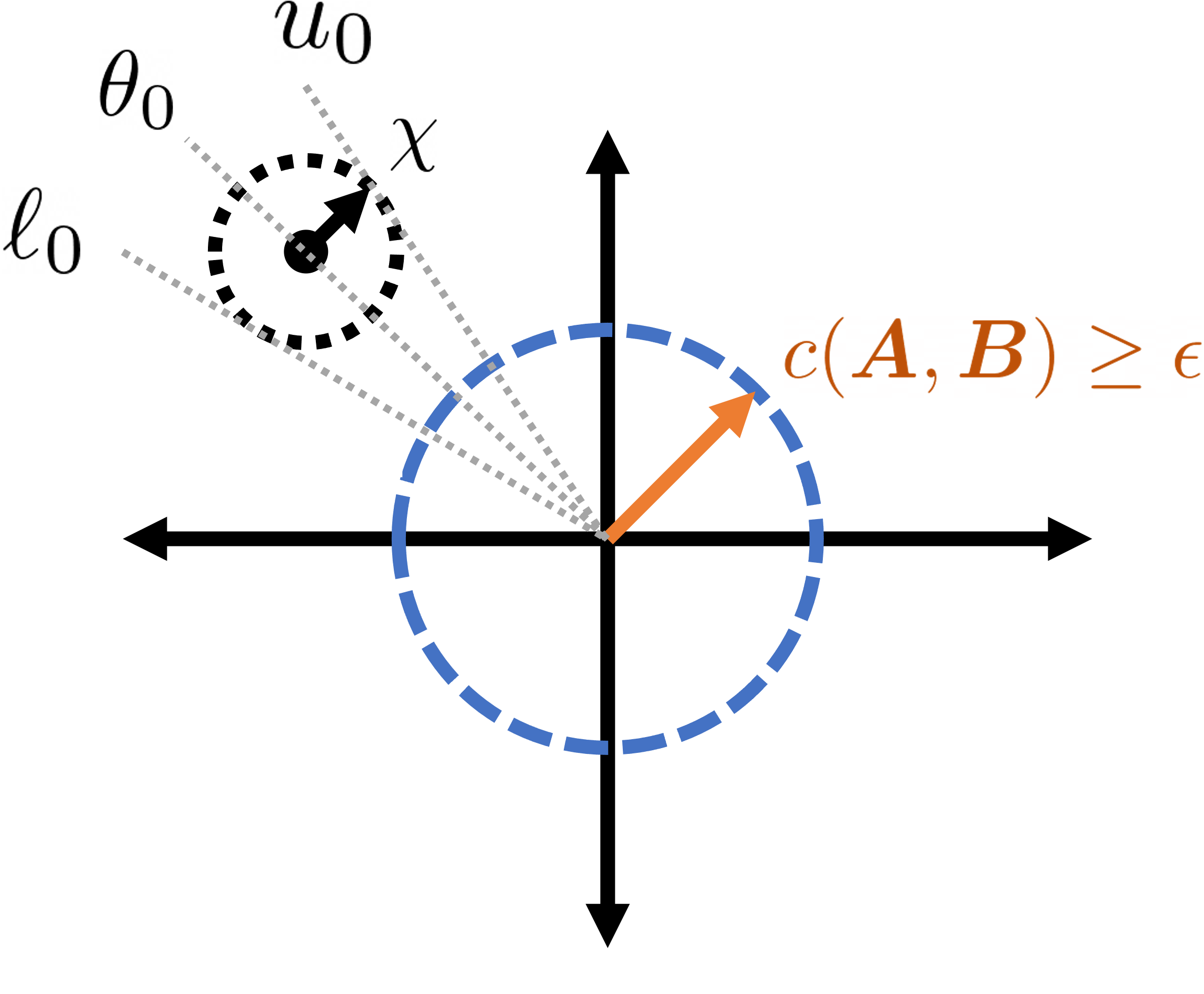}
    \caption{Mathias--Li}\label{fig:mathias_li}
  \end{subfigure}  

  \caption{Pictorial comparison of the proofs of Stewart's bound \eqref{eq:stewart} and Mathias and Li's bound \eqref{eq:mathias_li}.}
  \label{fig:perturbation_theorem_comparison}
\end{figure}

We show a pictorial comparison of the proof of Stewart's bound \eqref{eq:stewart} and the Mathias--Li bound \eqref{eq:mathias_li} in Figure~\ref{fig:perturbation_theorem_comparison}.
The perturbation theory of the definite generalized eigenvalues is naturally phrased in terms of the eigenangle $\theta_j := \tan^{-1} E_j$, which represents the angle of the ray through the complex number $\vec{x}_j^{\rev{*}}(\mat{A}+\iu \mat{B})\vec{x}_j$ and the positive imaginary axis.
In Stewart's theory, one argues that, for a unit vector $\vec{x}$, the complex number $z := \vec{x}^{\rev{*}}(\mat{A}+\iu \mat{B})\vec{x}$ must be a distance $c(\mat{A},\mat{B})$ from the origin and the perturbed point $\tilde{z} := \vec{x}^{\rev{*}}(\mat{\tilde{A}}+\iu \mat{\tilde{B}})\vec{x}$ is a distance at most $\chi$ from $z$.
In view of a variational characterization Stewart proved for the eigenangles \cite[Thm.~3.1]{Ste79}, it follows that the eigenangles change by at most $\sin^{-1} (\chi/c(\mat{A},\mat{B}))$.
Mathias and Li instead consider the points $z_j := \vec{x}_j^{\rev{*}}(\mat{A}+\iu \mat{B})\vec{x}_j$ for the unit-norm eigenvectors $\vec{x}_j$.
A disk of radius $\chi$ centered at $z_j$ is enclosed by rays with angles $\{\theta_j \pm \sin^{-1} (\chi/d_j)\} = \{\ell_j,u_j\}$.
Using Stewart's variational principle, Mathias and Li prove that, while the perturbed eigenangle $\tilde{\theta}_j := \tan^{-1} \tilde{E}_j$ need not lie within $[\ell_j,u_j]$, it must lie in $[\ell_j^\uparrow,u_j^\uparrow]$.

A consequence of Mathias and Li's analysis (see \cite[Eq.~(3.6)]{ML04}) is that $\ell_j \le \tilde{\theta}_j \le u_j$ does hold if there is a large enough gap between $\theta_j$ and other eigenangles relative to the size of the perturbation.
This bound $\ell_j \le \tilde{\theta}_j \le u_j$ is nearly as good a bound as one could hope since $d_j^{-1}$ is the condition number of $\theta_j$ \rev{\cite[Thm.~VI.2.2]{SS90}}.

\begin{corollary} \label{cor:mathias_li}
  Instate the notation and assumptions of Fact~\ref{fact:mathias_li}.
  Suppose that $E_j$ satisfies the gap condition
  \begin{equation} \label{eq:gap_condition}
    \min\mleft(\tan^{-1} E_j - \tan^{-1} E_{j-1}, \:\tan^{-1}E_{j+1} - \tan^{-1} E_j\mright) \ge \sin^{-1} \frac{q\chi}{\epsilon} - \sin^{-1} \frac{q\chi}{d_j},
  \end{equation}
  where the first or second term of the minimum can be ignored if $j = 0$ or $j = q-1$, respectively.
  Then
  \begin{equation} \label{eq:mathias_li}
    \mleft| \tan^{-1} \tilde{E}_j - \tan^{-1} E_j \mright| \le \sin^{-1} \frac{q\chi}{d_j}.
  \end{equation}
\end{corollary}

A couple comments are in order before we proceed with the proof of Corollary~\ref{cor:mathias_li}.
We have purged the suboptimal factor $\chi/\epsilon$ from the eigenangle bound \eqref{eq:mathias_li} and replaced it with the often much smaller quantity $\chi/d_j$ (tempered by a dimensional factor).
However, $\chi/\epsilon$ remains, just in the gap condition \eqref{eq:gap_condition} (and the hypothesis $q\chi \le \epsilon$).
If the eigenangle gap on the left-hand side of \eqref{eq:gap_condition} is reasonably large, then $\epsilon$ must merely be a modest multiple of $\chi$ for \eqref{eq:gap_condition} to be satisfied.

\begin{proof}[Proof of Corollary~\ref{cor:mathias_li}]
  Denote $\widetilde{\theta}_i := \tan^{-1} \tilde{E}_i$ and $\theta_i := \tan^{-1} E_i$ for every $i=0,1,\ldots,q-1$.
  We shall prove the upper bound $\tilde{\theta}_j - \theta_j \le \sin^{-1} (q\chi/d_j)$ with the corresponding lower bound being proven in exactly the same way.
  We shall do this by showing $u^\uparrow_j \le u_j$  under the gap assumption \eqref{eq:gap_condition}.
  To do this, we shall show that $u_i \le \theta_j + \sin^{-1}(q\chi/d_j)$ for every $0\le i < j$.
  If $j = 0$, then no such $i$ exists and the upper bound is automatically true.
  We thus continue in the case $j>0$.
  
  Fix $0\le i < j$.
  Since every eigenvalue of $\mat{B}$ is at least $\epsilon$, we have that $d_i = |\vec{x}_i^{\rev{*}}(\mat{A}+\iu \mat{B})\vec{x}_i| \ge \vec{x}_i^{\rev{*}}\mat{B}\vec{x}_i > \epsilon$.
  Thus, we have $u_i \le \theta_i + \sin^{-1} (q\chi/\epsilon)$.
  Since $E_i \le E_{j-1}$, we have $\theta_i = \tan^{-1} E_i \le \tan^{-1} E_{j-1}$.
  Thus,
  \begin{equation*}
    u_i \le \tan^{-1} E_i + \sin^{-1} \frac{q\chi}{\epsilon} \le \tan^{-1} E_{j-1} + \sin^{-1} \frac{q\chi}{\epsilon} \le \tan^{-1} E_j + \sin^{-1} \frac{q\chi}{d_j}
  \end{equation*}
  by \eqref{eq:gap_condition}.
  From this we conclude $u^\uparrow_j \le u_j$ so by Fact~\ref{fact:mathias_li}, $\tilde{\theta}_j \le u^\uparrow_j \le u_j = \theta_j + \sin^{-1} (q\chi/d_j)$.
\end{proof}

\subsection{How Do Perturbations Affect the Thresholded Problem?\nopunct}
\label{sec:how-do-perturbations}

In this section, we seek to understand how perturbations of the pair $(\mat{H},\mat{S})$ affect the thresholded problem $(\mat{A},\mat{B})$.
As the example \eqref{eq:perturbation_threshold} shows, it need not be the case that $(\mat{\tilde{A}},\mat{\tilde{B}})$ is close to $(\mat{A},\mat{B})$ if $(\mat{\tilde{H}},\mat{\tilde{S}})$ is close to $(\mat{H},\mat{S})$.
Thus, in view of the fact that \rev{$*$-}conjugations do not change the eigenvalues of a matrix pair \cite[Thm.~VI.1.8]{SS90}, our goal will be to show there exists a nonsingular matrix $\mat{W}$ such that $(\mat{W}^{\rev{*}}\mat{\tilde{A}}\mat{W},\mat{W}^{\rev{*}}\mat{\tilde{B}}\mat{W})$ is close to $(\mat{A},\mat{B})$.
An instrumental tool in this goal will be the Davis--Kahan $\sin\mat{\Theta}$ theorem \cite{DK70}, which we state a somewhat less general version here for reference.

\begin{fact}[Davis--Kahan $\sin \mat{\Theta}$ Theorem; see also {\cite[Thm.\ VII.3.1]{Bha97}}] \label{fact:davis_kahan}
  Consider Hermitian matrices $\mat{M}$ and $\mat{\tilde{M}}$ and let $\mat{\Pi}$ and $\mat{\tilde{\Pi}}$ be the spectral projectors of $\mat{M}$ and $\mat{\tilde{M}}$ associated with collections of eigenvalues of $\mat{M}$ within an interval $[a,b]$ and of $\mat{\tilde{M}}$ outside the interval $[a-\delta,b+\delta]$ respectively.
  Then $\norm{\mat{\Pi}\mat{\tilde{\Pi}}} \le \norm{\mat{\tilde{M}} - \mat{M}}/\delta$.
\end{fact}

\rev{We begin with} some notation.
Let $\vec{v}_1,\ldots,\vec{v}_n$ denote the eigenvectors of $\mat{S}$ with associated eigenvalues $\lambda_1,\ldots,\lambda_n$ \rev{and} $\mat{\Pi}$ denote the spectral projector associated with eigenvalues of $\mat{S}$ which are larger than $\epsilon$.
We denote by $m$ the critical index for which $\lambda_m > \epsilon \ge \lambda_{m+1}$.
All quantities with tildes shall denote quantities defined as above for the perturbed problem.

We begin with the following simple bound:

\begin{proposition} \label{prop:conjugation}
  Suppose that
  \begin{align}
      \norm{\mat{\Pi}\mat{H}\mat{\Pi} - \mat{\tilde{\Pi}}\mat{H}\mat{\tilde{\Pi}}} &\le \chi_{\mat{H}}, \label{eq:H_projection_error}\\
      \norm{\mat{\Pi}\mat{S}\mat{\Pi} - \mat{\tilde{\Pi}}\mat{S}\mat{\tilde{\Pi}}} &\le \chi_{\mat{S}}, \label{eq:S_projection_error}
  \end{align}
  and suppose that $\mat{\tilde{S}}$ and $\mat{S}$ have the same number of eigenvalues larger than $\epsilon$.
  Then for the nonsingular matrix $\mat{W} := \mat{\tilde{V}}^{\rev{*}}\mat{V}$,
  \begin{equation*}
    \sqrt{ \norm{\mat{W}^{\rev{*}}\mat{\tilde{A}}\mat{W} - \mat{A}}^2 + \norm{\mat{W}^{\rev{*}}\mat{\tilde{B}}\mat{W} - \mat{B}}^2 } \le \sqrt{ (\chi_{\mat{H}}+\eta_{\mat{H}})^2 + (\chi_{\mat{S}}+\eta_{\mat{S}})^2 }.
  \end{equation*}
\end{proposition}

\rev{This result gives} a bound on $\chi$ as defined in \eqref{eq:chi} if one redefines $(\mat{\tilde{A}},\mat{\tilde{B}})$ to its \rev{$*$-}conjugation by $\mat{W}$.

\begin{proof}
  First, note that $\mat{W}$ is nonsingular because it is the product of a matrix with full row rank and full column rank.
  Note also that $\norm{\mat{V}},\norm{\mat{\tilde{V}}},\norm{\mat{\Pi}},\norm{\mat{\tilde{\Pi}}} = 1$ since $\mat{V}$ and $\mat{\tilde{V}}$ have orthonormal columns.
  The result then follows immediately by the bound
  \begin{align*}
    \norm{\mat{W}^{\rev{*}}\mat{\tilde{A}}\mat{W} - \mat{A}}
    &= \norm{\mat{V}^{\rev{*}} \mat{\tilde{V}}\mat{\tilde{V}}^{\rev{*}}\mat{\tilde{H}}\mat{\tilde{V}}\mat{\tilde{V}}^{\rev{*}}\mat{V} - \mat{V}^{\rev{*}}\mat{H}\mat{V}} = \norm{\mat{V}^{\rev{*}}(\mat{\tilde{\Pi}}\mat{\tilde{H}}\mat{\tilde{\Pi}} - \mat{\Pi}\mat{H}\mat{\Pi})\mat{V}} \\
    &\le \norm{\mat{V}^{\rev{*}}(\mat{\tilde{\Pi}}\mat{H}\mat{\tilde{\Pi}} - \mat{\Pi}\mat{H}\mat{\Pi})\mat{V}} + \norm{\mat{V}^{\rev{*}}\mat{\tilde{\Pi}}(\mat{\tilde{H}} - \mat{H})\mat{\tilde{\Pi}}\mat{V}} \\
    &\le \norm{\mat{\tilde{\Pi}}\mat{H}\mat{\tilde{\Pi}} - \mat{\Pi}\mat{H}\mat{\Pi}} + \norm{\mat{\tilde{H}} - \mat{H}} \le \chi_{\mat{H}} + \eta_{\mat{H}}
  \end{align*}
  and similarly for $\norm{\mat{W}^{\rev{*}}\mat{\tilde{B}}\mat{W} - \mat{B}}$.
\end{proof}

Note that a necessary condition for the hypotheses of our error bound Corollary~\ref{cor:mathias_li} is for the distance $\chi$ \eqref{eq:chi} between $(\mat{\tilde{A}},\mat{\tilde{B}})$ and $(\mat{A},\mat{B})$ to be strictly smaller than $\epsilon$.
This forecloses our ability to use simple bounds for, e.g., \eqref{eq:S_projection_error} such as 
\begin{equation*}
  \norm{\mat{\Pi}\mat{S}\mat{\Pi} - \mat{\tilde{\Pi}}\mat{S}\mat{\tilde{\Pi}}} \le \norm{\mat{\Pi}\mat{S}\mat{\Pi} - \mat{S}} + \norm{\mat{S} - \mat{\tilde{S}}} + \norm{\mat{\tilde{S}}- \mat{\tilde{\Pi}}\mat{\tilde{S}}\mat{\tilde{\Pi}}} \le 2(\epsilon + \eta).
\end{equation*}
We thus seek bounds \rev{of the form} \eqref{eq:H_projection_error} and \eqref{eq:S_projection_error} without an additive $\order(\epsilon)$ term. 

We begin with the more challenging of the two bound\rev{s}, namely \eqref{eq:H_projection_error}.
Certainly, we should not expect a meaningful bound \eqref{eq:H_projection_error} if $\mat{H}$ and $\mat{S}$ have no relation to each other.
For the sake of generality, we shall perform analysis under the assumption that $(\mat{H},\mat{S})$ obey a weighted geometric mean \rev{inequality} of the form 
\begin{equation} \label{eq:geometric_mean}
  \mleft| \vec{v}_i^{\rev{*}} \mat{H}\vec{v}_j \mright| \le \mu\, \rev{\min(\lambda_i,\lambda_j)^{1-\alpha}\max(\lambda_i,\lambda_j)^{\alpha}} \quad \mbox{for all } 1\le i,j\le n,
\end{equation}
where $0\le \alpha\le 1/2$ and $\mu > 0$ are constants.
While this may appear strange, \eqref{eq:geometric_mean} necessarily holds with $\mu = \max |\Lambda(\mat{H},\mat{S})|$ and $\alpha = 1/2$ by a direct application of the Courant--Fischer principle for generalized eigenvalue problems \cite[Cor.~VI.1.16]{SS90}.
Numerical experiments suggest that $\alpha = 1/4$ and $\mu \approx \max |\Lambda(\mat{H},\mat{S})|$ appear to be more revealing of the empirically observed values of $|\vec{v}_i^{\rev{*}}\mat{H}\vec{v}_j|$ for $(\mat{H},\mat{S})$ computed by QSD for the physical models we tried; see section~\ref{sec:value-alpha}.

Our most \rev{challenging} technical result of this section will be a bound on the projection difference \eqref{eq:H_projection_error} under \rev{the} assumption \eqref{eq:geometric_mean}.

\begin{theorem} \label{thm:H_projection_error}
  Instate the prevailing notation and assume the bound \eqref{eq:geometric_mean} holds for $0\le \alpha \le 1/2$.
  Assume \rev{the eigenvalue gap condition} 
  \begin{equation} \label{eq:S_separation_gap}
  \lambda_{m+1} + \eta_{\mat{S}} \le \epsilon < (1+\rho)\epsilon \le \lambda_m
  \end{equation}
  for some $\rho > 0$.
  Suppose in addition that $\eta_{\mat{S}}$ is sufficiently small that $(1+\rho^{-1})\eta_{\mat{S}}/\epsilon \le 1$.
  Then the projection error \eqref{eq:H_projection_error} obeys the following bound
  \begin{equation}
    \norm{\mat{\Pi}\mat{H}\mat{\Pi} - \mat{\tilde{\Pi}\mat{H}\mat{\tilde{\Pi}}}} \le 3\mu n^3(1+\rho^{-1}) \mleft( \frac{\norm{\mat{S}}}{\epsilon} \mright)^\alpha \eta_{\mat{S}},
  \end{equation}
  where $\eta_{\mat{S}}$ is defined in \eqref{eq:eta}.
\end{theorem}

There are some unappealing features of this result, namely the cubic dependence on the problem size and the $\order(\eta_{\mat{S}}/\epsilon^\alpha)$ scaling.
The first of these, the cubic dependence on $n$, we believe to likely be an artifact of our proof technique, applying Davis--Kahan ``entry by entry''.
\rev{Fortunately,} numerical experiments do not suggest a dramatic dependence of the error on $n$.
The second effect---the $\eta_{\mat{S}}/\epsilon^\alpha$ dependence rather than a more desirable $\eta_{\mat{S}}$ dependence---appears to be a genuine feature of this problem, at least without additional assumptions; see section~\ref{sec:evid-tightn-theor}.\footnote{Our evidence for this, presented in section~\ref{sec:evid-tightn-theor}, is a synthetically generated pair $(\mat{H},\mat{S})$ obeying the geometric mean condition \eqref{eq:geometric_mean}; we did not obtain this pair from QSD.}
Fortunately, we have yet to find an instance of a pair $(\mat{H},\mat{S})$ generated by the QSD algorithm for which the $\epsilon^\alpha$ factor appears to be necessary to understand the true error.

Finally, we note that the separation hypothesis \eqref{eq:S_separation_gap} is relatively mild.
The first inequality of \eqref{eq:S_separation_gap} is necessary just to assure that $\mat{A}$ and $\mat{\tilde{A}}$ have the same size.
If we assume just a little more of a spectral gap around the thresholding level, quantified by the requirement that $\rho$ is bounded away from zero,  then we get a nice bound.
A less careful application of Davis--Kahan would require that \emph{all} the eigenvalues of $\mat{S}$ are well-separated, so we consider a modest gap at the thresholding level to be a fairly mild requirement.

\begin{proof}[Proof of Theorem~\ref{thm:H_projection_error}]
  The proof shall be an enthusiastic exercise in applying the Davis--Kahan $\sin \mat{\Theta}$ theorem, Fact~\ref{fact:davis_kahan}.
  We begin by bounding
  \begin{equation} \label{eq:entries}
    \begin{split}
      \norm{\mat{\Pi}\mat{H}\mat{\Pi} - \mat{\tilde{\Pi}}\mat{H}\mat{\tilde{\Pi}}}^2 &\le \norm{\mat{\Pi}\mat{H}\mat{\Pi} - \mat{\tilde{\Pi}}\mat{H}\mat{\tilde{\Pi}}}_{\rm F}^2 \\
      &= \sum_{i,j=1}^n \mleft( \vec{v}_i^{\rev{*}}\mleft(\mat{\Pi}\mat{H}\mat{\Pi} - \mat{\tilde{\Pi}}\mat{H}\mat{\tilde{\Pi}}\mright)\vec{v}_j \mright)^2 =: \sum_{i,j=1}^n I_{ij}^2.
      \end{split}
  \end{equation}
  Our strategy will be to bound each of the terms $I_{ij}$.
  
  For each $i$, we can expand $\mat{\tilde{\Pi}}\vec{v}_i = \sum_{k=1}^n c_{ik} \vec{v}_k$.
  Multiplying through by $\vec{v}_i^{\rev{*}}$ then gives that $c_{ik} = \vec{v}_i^{\rev{*}}\mat{\tilde{\Pi}}\vec{v}_k$, which shows in particular that $c_{ik} = c_{ki}$.
  Our first goal will be to bound $|c_{ik}|$.
  We break into two cases, $k\le m$ and $k > m$.

  For case one, assume that $k\le m$.
  By Weyl's inequality \cite[Cor.~IV.4.9]{SS90}, the $(m+1)$st largest eigenvalue of $\mat{\tilde{S}}$ satisfies $\tilde{\lambda}_{m+1} \le \lambda_{m+1} + \eta_{\mat{S}} \le \epsilon$.
  The Davis--Kahan theorem shows the difference $\vec{\delta}_k := \mat{\tilde{\Pi}}\vec{v}_k - \vec{v}_k$ satisfies
  \begin{equation*}
    \norm{\vec{\delta}_k} \le \frac{\eta_{\mat{S}}}{\lambda_k - \tilde{\lambda}_{m+1}} \le \frac{\eta_{\mat{S}}}{\lambda_k - \epsilon} \le \frac{(1+\rho^{-1})\eta_{\mat{S}}}{\lambda_k}.
  \end{equation*}
  This gives a bound on the coefficients $c_{ik}$ for $i\ne k$:
  \begin{equation} \label{eq:c_ik_bound}
    |c_{ik}| = \mleft| \vec{v}_i^{\rev{*}}\mat{\tilde{\Pi}}\vec{v}_k \mright| = \mleft|\vec{v}_i^{\rev{*}}\vec{v}_k + \vec{v}_i^{\rev{*}}\vec{\delta}_k\mright| = \mleft|\vec{v}_i^{\rev{*}}\vec{\delta}_k\mright| \le \norm{\vec{\delta}_k} \le \frac{(1+\rho^{-1})\eta_{\mat{S}}}{\lambda_k}.
  \end{equation}
  For $i = k$, we have
  \begin{equation} \label{eq:c_ii_bound}
    |1 - c_{ii}| = |1 - \vec{v}_i^{\rev{*}}(\vec{v}_i + \vec{\delta}_i)| \le \norm{\vec{\delta}_i} \le \frac{(1+\rho^{-1})\eta_{\mat{S}}}{\lambda_i}.
  \end{equation}

  Now consider case two where $k>m$.
  Since $\tilde{\lambda}_{m+1} \le \epsilon$ as argued above, $\mat{\Pi}$ and $\mat{\tilde{\Pi}}$ are projections onto subspaces of the same dimensions so $\norm{\mat{\Pi} - \mat{\tilde{\Pi}}} = \norm{\mat{\Pi}(\mat{I} - \mat{\tilde{\Pi}})}$ by \cite[Thm.~I.5.5]{SS90}.
  Applying Davis--Kahan then gives
  \begin{equation*} 
    \norm{\mat{\Pi} - \mat{\tilde{\Pi}}} = \norm{\mat{\Pi}(\mat{I} - \mat{\tilde{\Pi}})} \le \frac{\eta_{\mat{S}}}{\rho\epsilon}.
  \end{equation*}
  Then
  \begin{equation} \label{eq:c_ik_small_bound}
    |c_{ik}| = \mleft| \vec{v}_i^{\rev{*}}\mat{\tilde{\Pi}}\vec{v}_k \mright| \le \mleft| \vec{v}_i^{\rev{*}}(\mat{\tilde{\Pi}} - \mat{\Pi})\vec{v}_k \mright| + \mleft| \vec{v}_i^{\rev{*}}\mat{\Pi}\vec{v}_k \mright| \le \norm{\mat{\Pi} - \mat{\tilde{\Pi}}} \le \frac{\eta_{\mat{S}}}{\rho\epsilon}.
  \end{equation}
  
  With these bounds in hand, we return to bounding $I_{ij}$ as defined in \eqref{eq:entries}.
  \rev{Let us introduce shorthand notation $a\wedge b$ and $a\vee b$ for the minimum and maximum of $a$ and $b$ respectively.}
  Expanding $\mat{\tilde{\Pi}}\vec{v}_i$ and $\mat{\tilde{\Pi}}\vec{v}_j$ and using the bound \eqref{eq:geometric_mean}, we obtain
  \begin{equation} \label{eq:I_ij}
    \begin{split}
    I_{ij}
    &= \vec{v}_i^\dagger (\mat{\Pi}\mat{H}\mat{\Pi} - \mat{\tilde{\Pi}}\mat{H}\mat{\tilde{\Pi}}) \vec{v}_j =(1-\overline{c_{ii}}c_{jj}) \cdot \vec{v}_i^\dagger\mat{H}\vec{v}_j-\sum_{\substack{k,\ell=1 \\ k\ne i \textrm{ or } \ell\ne j}}^n \overline{c_{ik}} c_{j\ell} \vec{v}_k^\dagger\mat{H}\vec{v}_\ell  \\
    &\le \mu \mleft[ |1-\overline{c_{ii}}c_{jj}| (\lambda_i\wedge \lambda_j)^{1-\alpha}(\lambda_i \vee \lambda_j)^\alpha + \mkern-18mu\sum_{\substack{k,\ell=1 \\ k\ne i \textrm{ or } \ell\ne j}}^n \mkern-18mu|c_{ik} c_{j\ell}| (\lambda_k\vee \lambda_\ell)^{1-\alpha}(\lambda_k \wedge \lambda_\ell)^\alpha\mright].
    \end{split}
  \end{equation}
  We thus turn our attention to bounding the summands in the second term of \eqref{eq:I_ij}  for $k\ne i$ or $j\ne \ell$.
  First, assume that $k\ne i$ and suppose that $k \le m$, we bound using \eqref{eq:c_ik_bound}:
  \begin{equation*}
    \begin{split}
      |c_{ik}c_{j\ell}|(\lambda_k\wedge \lambda_\ell)^{1-\alpha}(\lambda_k \vee \lambda_\ell)^\alpha &\le |c_{ik}|\lambda_k^{1-\alpha}\norm{\mat{S}}^\alpha \le 
      \frac{(1+\rho^{-1})\eta_{\mat{S}} \lambda_k^{1-\alpha}\norm{\mat{S}}^\alpha}{\lambda_k} \\
      &\le (1+\rho^{-1})\eta_{\mat{S}} (\norm{\mat{S}}/\epsilon)^\alpha.
    \end{split}
  \end{equation*}
  Next assuming $k>m$, \eqref{eq:c_ik_small_bound} yields
  \begin{equation*}
    |c_{ik}c_{j\ell}|(\lambda_k\wedge \lambda_\ell)^{1-\alpha}(\lambda_k \vee \lambda_\ell)^\alpha \le |c_{ik}|\epsilon^{1-\alpha}\norm{\mat{S}}^\alpha \le \eta_{\mat{S}} \rho^{-1} (\norm{\mat{S}}/\epsilon)^\alpha.
  \end{equation*}
  Turning our attention to the first term of the final bound in \eqref{eq:I_ij}, \eqref{eq:c_ii_bound} and the assumption $(1+\rho^{-1})\eta_{\mat{S}} / \epsilon \le 1$ give
  \begin{align*}
    &|1-\overline{c_{ii}}c_{jj}|(\lambda_i\wedge \lambda_j)^{1-\alpha}(\lambda_i\vee \lambda_j)^\alpha\\
    &\qquad\le \mleft( |1-c_{ii}| + |1-c_{jj}| + |1-c_{ii}||1-c_{jj}| \mright) (\lambda_i\wedge\lambda_j)^{1-\alpha}\norm{\mat{S}}^\alpha \\
    &\qquad\le 3\eta_{\mat{S}}(1+\rho^{-1}) (\norm{\mat{S}}/\epsilon)^\alpha.
  \end{align*}
  Using the three previous displays to bound each of the $n^2$ summands in \eqref{eq:I_ij}, we obtain
  \begin{equation*}
    I_{ij} \le 3\mu n^2(1+\rho^{-1})(\norm{\mat{S}}/\epsilon)^\alpha \eta_{\mat{S}},
  \end{equation*}
  which then leads to the stated bound.
\end{proof}

A bound for \eqref{eq:S_projection_error} is entirely analogous.
The analysis is made significantly easier by the fact that the spectral projector $\mat{\Pi}$ is defined in terms of the matrix $\mat{S}$ itself.

\begin{theorem} \label{thm:S_projection_error}
  Instate the prevailing notation.
  Assume that \eqref{eq:S_separation_gap} holds for some $\rho > 0$.
  The projection error \eqref{eq:S_projection_error} satisfies the bound
  \begin{equation*}
    \norm{\mat{\Pi}\mat{S}\mat{\Pi} - \mat{\tilde{\Pi}}\mat{S}\mat{\tilde{\Pi}}} \le \norm{\mat{\Pi}\mat{S}\mat{\Pi} - \mat{\tilde{\Pi}}\mat{S}\mat{\tilde{\Pi}}}_{\rm F} \le 2 (1+\rho^{-1})\eta_{\mat{S}} n + \epsilon^{-1} \mleft[ (1+\rho^{-1})\eta_{\mat{S}} n\mright]^2.
  \end{equation*}
  In particular if $(1+\rho^{-1})\eta_{\mat{S}} n / \epsilon \le 1$, we have
  \begin{equation*}
    \norm{\mat{\Pi}\mat{S}\mat{\Pi} - \mat{\tilde{\Pi}}\mat{S}\mat{\tilde{\Pi}}} \le \norm{\mat{\Pi}\mat{S}\mat{\Pi} - \mat{\tilde{\Pi}}\mat{S}\mat{\tilde{\Pi}}}_{\rm F} \le 3 (1+\rho^{-1})\eta_{\mat{S}} n.
  \end{equation*}
\end{theorem}

\begin{proof}
  The proof is quite similar to Theorem~\ref{thm:H_projection_error} and we shall thus proceed more quickly.
  First, we bound
  \begin{equation*}
    \begin{split}
      \norm{\mat{\Pi}\mat{S}\mat{\Pi} - \mat{\tilde{\Pi}}\mat{S}\mat{\tilde{\Pi}}}^2 \le \norm{\mat{\Pi}\mat{S}\mat{\Pi} - \mat{\tilde{\Pi}}\mat{S}\mat{\tilde{\Pi}}}_{\rm F}^2
      = \sum_{i,j=1}^n \mleft( \vec{v}_i^{\rev{*}}\mleft(\mat{\Pi}\mat{S}\mat{\Pi} - \mat{\tilde{\Pi}}\mat{S}\mat{\tilde{\Pi}}\mright)\vec{v}_j \mright)^2 =: \sum_{i,j=1}^n I_{ij}^2.
      \end{split}
  \end{equation*}
  Consider the expansion $\mat{\tilde{\Pi}}\vec{v}_i = \sum_{k=1}^n c_{ik} \vec{v}_k$ as in the proof of Theorem~\ref{thm:H_projection_error}.
  By the same arguments, the bounds \eqref{eq:c_ik_bound}, \eqref{eq:c_ii_bound}, and \eqref{eq:c_ik_small_bound} hold under the respective hypotheses that $i\le m$ and $i\ne k$, $i = k$, and $i > m$ respectively.

  We now compute $I_{ij}$ using the fact that $\vec{v}_i^{\rev{*}}\mat{S}\vec{v}_j = \lambda_i\delta_{ij}$ where $\delta_{ij}$ denotes the Kronecker delta:
  \begin{equation*}
    \begin{split}
      I_{ij}
      &= \vec{v}_i^{\rev{*}} (\mat{\Pi}\mat{S}\mat{\Pi} - \mat{\tilde{\Pi}}\mat{S}\mat{\tilde{\Pi}}) \vec{v}_j =(1-\overline{c_{ii}}c_{jj}) \cdot \vec{v}_i^{\rev{*}}\mat{S}\vec{v}_j-\sum_{\substack{k,\ell=1 \\ k\ne i \mbox{ or } \ell\ne j}}^n \overline{c_{ik}} c_{j\ell} \vec{v}_k^{\rev{*}}\mat{H}\vec{v}_\ell \\
      &= \lambda_i\delta_{ij} - \sum_{k=1}^n \overline{c_{ik}} c_{jk} \lambda_k.
    \end{split}
  \end{equation*}
  We now distinguish two cases.
  First suppose $i \ne j$.
  Then we bound using \eqref{eq:c_ik_bound} and \eqref{eq:c_ik_small_bound}:
  \begin{equation*}
    \begin{split}
      |I_{ij}| &\le \sum_{k=1}^n |c_{ik}c_{jk}| \lambda_k \le |c_{ij}|(\lambda_i + \lambda_j) + \sum_{\substack{k=1 \\ k\notin\{i,j\}}}^m |c_{ik}c_{jk}| \lambda_k + \sum_{m+1}^n |c_{ik}c_{jk}| \lambda_k \\
      &\le 2\eta_{\mat{S}}(1+\rho^{-1})+\frac{n(1+\rho^{-1})^2\eta_{\mat{S}}^2}{\epsilon}.
      \end{split}
  \end{equation*}
  Next suppose $i = j$.
  Then applying \eqref{eq:c_ik_bound}, \eqref{eq:c_ii_bound}, and \eqref{eq:c_ik_small_bound} gives the same bound
  \begin{equation*}
    |I_{ij}| \le 2\eta_{\mat{S}}(1+\rho^{-1})+\frac{n(1+\rho^{-1})^2\eta_{\mat{S}}^2}{\epsilon}.
  \end{equation*}
  This entrywise bound immediately gives the desired \rev{result}.
\end{proof}

\subsection{Main Result}
\label{sec:putting_pieces_together}

We conclude this section by combining Corollary~\ref{cor:mathias_li}, Proposition~\ref{prop:conjugation}, and Theorems~\ref{thm:H_projection_error} and \ref{thm:S_projection_error} into our main result, which provides a formal statement of Informal Theorem~\ref{thm:main_theorem} from the introduction.

\begin{theorem} \label{thm:main_theorem_formal}
  Let $(\mat{H},\mat{S})$ be a pair of $n\times n$ Hermitian matrices perturbed to a pair $(\mat{\tilde{H}},\mat{\tilde{S}})$ by perturbations $\mat{\Delta}_{\mat{H}}$ and $\mat{\Delta}_{\mat{S}}$ of spectral norms $\eta_{\mat{H}}$ and $\eta_{\mat{S}}$.
  Assume the following:
  \begin{itemize}
    \item The pair $(\mat{H},\mat{S})$ satisfies the geometric mean bound \eqref{eq:geometric_mean} for some parameters $\mu > 0$ and $0\le \alpha \le 1/2$.
    \item There exists an index $m$ for which \eqref{eq:S_separation_gap} holds for some $\rho > 0$.
    \item The noise $\eta_{\mat{S}}$ is sufficiently small so that $(1+\rho^{-1})\eta_{\mat{S}}/\epsilon \le 1$.
  \end{itemize}
  Let $(\mat{A},\mat{B}) = (\mat{V}^{\rev{*}}\mat{H}\mat{V},\mat{V}^{\rev{*}} \mat{S}\mat{V})$ denote the thresholded matrix pair.
  The eigenvalues recovered by the thresholding procedure applied to the noise-perturbed pair $(\mat{\tilde{H}},\mat{\tilde{S}})$ are the same as the eigenvalues of a pair $(\mat{\tilde{A}},\mat{\tilde{B}})$ satisfying
  \begin{equation*}
    \sqrt{\norm{\mat{\tilde{A}} - \mat{A}}^2 + \norm{\mat{\tilde{B}} - \mat{B}}^2} \le 3(2+\mu) n^3(1+\rho^{-1}) \mleft( \frac{\norm{\mat{S}}}{\epsilon} \mright)^\alpha \eta_{\mat{S}} + \eta_{\mat{H}} =: \chi.
  \end{equation*}
  Let $E_0$ and $E_1$ denote the least and second-to-least eigenvalues of $(\mat{A},\mat{B})$.
  Suppose further that
  \begin{itemize}
      \item The error bound $\chi$ is sufficiently small: $n\chi \le \epsilon$.
      \item The gap condition $\tan^{-1} E_1 - \tan^{-1} E_0 \ge \sin^{-1} (n\chi/\epsilon)$ holds.
  \end{itemize}
  Then with $d_0^{-1}$ the condition number of the eigenangle $\tan^{-1} E_0$ and $\tilde{E}_0$ the eigenvalue recovered by thresholding applied to $(\mat{\tilde{H}},\mat{\tilde{S}})$,
  \begin{equation*}
      \mleft| \tan^{-1} E_0 - \tan^{-1}\tilde{E}_0 \mright| \le \sin^{-1} \frac{n\chi}{d_0}.
  \end{equation*}
\end{theorem}

In particular, for the theorem to hold, the threshold parameter must be chosen 
\begin{equation*}
    \epsilon > n\chi = 3(2+\mu) n^4(1+\rho^{-1}) \mleft( \frac{\norm{\mat{S}}}{\epsilon} \mright)^\alpha \eta_{\mat{S}} + n\eta_{\mat{H}}.
\end{equation*}
This leads to the claimed value $\epsilon = \Theta\mleft(\eta^{\tfrac{1}{1+\alpha}}\mright)$ and bound $\mleft| \tan^{-1} E_0 - \tan^{-1}\tilde{E}_0 \mright| \le \order\mleft(d_0^{-1}\eta^{\tfrac{1}{1+\alpha}}\mright)$ in Informal Theorem~\ref{thm:main_theorem}.

\section{Analysis of \rev{unitary Krylov subspace approximation with thresholding}}
\label{sec:analys-quant-subsp}

\rev{Having studied the errors due to noise in the previous section, we now turn to analyzing the Rayleigh--Ritz errors in computing the ground-state eigenvalue $E_0$ of $\oper{H}$ using the unitary Krylov space \eqref{eq:quantum_subspace} and the thresholding procedure Algorithm~\ref{alg:thresholding}.
Specifically, we shall bound the difference between $E_0$ and the result $\tilde{E}_0$ of applying thresholding to the pair $(\mat{H},\mat{S})$ \eqref{eq:HS} computed on an error-free quantum computer.}

If one considers our analysis with the threshold parameter $\epsilon$ set to zero, one obtains an analysis of the QSD method \rev{with no truncation}. 
To our knowledge, this is the first quantitative error analysis of the QSD method even in the noise-free setting.
This builds on two earlier explanations of the success of QSD.
The first, by Stair, Huang, and Evangelista \cite{SHE20}, argues based on Taylor series that the QSD subspaces approximately coincides with the classical polynomial Krylov space.
This explanation has two drawbacks: (1) small timesteps make the ill-conditioning of $\mat{H}$ and $\mat{S}$ worse \cite[\S2.1]{SHE20} and (2) QSD often performs better with larger time steps \cite[\S{II}.B]{KMC+21}.
An alternate analysis based on filter diagonalization is provided by Klymko et al.\ \cite{KMC+21}, which provides an overcomplete set of \emph{phase cancellation conditions} under which QSD computes the eigenvalues of interest with zero error.
They then argue that these conditions hold approximately in the long-time limit for a randomly chosen timestep.

Our analysis is more direct than the two previous, emulating the classical analysis of the Lanczos method by Saad \cite{Saa80}.
This leads to a quantitative error bound in terms of the distribution of the spectra which decreases exponentially in the number of timesteps used.
The basic idea will be to use a linear combination of the QSD basis states $\rev{\vec{\varphi}_j}$ which has the effect of applying a trigonometric polynomial to the eigenvalues $E_0,\ldots,E_{N-1}$ of $\oper{H}$.
If $\rev{\vec{\varphi}_j}$ approximately lies in the span of the first $M+1$ eigenstates, we shall choose this trigonometric polynomial to be large at the eigenvalue of interest and exponentially small (in $n$) at eigenvalues $E_1,\ldots,E_M$.
The trigonometric polynomial will be bounded so it won't amplify any components of the eigenvector in the direction of any of the remaining eigenvectors.
As an additional feature of this analysis, we are able to directly analyze thresholding ``for free'' in the noiseless setting, where thresholding has the effect of perturbing this trigonometric polynomial.
The main theorem of this section is as follows:

\begin{theorem} \label{thm:a_priori_bound}
  Let $\rev{\vec{\psi}_0},\ldots,\rev{\vec{\psi}_{N-1}}$ be the eigenvectors of a Hermitian operator $\oper{H}$ with eigenvalues $E_0,\ldots,E_{N-1}$.
  Suppose the initial vector is expanded as
  \begin{equation} \label{eq:initial_guess_expansion}
    \rev{\vec{\varphi}_0} = \sum_{j=0}^{N-1} \gamma_i \rev{\vec{\psi}_i}.
  \end{equation}
  Let $\Delta E_j := E_j - E_0$ and choose an index $0\le M\le N-1$.
  Suppose the QSD algorithm is implemented with time sequence $\{ t_j \}_{j=-k}^k$ for $t_j = \pi j / \Delta E_M$.
  Suppose the generalized eigenvalue problem 
 \eqref{eq:gep} is solved with thresholding parameter $\epsilon$ and let $\epsilon_{\rm total}$ be the sum of the eigenvalues of $\mat{S}$ discarded by thresholding.
  Then
  \begin{equation} \label{eq:a_priori_bound}
    0 \le \tilde{E}_0 - E_0 \le \frac{2\mleft[\Delta E_{N-1}\epsilon_{\rm total} + 4(1+\tfrac{\pi \Delta E_1}{\Delta E_M})^{-2k}\sum\limits_{i=1}^M \Delta E_i|\gamma_i|^2 + \sum\limits_{i=M+1}^{N-1} \Delta E_i|\gamma_i|^2\mright]}{|\gamma_0|^2 - 2|\gamma_0|\sqrt{(2k+1)\epsilon}}.
  \end{equation}
\end{theorem}

\begin{remark}
The value $|\gamma_0|^2=|\braket*{\varphi_0}{\psi_0}|^2$ is referred to as the initial overlap. 
In order for the QSD algorithm to succeed with a relatively small number of Krylov steps, $|\gamma_0|^2$ must be sufficiently large (for instance, $\ge0.1$). This is qualitatively different from the assumption of classical Krylov subspace methods for solving eigenvalue problems, where the initial overlap $|\gamma_0|^2$ should also be nonzero but can be very small. 
\end{remark}

Let us note some salient features of this result.
First, if we consider the noise-free case without thresholding, we obtain the bound
\begin{equation*}
  0 \le \tilde{E}_0 - E_0 \le 8\mleft(1+\frac{\pi \Delta E_1}{\Delta E_M}\mright)^{-2k} \sum_{i=1}^M \Delta E_i\frac{|\gamma_i|^2}{|\gamma_0|^2} + 2\sum_{i=M+1}^{N-1} \Delta E_i\frac{|\gamma_i|^2}{|\gamma_0|^2}.
\end{equation*}
The error bound has two terms, a term concerning the eigenvalues $E_1,\ldots,E_M$ which is damped exponentially fast with rate $\Delta E_1/\Delta E_M$ and an undamped (but also unamplified) term with the eigenvalues $E_{M+1},\ldots,E_{N-1}$.
If the components of $\rev{\vec{\varphi}_0}$ in the directions of the $(M+1)$st to $(N-1)$st eigenvectors are small in the sense that $\Delta E_i |\gamma_i|^2 \ll 1$, this second term will be small.
This bound thus has a tradeoff: If $M$ is chosen larger (i.e., the timestep becomes smaller), more terms will be exponentially damped but at a slower rate as $\Delta E_1/\Delta E_M$ will decrease.
One can obtain the simplest bound by choosing $M = N-1$ and bounding $\Delta E_i \le \Delta E_{N-1}$, leading to 
\begin{equation} \label{eq:a_priori_bound_simplified}
  0 \le \tilde{E}_0 - E_0 \le 8\Delta E_{N-1}\frac{1-|\gamma_0|^2}{|\gamma_0|^2} \mleft(1+\frac{\pi \Delta E_1}{\Delta E_{N-1}}\mright)^{-2k}.
\end{equation}
Unfortunately, this simplified bound often grossly overestimates the error.

Our analysis also indicates that thresholding has only a mild effect on the accuracy. We pick up a $1 + \order(\sqrt{k\epsilon})$ prefactor and an additional term proportional to the sum of the discarded eigenvalues of $\mat{S}$, which in turn can be bounded $\epsilon_{\rm total} \le (2k+1) \epsilon$.
In practice, we usually have $\epsilon_{\rm total} \approx \epsilon$ due to rapid spectral decay.

Combining Theorem~\ref{thm:a_priori_bound} with Informal Theorem~\ref{thm:main_theorem} (formalized as Theorem~\ref{thm:main_theorem_formal}) leads directly to Informal Theorem~\ref{thm:main_theorem_big}.
More precise but also more complex error bounds can be obtained by using the full power \eqref{eq:a_priori_bound} of Theorem~\ref{thm:a_priori_bound} directly with Theorem~\ref{thm:main_theorem_formal}.

\subsection{Proof of Theorem~\ref{thm:a_priori_bound}}
\label{sec:proof-theorem}

Our proof is based on the observation that thresholding is equivalent to applying the Rayleigh--Ritz procedure with a subspace spanned by the dominant left singular vectors of the Krylov matrix.
Consider the Krylov matrix $\mat{K}$ defined and factorized as
\begin{equation} \label{eq:krylov_matrix}
  \begin{split}
    \mat{K} &:= \begin{bmatrix} \rev{\vec{\varphi_{-k}}} & \cdots & \rev{\vec{\varphi_{k}}} \end{bmatrix} \\
    &= \underbrace{\begin{bmatrix} \rev{\vec{\psi}_0} & \cdots & \rev{\vec{\psi}_{N-1}} \end{bmatrix}}_{:=\mat{\Psi}} \underbrace{\begin{bmatrix} \gamma_0 \\ & \ddots \\ && \gamma_{N-1} \end{bmatrix}}_{:=\mat{\Gamma}} \underbrace{\begin{bmatrix} \e^{-\iu k E_0} & \cdots & \e^{\iu k E_0} \\
        \vdots &\ddots & \vdots \\
        \e^{-\iu kE_{N-1}} & \cdots & \e^{\iu k E_{N-1}} \end{bmatrix}}_{:=\mat{F}}.
  \end{split}
\end{equation}
Then we easily see that $\mat{H} = \mat{K}^{\rev{*}}\oper{H}\mat{K}$ and $\mat{S} = \mat{K}^{\rev{*}}\mat{K}$.
Since the eigenvalues of $\mat{K}^{\rev{*}}\mat{K}$ are the squares of singular values of $\mat{K}$ with eigenvectors equal to the right singular vectors of $\mat{K}$, it follows that the thresholded problem is precisely the Rayleigh--Ritz procedure applied to the left singular subspace of $\mat{K}$ with singular values \rev{larger than} $\sqrt{\epsilon}$.
From these left singular vectors, we are able to reconstruct the matrix $\mat{K}$ up to a Frobenius norm error $\sqrt{\epsilon_{\rm total}}$.
We thus can analyze QSD with thresholding in much the same way as Saad's analysis of the Lanczos method \cite{Saa80} with two twists.
First, we have trigonometric polynomials in place of polynomials owing to the QSD algorithm's use of the unitary time-evolution operator.
Second, our trigonometric basis functions are perturbed as a result of the truncated singular value decomposition.

With this roadmap in mind, we begin with a trigonometric version of a classic result in polynomial approximation theory.

\begin{lemma}
  Let $0 < a < \pi$ and denote by $\trigpoly_k$ the space of degree $\le k$ trigonometric polynomials.
  The trigonometric polynomial minimax approximation problem
  \begin{equation*}
    \beta(a,k) = \min_{\substack{p \in \trigpoly_k \\ p(0) = 1}} \max_{t\in(-\pi,\pi) \setminus (-a,a)} |p(t)|
  \end{equation*}
  is solved by
  \begin{equation} \label{eq:trig_optimum_solution}
    p^*(\theta) = \frac{T_k(1+2\tfrac{\cos\theta-\cos a}{\cos a+1})}{T_k(1+2\tfrac{1-\cos a}{\cos a+1})}
  \end{equation}
  where $T_k$ denotes the $k$th Chebyshev polynomial. 
  The optimum value is
  \begin{equation}
    \label{eq:trig_optimal_value}
    \beta(a,k) = \mleft( T_k\mleft( 1+ 2\tfrac{1-\cos a}{\cos a+1} \mright) \mright)^{-1} \le 2\mleft( 1 + 2\sqrt{\frac{1-\cos a}{\cos a+1}} \mright)^{-k} \le 2(1+a)^{-k}.
  \end{equation}
\end{lemma}

\begin{proof}
  Once one convinces oneself that an optimal solution can be taken to be real and even, this result follows immediately from the analogous result for polynomial approximation \cite[Thm.~4.1.11]{Bjo15} together with the standard reparametrization $f(x) \mapsto f^\circ(\theta) := f(\cos \theta)$ which puts into bijection algebraic and even trigonometric polynomials.
\end{proof}

We shall need a bound for the optimal trigonometric polynomial \eqref{eq:trig_optimum_solution}.

\begin{proposition}
  \label{prop:optimal_poly_bound}
  The trigonometric polynomial $p^*$ defined in \eqref{eq:trig_optimum_solution} is bounded in absolute value by $1$ and satisfies the $L^2$ bound
  \begin{equation} \label{eq:optimal_poly_bound}
    \int_{-\pi}^\pi \mleft|p^*(\theta)\mright|^2 \, d\theta \le 2a + (2\pi-2a) \mleft(\beta(a,k)\mright)^2 \le 2\pi.
  \end{equation}
\end{proposition}

\begin{proof}
  First, we show that $p^*$ is monotone on $[0,a]$ by showing its derivative can't have a zero on $(0,a)$.
  Up to a scaling factor and a change of variables on the input, $p^*$ coincides with $T_k$ on $[a,\pi]$.
  Thus, $p^*$ has $k-1$ local extrema on $(a,\pi)$ and symmetrically $k-1$ on $(-\pi,-a)$.
  Since $p^*$ is even and $2\pi$-periodic, $p^*$ has local extrema at $0$ and $\pi$.
  Since $(p^*)'$ is a degree-$k$ trigonometric polynomial, it has at most $2k$ zeros, all of which have already been accounted for.
  Thus, $p^*$ is monotone on $[0,a]$.

  Since $p^*(a) \le \beta(a,k) < 1$, $p^*$ is monotone decreasing on $[0,a]$ and thus achieves its maximum value on $[-a,a]$ of $1$ at $0$.
  On $(-\pi,\pi) \setminus [-a,a]$, $|p^*| \le \beta(a,k)$, from which the bound \eqref{eq:optimal_poly_bound} follows.
\end{proof}

We shall be content with using the looser upper bound $2\pi$ in \eqref{eq:optimal_poly_bound} in our subsequent analysis.
With these approximation results in hand, we prove Theorem~\ref{thm:a_priori_bound}.

\begin{proof}[Proof of Theorem~\ref{thm:a_priori_bound}]
  Let $\rev{\vec{\varphi}_0}$ be expanded as \eqref{eq:initial_guess_expansion} and consider the Krylov matrix and its factorization defined in \eqref{eq:krylov_matrix}.
  Let $\mat{\tilde{K}}$ represent the truncation of $\mat{K}$ by settings its singular values \rev{which are at most} $\sqrt{\epsilon}$ to zero and factor it as $\mat{\tilde{K}} = \mat{\Psi}\mat{\Gamma}\mat{\tilde{F}}$ so that
  \begin{equation} \label{eq:epsilon_total}
    \epsilon_{\rm total} = \norm{\mat{\tilde{K}} - \mat{K}}_{\rm F}^2 = \norm{\mat{\Gamma}(\mat{\tilde{F}} - \mat{F})}_{\rm F}^2 = \sum_{i=0}^{N-1}\sum_{j=-k}^k |\gamma_i|^2 \underbrace{\mleft| \tilde{f}_{ij} - \e^{\iu j E_i} \mright|^2}_{:= \alpha_{ij}^2},
  \end{equation}
  where $\tilde{f}_{ij}$ denotes the $ij$ entry of $\mat{\tilde{F}}$.
  
  Let $\mathcal{R}_\epsilon$ denote the range of $\mat{\tilde{K}}$.
  By the Courant--Fischer theorem \cite[Cor.~IV.4.7]{SS90},
  \begin{equation} \label{eq:courant_fischer}
  \tilde{E}_0 - E_0 = \min_{\rev{\ket{\xi}} \in \mathcal{R}_\epsilon \setminus \{\vec{0}\}} \frac{\rev{\vec{xi}^*} (\oper{H} - E_0 \rev{\mat{I}})\rev{\vec{\xi}}}{\rev{\vec{\xi}^*\vec{\xi}}},
  \end{equation}
  where \rev{$\mat{I}$} denotes the identity.
  Since $\oper{H} - E_0\mat{I}$ is positive \rev{semidefinite}, we have $\tilde{E}_0 - E_0 \ge 0$.
  The remainder of our effort will be dedicated to establishing an upper bound.

  We shall use the minimax optimal trigonometric polynomial $p^*$ \eqref{eq:trig_optimum_solution} to construct an ansatz $\rev{\vec{\xi}_\star}$ to plug into \eqref{eq:courant_fischer}.
  Let $p^*(\theta-\pi E_0/\rev{\Delta E_{M-1}}) = \sum_{j=-k}^k c_j \e^{\iu j \theta}$ be the Fourier series of the $\pi E_0/\Delta E_{M-1}$-translate of $p^*$.
  Choose as ansatz
  \begin{equation*}
    \rev{\vec{\xi}_\star} := \sum_{i=0}^{N-1} \sum_{j=-k}^k \gamma_i \tilde{f}_{ij} c_j \rev{\vec{\psi}_i} \in \mathcal{R}_\epsilon.
  \end{equation*}
  Plugging this into \eqref{eq:courant_fischer}, we obtain an upper bound
  \begin{equation} \label{eq:first_upper_bound}
    \tilde{E}_0 - E_0 \le \rev{\frac{\vec{\xi}_\star^*(\oper{H} - E_0\rev{\mat{I}})\vec{\xi}_\star}{\vec{\xi}_\star^*\vec{\xi}_\star}} = \frac{\sum_{i=1}^{N-1} \Delta E_i |\gamma_i|^2 \mleft| \sum_{j=-k}^k \tilde{f}_{ij}c_j \mright|^2}{\sum_{i=0}^{N-1} |\gamma_i|^2 \mleft| \sum_{j=-k}^k \tilde{f}_{ij} c_j \mright|^2}.
  \end{equation}
  First, focus on the numerator of the final bound in \eqref{eq:first_upper_bound}.
  We bound
  \begin{equation*}
    \mleft| \sum_{j=-k}^k \tilde{f}_{ij}c_j \mright|^2 \le \mleft| \sum_{j=-k}^k (\tilde{f}_{ij} - \e^{\iu j E_i})c_j + \sum_{j=-k}^k \e^{\iu jE_i}c_j \mright|^2 \le \mleft( \sum_{|j|\le k} \alpha_{ij}c_j + p^*(E_i - E_0) \mright)^2 ,
  \end{equation*}
  where $\alpha_{ij} \ge 0$ is defined in \eqref{eq:epsilon_total}.
  
  First suppose that $1\le i \le M$.
  Then by the fact that $|p^*(E_i - E_0)| \le \beta(a,k)$ as defined in \eqref{eq:trig_optimal_value} with $a := \pi \Delta E_1/\Delta E_M$, the Parseval theorem (i.e., $2\pi\sum_{j=-k}^k |c_j|^2 = \int_{-\pi}^\pi |p^*(\theta)|^2\, d\theta$), and the bound \eqref{eq:optimal_poly_bound} we obtain
  \begin{equation} \label{eq:fub_numerator}
    \mleft| \sum_{j=-k}^k \tilde{f}_{ij}c_j \mright|^2 \le 2\mleft(\sum_{j=-k}^k |c_j|^2 \mright)\mleft(\sum_{j=-k}^k \alpha_{ij}^2 \mright) + 2(\beta(a,k))^2 \le 2 \sum_{j=-k}^k \alpha_{ij}^2 + 2(\beta(a,k))^2.
  \end{equation}
  For $M < i < N$, we get the same bound except with $1$ in place of $\beta(a,k)$ since $|p^*| \le 1$ by Proposition~\ref{prop:optimal_poly_bound}.
  
  For the denominator of the final bound in \eqref{eq:first_upper_bound}, we bound
  \begin{equation} \label{eq:fub_denominator}
    \mleft| \sum_{j=-k}^k \tilde{f}_{0j} c_j \mright| \ge 1 - \sum_{j=-k}^k \alpha_{0j}  \ge 1 - \sqrt{(2k+1)\sum_{j=-k}^k \alpha_{0j}^2} \ge 1 - \frac{1}{|\gamma_0|} \sqrt{(2k+1)\epsilon},
  \end{equation}
  where we used a spectral norm bound similar to \eqref{eq:epsilon_total}:
  \begin{equation*}
      \epsilon \ge \norm{\mat{\tilde{K}} - \mat{K}} = \norm{\mat{\Gamma}(\mat{\tilde{F}} - \mat{F})} \ge \sqrt{|\gamma_0|^2\sum_{j=-k}^k \mleft|\tilde{f}_{ij} - \e^{\iu jE_i}\mright|^2} = |\gamma_0|\sqrt{\sum_{j=-k}^k \alpha_{0j}^2}.
  \end{equation*}
  Plugging \eqref{eq:fub_numerator} and \eqref{eq:fub_denominator} into \eqref{eq:first_upper_bound},
  \begin{equation*}
    \begin{split}
      \tilde{E}_0 - E_0 &\le \frac{2\mleft[\sum\limits_{i=1}^M \Delta E_i \mleft(\sum_{j=-k}^k \alpha_{ij}^2 + (\beta(a,k))^2\mright) + \sum\limits_{i=M+1}^{N-1} \Delta E_i \mleft(\sum_{j=-k}^k \alpha_{ij}^2 + 1\mright)\mright]}{\mleft(|\gamma_0| - \sqrt{(2k+1)\epsilon}\mright)^2} \\
      &\le \frac{2\mleft[ \Delta E_{N-1}\epsilon_{\rm total} + (\beta(a,k))^2 \sum_{i=1}^M \Delta E_i |\gamma_i|^2 + \sum_{i=M+1}^{N-1} \Delta E_i |\gamma_i|^2\mright]}{|\gamma_0|^2 - 2|\gamma_0|\sqrt{(2k+1)\epsilon}}.
    \end{split}
  \end{equation*}
  Using the bound \eqref{eq:trig_optimal_value} with $a = \pi \Delta E_1 / \Delta E_M$ leads precisely to \eqref{eq:a_priori_bound}.
\end{proof}

\section{Additional Results and Discussions}
\label{sec:addit-resul}

In this section, we include some additional results and discussions which follow from our analysis but are not directly germane to the main analysis of the QSD algorithm comprising Theorems~\ref{thm:main_theorem_formal} and \ref{thm:a_priori_bound}.

\subsection{On the Toeplitz Structure of \texorpdfstring{$\mat{H},\mat{S}$}{the projected Hamiltonian and overlap matrix}}\label{sec:toeplitz}

With the choice of the basis vectors $\rev{\vec{\varphi}_j} = \e^{\iu t_j \oper{H}} \rev{\vec{\varphi}_0}$ as in the Parrish--McMahon QSD procedure, the matrix elements of the projected matrices satisfy
\begin{equation*}
\mat{H}_{jk} = \rev{\vec{\varphi}_j^*\oper{H}\vec{\varphi}_k}=
\rev{\vec{\varphi}_0^*\oper{H}\e^{\iu (t_k-t_j) \oper{H}}\vec{\varphi}_0} \quad \rev{\textnormal{and}}\quad
\mat{S}_{jk} = \rev{\vec{\varphi}_j^*\vec{\varphi}_k}=
\rev{\vec{\varphi}_0^*\e^{\iu (t_k-t_j) \oper{H}}\vec{\varphi}_0}.
\end{equation*}
Therefore, both $\mat{H},\mat{S}$ are Hermitian--Toeplitz matrices. 
Unfortunately, the Toeplitz structure relies on the assumption that the Hamiltonian simulation problem (i.e., $\e^{\iu t_j \oper{H}} \rev{\vec{\varphi}_0}$) is computed exactly. 
In practice, the Hamiltonian simulation problem is often performed with approximate techniques (such as Trotter splitting), and the resulting projected matrices $\mat{\tilde{H}}$ and $\mat{\tilde{S}}$ may not have the Toeplitz structure. 
If this is the case, all $n^2$ entries of $\mat{\tilde{H}}$ (and perhaps $\mat{\tilde{S}}$ as well) need to be computed to apply the Rayleigh--Ritz procedure to the computed basis states ${\rev{\vec{\varphi}_0}},\ldots,\rev{\vec{\varphi}_{n-1}}$ in earnest.
However, if one measures only the first row of $\mat{H}$ and $\mat{S}$ and imputes the remaining entries from the Hermitian--Toeplitz structure, then resulting recovered matrices $\mat{\tilde{H}}$ and $\mat{\tilde{S}}$ represent the true $\mat{H}$ and $\mat{S}$ corrupted by both Monte Carlo and discretization errors.
Our main analysis makes no use of the Toeplitz structure.

An important question for the QSD procedure is how entrywise errors in the entries $\mat{H}$ and $\mat{S}$ correspond to errors in the spectral norm.
\rev{For the standard QSD estimation procedure,} the \rev{entries} of $\mat{M} \in \{ \mat{H}, \mat{S} \}$ are approximated by averaging $m$ unbiased estimators each with maximum error $B$ ($B = \order(1)$ for $\mat{S}$ and $B = \order(\norm{\mat{H}})$ for $\mat{H}$).
Consider the case where the Hamiltonian simulation problem is solved exactly and we compute estimates for $\mat{M} \in \{ \mat{H}, \mat{S} \}$ \rev{by measuring the first row of $\mat{M}$ and computing} the remaining entries from the Hermitian--Toeplitz structure.
\rev{Straightforward application of matrix concentration inequalities then shows that the approximation is $\order(B\sqrt{(n\log n)/m})$-close to $\mat{M}$ (see, e.g., \cite[Thms.~3.6.1 and 4.6.1]{Tro15}).}

\subsection{Stability of Best Low-rank Approximation}

Theorem~\ref{thm:S_projection_error} constitutes a stability result for the \rev{Eckart--Young} best rank-$m$ approximation $\lowrank{\mat{S}}_m = \mat{\Pi}\mat{S}\mat{\Pi}$ to $\mat{S}$.
Since this result may be of independent interest, we state it here unburdened by the particularities of the QSD context.

\begin{theorem} \label{thm:low_rank_approximation}
  Let $\mat{A}\in\complex^{n\times n}$ be a positive semidefinite matrix with eigenvalues $\lambda_1 \ge \lambda_2 \ge \cdots \ge \lambda_n \ge 0$.
  Let $\lowrank{\mat{A}}_m$ denote the \rev{Eckart--Young} best rank-$m$ approximation to $\mat{A}$.
  Then, for any quadratic unitarily invariant norm $\norm{\cdot}_{\rm QUI}$ (such as the spectral or Frobenius norms; see \cite[Def.~IV.2.9]{Bha97}), 
  \begin{equation*}
    \norm*{\lowrank{\mat{A}+\mat{\Delta}}_m - \lowrank{\mat{A}}_m}_{\rm QUI} \le \norm{\mat{\Delta}}_{\rm QUI} + \frac{2n\lambda_m\norm{\mat{\Delta}}}{\lambda_m - \lambda_{m+1} - \norm{\mat{\Delta}}} \mleft( 1 + \frac{0.5\, n\norm{\mat{\Delta}}}{\lambda_m - \lambda_{m+1} - \norm{\mat{\Delta}}} \mright).
  \end{equation*}
\end{theorem}

\begin{proof}
  This follows immediately from Theorem~\ref{thm:S_projection_error} and the maximality of the Frobenius norm among quadratic unitarily invariant norms \cite[Eq.~(IV.38)]{Bha97}.
\end{proof}

This bound is interesting because, in the setting where the gap $\lambda_m - \lambda_{m+1} - \norm{\mat{\Delta}}$ is comparable in size to $\lambda_m$, the best rank-$m$ approximation changes only by $\approx n\norm{\mat{\Delta}}_{\rm QUI}$ independent of the approximation error $\norm*{\mat{A} - \lowrank{\mat{A}}_m}_{\rm QUI}$.
This can be a large improvement over the general-purpose bound (similar to \cite[Cor.~2.4]{DI19}) which holds for general unitarily invariant norms $\norm{\cdot}_{\rm UI}$:
\begin{equation*}
  \norm*{\lowrank{\mat{A}+\mat{\Delta}}_m - \lowrank{\mat{A}}_m}_{\rm UI} \le 2(\norm*{\mat{A} - \lowrank{\mat{A}}_m}_{\rm UI} + \norm{\mat{\Delta}}_{\rm UI}),
\end{equation*}
which does depend on the approximation error $\norm*{\mat{A} - \lowrank{\mat{A}}_m}_{\rm UI}$.
 We believe the dimensional factor $n$ is likely quite pessimistic, and can hopefully be replaced by a small constant (at least) if the matrix enjoys a rapidly decaying spectrum.

\subsection{General Analysis of Thresholding}
\label{sec:gener-analys-thresh}

We have presented an analysis in Theorem~\ref{thm:a_priori_bound} of thresholding for the QSD problem, which simultaneously treats the Rayleigh--Ritz error from approximation from the titular quantum subspace and the thresholding procedure together.
This is quite natural as the total error is what is most important to the practitioner, and one should in principle be able to obtain a more precise error bound by not decoupling these pieces.
However, it remains a question of mathematical interest and of interest to the broader uses of thresholding beyond QSD to provide an analysis of thresholding for general matrices.

As the bad example \eqref{eq:thresholding_bad} shows, thresholding may not work for general matrices, with thresholding with parameter $\epsilon$ able to introduce errors $\gg \epsilon$.
This behavior is less surprising, however, when one notes that neither eigenvalue of \eqref{eq:thresholding_bad} is well-conditioned, with both having a condition number $\Theta(\epsilon^{-1})$.
In fact, this is the only obstruction to thresholding working (at least for recovering the smallest eigenvalue), as we shall show that the least eigenvalue is recovered accurately by thresholding if it is well-conditioned.
(We recall \cite[Thm.~VI.2.2]{SS90} that the condition number of the eigenangle $\tan^{-1} E_0$ associated with the least eigenvalue $E_0$ of $(\mat{H},\mat{S})$ is $\norm{\vec{c}_0}^2\sqrt{1+E_0^2}$ where $\vec{c}_0$ is the $\mat{S}$-normalized eigenvector, $\vec{c}_0^{\rev{*}}\mat{S}\vec{c}_0 = 1$, associated with $E_0$.)

\begin{theorem} \label{thm:thresholding}
  Let $\vec{c}_0$ be the $\mat{S}$-normalized eigenvector associated with the least eigenvalue $E_0$ of the generalized eigenvalue problem \eqref{eq:gep} for a Hermitian and Hermitian positive matrix $\mat{H}$ and $\mat{S}$.
  Then, for $\tilde{E}_0$ the least eigenvalue recovered by thresholding with parameter $\epsilon$ and provided $2\sqrt{\epsilon}\norm{\vec{c}_0} < 1$, 
  \begin{equation*}
    0 \le \tilde{E}_0 - E_0 \le \frac{\Delta E \, \epsilon\norm{\vec{c}_0}^2}{1 - 2\sqrt{\epsilon}\norm{\vec{c}_0}},
  \end{equation*}
  where $\Delta E$ is the difference between the largest and smallest eigenvalue of $(\mat{H},\mat{S})$.
\end{theorem}

\begin{proof}
  Let $\mathcal{R}_\epsilon$ be the span of the eigenvectors of $\mat{S}$ with eigenvalue \rev{greater than} $ \epsilon$.
  Using the same observation which motivated the proof of Theorem~\ref{thm:a_priori_bound}, we have
  \begin{equation} \label{eq:thresholding_only}
    \tilde{E}_0 - E_0 = \min_{\vec{c} \in \mathcal{R}_\epsilon \setminus \{\vec{0}\}} \frac{\vec{c}^{\rev{*}}(\mat{H} - E_0 \, \mat{S})\vec{c}}{\vec{c}^{\rev{*}}\mat{S}\vec{c}}.
  \end{equation}
  In particular, since $\mat{H} - E_0\mat{S}$ is positive semidefinite, this implies $\tilde{E}_0 - E_0 \ge 0$ so we just need to concern ourselves with obtaining an upper bound.
  Letting $\mat{\tilde{S}}$ be the $\mat{S}$ matrix with all its eigenvalues \rev{at most} $\epsilon$ set to zero, we shall evaluate \eqref{eq:thresholding_only} at $\vec{\tilde{c}}_0 := \mat{S}^{-1/2}\mat{\tilde{S}}{}^{1/2}\vec{c}_0$, obtaining an upper bound.
  Defining the error $\vec{\delta} := \vec{\tilde{c}}_0 - \vec{c}_0$, we have that $\vec{\delta}$ satisfies the bound $\big\|\mat{S}^{1/2}\vec{\delta}\big\| \le \big\| \mat{S}^{1/2} - \mat{\tilde{S}}{}^{1/2}\big\| \norm{\vec{c}_0} \le \sqrt{\epsilon} \norm{\vec{c}_0}$.
  \rev{Thus}
  \begin{equation*}
    \vec{\tilde{c}}_0^{\rev{*}}\mat{S}\vec{\tilde{c}}_0 \ge \vec{c}_0^{\rev{*}}\mat{S}\vec{c}_0 - 2\mleft| \vec{c}_0^{\rev{*}} \mat{S}\vec{\delta} \mright| \ge 1 - 2\mleft( \vec{c}_0^{\rev{*}}\mat{S}\vec{c}_0\mright)^{1/2}\mleft( \vec{\delta}^{\rev{*}}\mat{S}\vec{\delta}\mright)^{1/2} \ge 1-2\sqrt{\epsilon}\norm{\vec{c}_0}
  \end{equation*}
  and
  \begin{align*}
    \vec{\tilde{c}}_0^{\rev{*}}(\mat{H} - E_0 \mat{S})\vec{\tilde{c}}_0 = \vec{\delta}^{\rev{*}}(\mat{H} - E_0 \mat{S})\vec{\delta} \le \big\| \mat{S}^{-1/2}\mat{H}\mat{S}^{-1/2} \!-\! E_0 \mat{I} \big\| \, \big\| \mat{S}^{1/2}\vec{\delta} \|^2
    \le \Delta E \, \epsilon\norm{\vec{c}_0}^2.
  \end{align*}
  Plugging $\vec{\tilde{c}}_0$ into \eqref{eq:thresholding_only} and applying the previous two displays leads immediately to the stated result.
\end{proof}

Theorem~\ref{thm:thresholding} yields an alternative analysis of the QSD algorithm with thresholding: Combine Theorem~\ref{thm:a_priori_bound} with $\epsilon = 0$ (to measure the Rayleigh--Ritz error in isolation) together with Theorem~\ref{thm:thresholding} (to measure the thresholding error).
Compared to using Theorem~\ref{thm:a_priori_bound} alone, this alternative analysis has the advantage that the two types of error (Rayleigh--Ritz and thresholding) can be bounded independently of each other.
However, to use Theorems~\ref{thm:a_priori_bound} and \ref{thm:thresholding} in this way, an additional piece of information is needed beyond what is required by Theorem~\ref{thm:a_priori_bound} alone, namely the norm $\norm{\vec{c}_0}$.\footnote{The norm of the smallest eigenvector of the perturbed thresholded pair $(\mat{\tilde{H}},\mat{\tilde{S}})$ \rev{can provide} a good \emph{a posteriori} estimate of this quantity in the limit of a small perturbation (cf.\ \cite[\S6.2]{LN20}).}
See \ref{sec:noise-less-case} for a comparison of these two approaches on some numerical examples.

\section{Numerical Experiments}
\label{sec:numer-exper}

In this section, we present some numerical experiments demonstrating the success of the theory in explaining the performance of QSD and other features of the numerical performance of the QSD method.
We consider two sets of examples: the one-dimensional transverse field Ising model (TFIM) and the one-dimensional Hubbard model.

The Hamiltonian of the TFIM with $L$ spins is 
\begin{equation}\label{eqn:ham_tfim}
\oper{H}=-\sum_{i=1}^{L} \oper{Z}_i \oper{Z}_{i+1}-g\sum_{i=1}^L \oper{X}_i,
\end{equation}
where $\oper{X}_i,\oper{Z}_i$ are the Pauli $X,Z$ operators acting on the $i$th spin, respectively. We use the periodic boundary condition, and therefore  $\oper{Z}_{L+1}$ \rev{is identified with} $\oper{Z}_{1}$. 

The Hamiltonian for the (spinful) Hubbard model of $L$ sites is
\begin{equation}
\oper{H}=-\sum_{i=1}^{L} \sum_{\sigma\in\{\uparrow,\downarrow\}}\oper{a}_{i\sigma}^{{\rev{*}}} \oper{a}_{i+1,\sigma}+U\sum_{i=1}^{L}  \oper{a}_{i\uparrow}^{{\rev{*}}}\oper{a}_{i\uparrow} \oper{a}_{i\downarrow}^{{\rev{*}}} \oper{a}_{i\downarrow}.
\label{eqn:fermion_hamiltonian}
\end{equation}
Here $\oper{a}^{{\rev{*}}}_{i\sigma},\oper{a}_{i\sigma}$ are the fermionic creation and annihilation operators at site $i$ with spin $\sigma$, which can be expressed in terms of spin operators following the Jordan--Wigner transformation (see, e.g., \cite{NegeleOrland1988}). 
Similarly due to periodic boundary conditions, $\oper{a}^{{\rev{*}}}_{L+1,\sigma},\oper{a}_{L+1,\sigma}$ \rev{are} identified with $\oper{a}^{{\rev{*}}}_{1\sigma},\oper{a}_{1\sigma}$, respectively.
We are interested in finding the ground state energy of $\oper{H}$. 
The dimension of $\oper{H}$ of the TFIM is $2^L$ and that for the Hubbard model is $2^{2L}$ (due to spin degrees of freedom). 
Due to the high dimensionality, these models are generally intractable to be solved directly when $L$ is large.\footnote{There exists special techniques that are particularly efficient for handling one-dimensional quantum systems. The main point of our numerical results is to demonstrate the performance of QSD algorithms, which does not rely on such special properties.}
In all examples below, we set $L=10$, $g=-\sqrt{2}$ for the TFIM model and $L=10$, $U=8$ for the Hubbard model.\footnote{In particular, $U=8.0$ for the Hubbard model corresponds to a strongly correlated quantum system and is typically considered to be difficult.} The Hubbard model also has an extra parameter called the total number of fermions denoted by $N_e$, which constrains $\oper{H}$ to a smaller diagonal block, and its value is set to the half filling with $N_e=L=10$. We find that the numerical results do not depend sensitively to these parameters. Additional numerical results with other values of $L$, $U$, etc.\ can be found in the Supplementary Materials.

The number of time steps used shall be denoted by $n$, and the time grid is $t_j=j\Delta t$ where $j=0,\ldots,n-1$. The initial vector ${\rev{\vec{\varphi}_0}}$ of the TFIM model is taken to be a product state (an eigenstate with $g=0$), and that of the Hubbard model is taken to be a Slater determinant state (an eigenstate with $U=0$), respectively.
We find that such a setup leads to a sufficiently large initial overlap $|\gamma_0|^2=|\braket*{\varphi_0}{\psi_0}|^2$. This ensures that in the noiseless setting, the ground state energy of $\oper{H}$ can be estimated to high accuracy with a very modest value of $n$ (see Table~\ref{tab:param_model}). 

\begin{table}[t]
\begin{center}
\begin{tabular}{lcc}\toprule
 & TFIM $(L=10)$ & Hubbard $(L=10)$ \\\midrule
Time step $\Delta t$ & 1.0 & 0.1 \\
Initial overlap $|\gamma_0|^2$ & 0.079 & 0.122 \\
Exact $E_0$ & $-15.97997\mathbf{5}0$ &  $-3.3149967\mathbf{3}$ \\
Noiseless $E_0$ with $n=40$ & $-15.97997\mathbf{4}8$ & $-3.3149967\mathbf{0}$ \\\bottomrule
\end{tabular}
\end{center}
\caption{Some parameters for the TFIM and the Hubbard models. Exact $E_0$ is obtained by diagonalizing $\oper{H}$. Noiseless $E_0$ is obtained by solving the projected generalized eigenvalue problem (of size $n\times n$) using the QSD algorithm in the noiseless setting.}
\label{tab:param_model}
\end{table}

The Hamiltonian simulation and the computation of the projected matrix elements are performed using the QuSpin package~\cite{WeinbergBukov2017,WeinbergBukov2019} in Python on a classical computer.%
All further experiments are performed in MATLAB, with the noise matrices $\mat{\Delta}_{\mat{H}}$ and $\mat{\Delta}_{\mat{S}}$ modeled as complex Gaussian Hermitian--Toeplitz matrices with the entries in the first rows independent with specified variances.

\subsection{The Need for Thresholding}
\label{sec:need-thresholding}

\begin{figure}[t]
  \centering
    \begin{subfigure}[b]{0.47\textwidth}
    \centering
    \includegraphics[width=\textwidth]{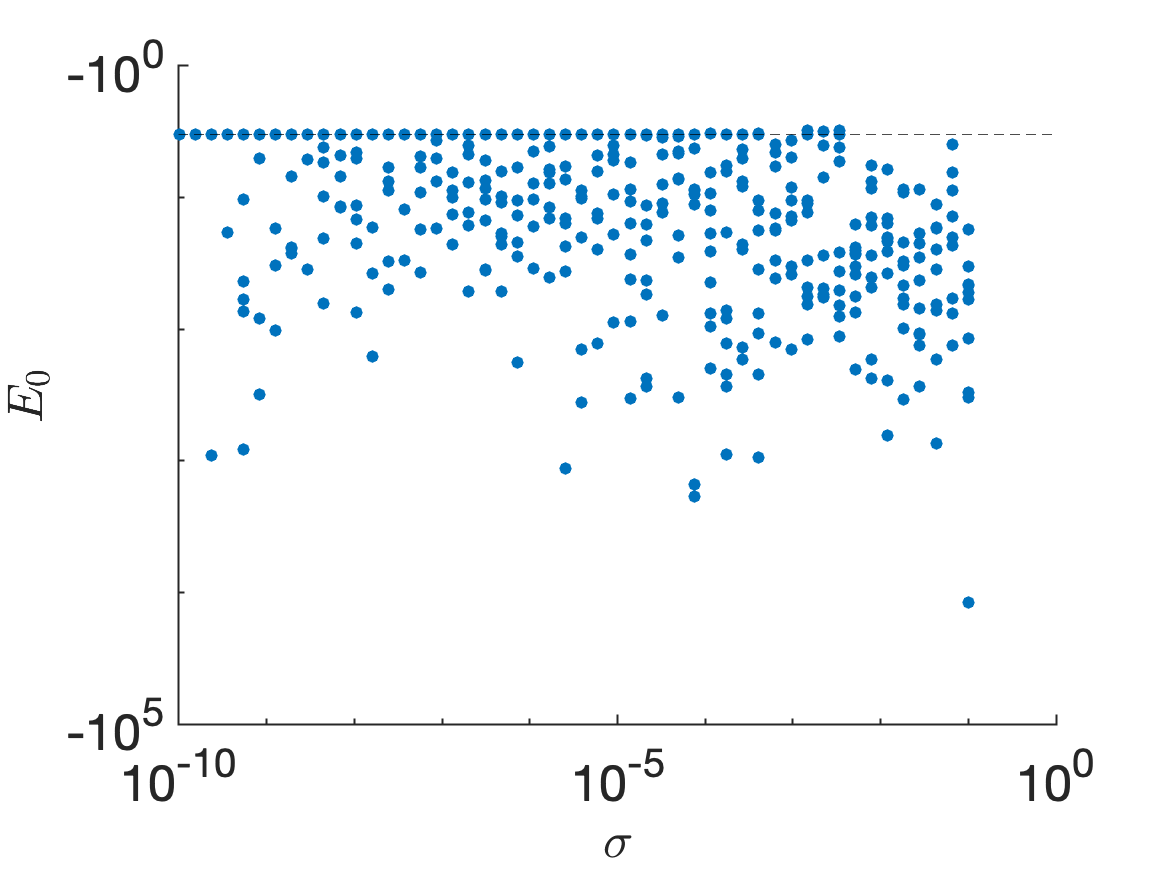}
    \caption{$n = 20$}\label{fig:doing_nothing_q20}
  \end{subfigure}
  ~
  \begin{subfigure}[b]{0.47\textwidth}
    \centering
    \includegraphics[width=\textwidth]{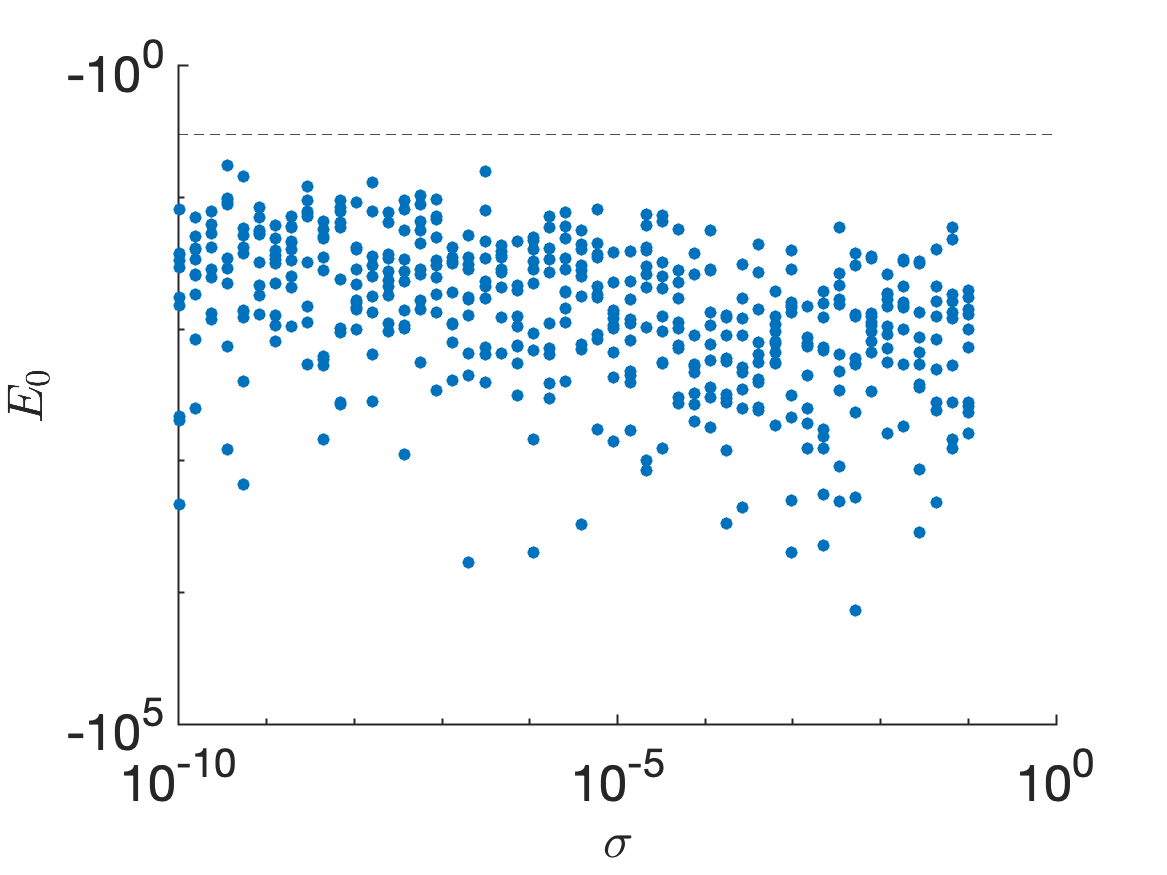}
    \caption{$n = 80$}\label{fig:doing_nothing_q80}
  \end{subfigure}  

  \caption{Least eigenvalues computed from the perturbed pair $(\mat{\tilde{H}},\mat{\tilde{S}})$ without any procedure to ameliorate the affects of noise. Shown are 10 random initializations of the noise for several random noise levels $\sigma$ for the Hubbard example with $n = 20$ (left) and $n = 80$ (right).
  The true eigenvalue is shown as a horizontal dashed line.}
  \label{fig:doing_nothing}
\end{figure}

We first demonstrate why we advocate the use of thresholding by showing the potential pitfalls of some other strategies.
Naturally, the first strategy one might attempt would be to do nothing at all: just solve the generalized eigenvalue problem
\begin{equation} \label{eq:perturbed_gep}
  \mat{\tilde{H}}\vec{\tilde{c}} = \tilde{E} \, \mat{\tilde{S}}\vec{\tilde{c}}
\end{equation}
and return the least computed eigenvalue.
The futility of this strategy is shown in Figure~\ref{fig:doing_nothing}.
Even for extremely low noise levels ($\sigma \approx 10^{-10}$), we see that the recovered least eigenvalue can deviate quite far from the genuine least eigenvalue with high probability.
For nicer problem instances (see, e.g., Figure~\ref{fig:doing_nothing_q20}), characterized by an only modestly ill-conditioned (e.g., $\kappa(\mat{S}) = \norm{\mat{S}}\norm{\mat{S}^{-1}} \lesssim 10^{12}$) $\mat{S}$ matrix, the eigenvalue of interest could perhaps be reliably recovered by taking a median over multiple trials (each of which requiring the quantum computation to be re-run).
However, for problem instances with a more ill-conditioned $\mat{S}$ matrix, such as is shown in Figure~\ref{fig:doing_nothing_q80}, the probability of finding an eigenvalue close to the genuine smallest eigenvalue appears to occur with vanishingly small probability.

Alternately, one might try to apply ``just a bit of thresholding'' by setting the threshold parameter at a small constant value, independent of the noise level.
This can be modestly effective for a well-conditioned $\mat{S}$ matrix (particularly if combined with a median of multiple trials), but it falls down as soon as $\sigma \gtrapprox \epsilon$ in general.
See Figure~\ref{fig:fixed_threshold} for a demonstration of this.
If one is to rely on thresholding alone to deal with the noise, then threshold parameter must be chosen large enough.

As another alternative to thresholding, one might attempt to solve the problem without explicitly filtering out the noise by thresholding (or only using a tiny threshold much smaller than the noise level) and attempting to systematically determine which eigenvalues are ``real''.
We investigate such strategies in section~\ref{sec:failure-heuristics} and ultimately conclude that these are less robust and less accurate than thresholding.

\subsection{Choice of the Thresholding Parameter}
\label{sec:choice-thresh-param}

\begin{figure}[t]
  \centering
    \begin{subfigure}[b]{0.47\textwidth}
    \centering
    \includegraphics[width=\textwidth]{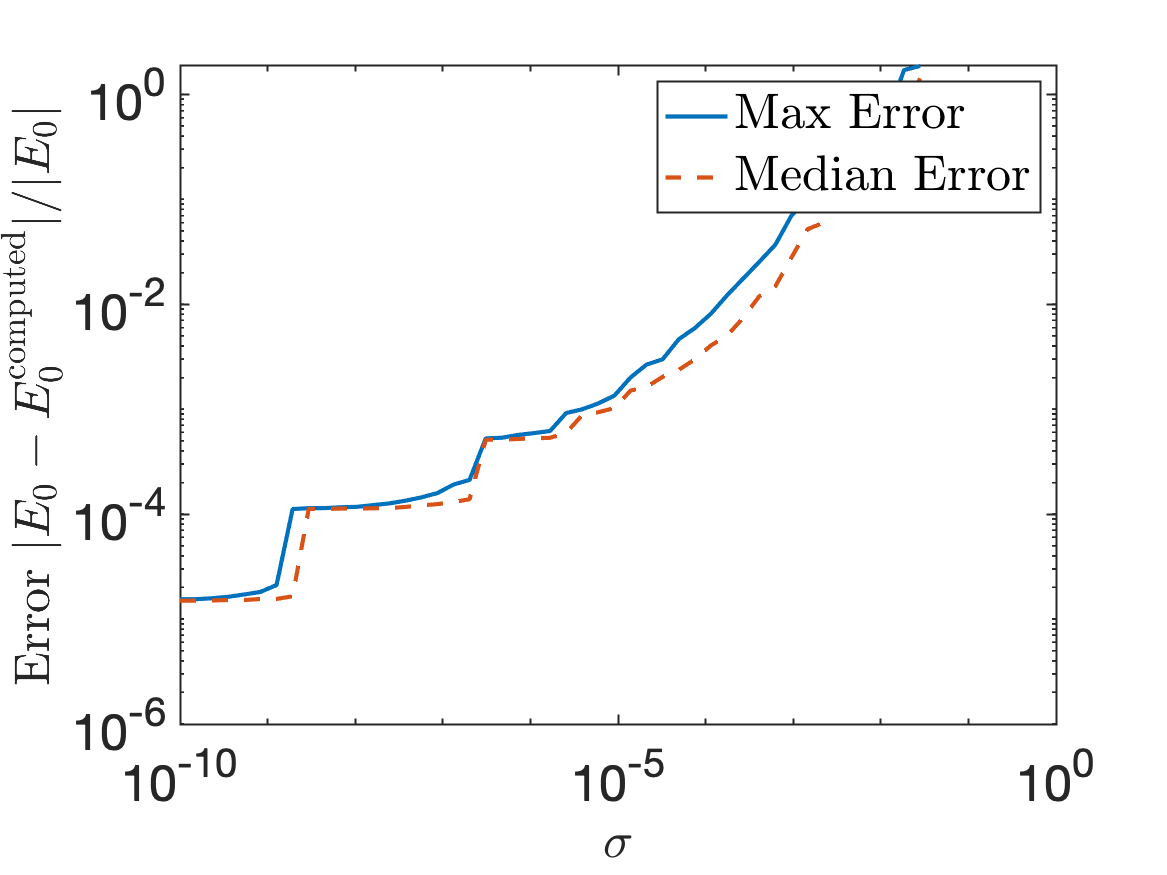}
    \caption{Hubbard, $n = 20$}\label{fig:thresholding_hubbard_l10_u8_q20}
  \end{subfigure}
  ~
  \begin{subfigure}[b]{0.47\textwidth}
    \centering
    \includegraphics[width=\textwidth]{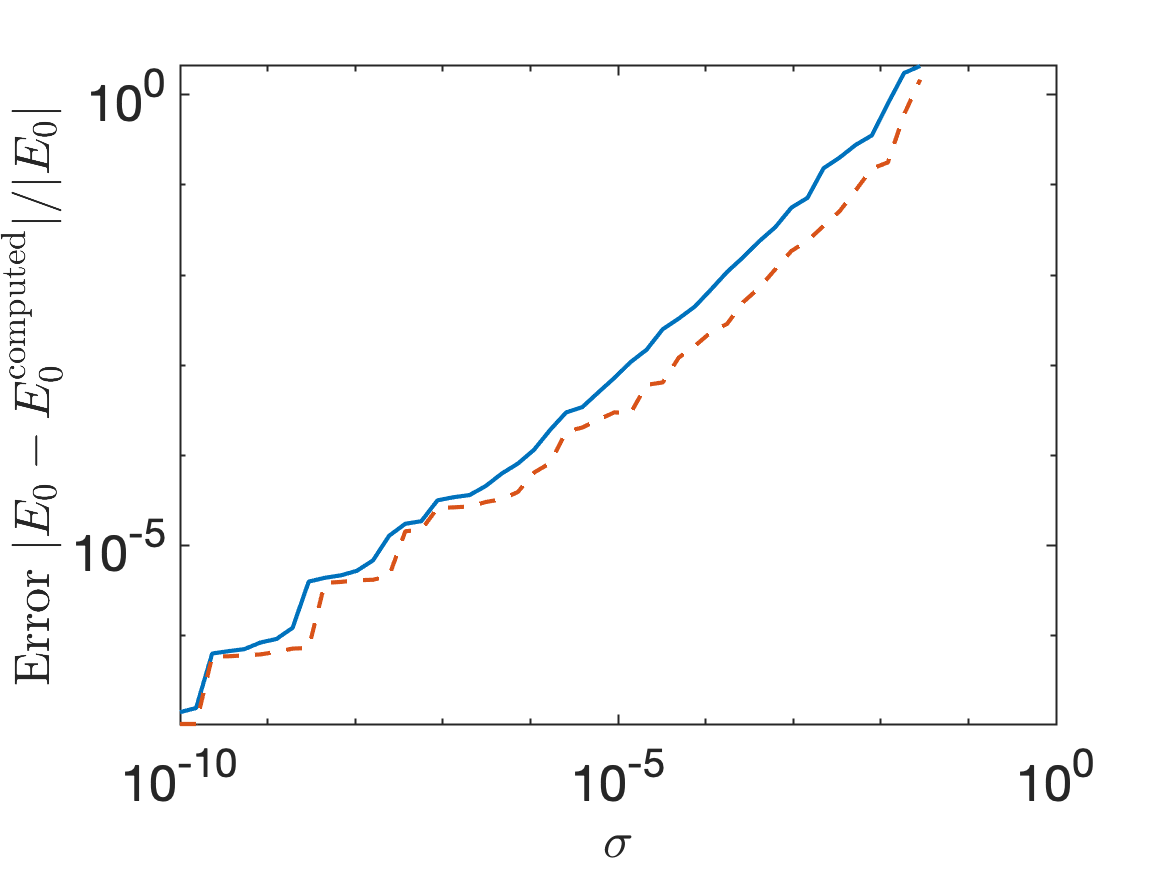}
    \caption{Hubbard, $n = 40$}\label{fig:thresholding_hubbard_l10_u8_q40}
  \end{subfigure}  

  \begin{subfigure}[b]{0.47\textwidth}
    \centering
    \includegraphics[width=\textwidth]{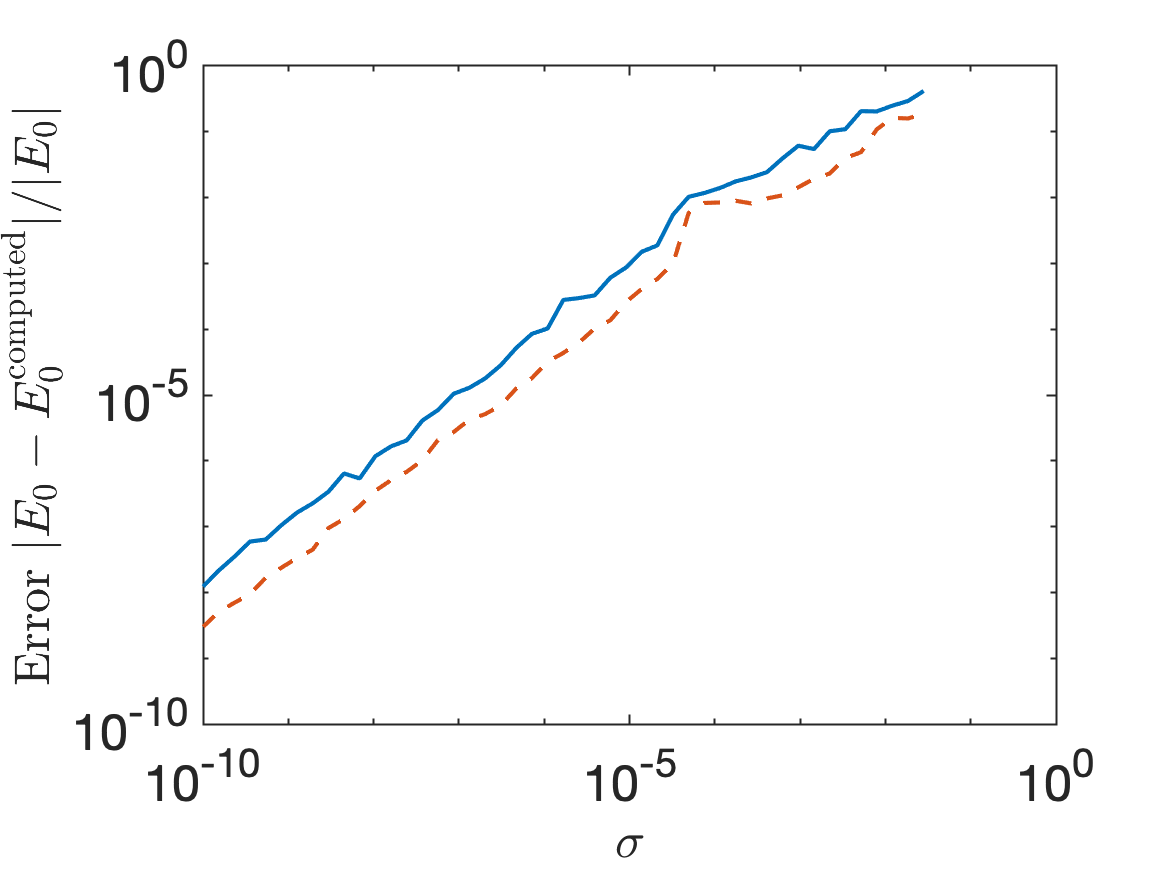}
    \caption{Ising, $n = 20$}\label{fig:thresholding_ising_l10_q20}
  \end{subfigure}
  ~
  \begin{subfigure}[b]{0.47\textwidth}
    \centering
    \includegraphics[width=\textwidth]{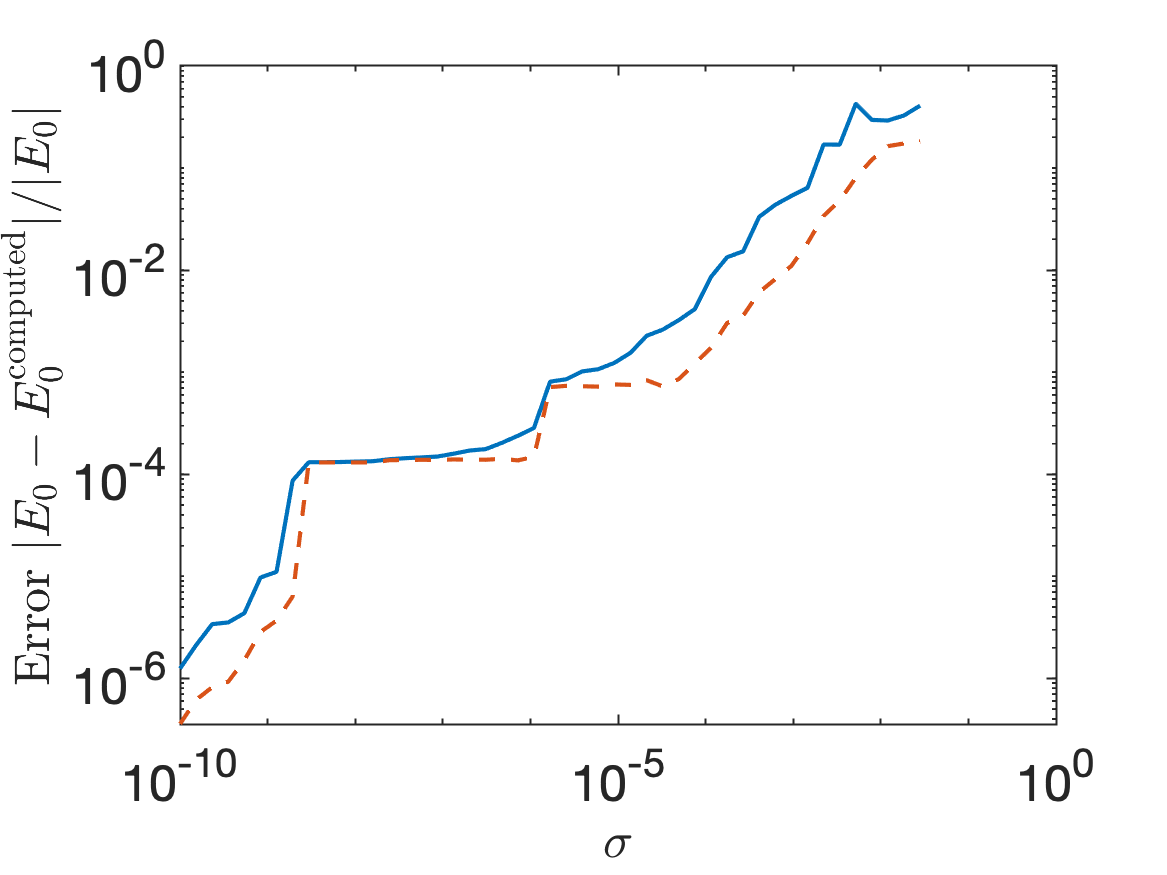}
    \caption{Ising, $n = 40$}\label{fig:thresholding_ising_l10_q40}
  \end{subfigure}
  
  \caption{Maximum (blue solid) and median (red dashed) error over 100 initializations for eigenvalues computed from the noise-perturbed pair $(\mat{\tilde{H}},\mat{\tilde{S}})$ using thresholding with threshold parameter $25\sigma\norm{\mat{\tilde{S}}}$ for Hubbard model (top) and Ising model (bottom) with $n = 20$ (left) and $40$ (right).}  
  \label{fig:thresholding}
\end{figure}

Now that we have demonstrated why we prefer thresholding for solving the noise-perturbed QSD generalized eigenvalue problem \eqref{eq:perturbed_gep}, let us demonstrate the success of thresholding.
In Figure~\ref{fig:thresholding}, we demonstrate the error for the thresholding procedure with a threshold parameter $\epsilon$ proportional to the noise level.\footnote{This is smaller than the theory (Theorem~\ref{thm:H_projection_error} and Corollary~\ref{cor:mathias_li}) predicts $\epsilon$ should be taken as, which suggests that one should choose $\epsilon \propto \sigma^{1/(1+\alpha)}$ ($\alpha \le 1/2$ is the value for which \eqref{eq:geometric_mean} holds).}
(See also Figure~\ref{fig:thresholding_more} for more examples.)
As the plots show, thresholding is robust on these examples in the sense that the maximum error over multiple trials is similar to the median, showing that thresholding is reliably able to filter out the noise over different random initializations.\footnote{This is seen to be not true for one example (Figure~\ref{fig:thresholding_hubbard_l10_u10_q80}) in the Supplementary Material. This shows that thresholding is not infallible, but is still better than alternate strategies within our knowledge (section~\ref{sec:failure-heuristics}).}
Since the norm of the noise matrix is a random quantity prone to occasionally being appreciably larger than its average value,%
the threshold parameter must be somewhat larger than the expected noise level to achieve good performance with high probability.
These plots demonstrate that, for these examples at least, this multiple can be quite modest---just $25$ is enough.
In the error plots in Figure~\ref{fig:thresholding} we see two types of behavior: Sometimes the error increases relatively continuously with the thresholding parameter (e.g., most of Figure~\ref{fig:thresholding_ising_l10_q20}) whereas other times it increases in a more stepwise fashion (e.g., low noise levels in Figure~\ref{fig:thresholding_hubbard_l10_u8_q20}).
The first behavior is indicative of the error being dominated by the noise level (with the slope in log-log space being $\approx 1$ demonstrating a linear dependence of the error on the noise level) with the second exemplifying the thresholding error being the dominant contribution.
Behavior of the second type might suggest that the threshold parameter is being chosen conservatively and lower error could be achieved with a smaller threshold value.

\begin{figure}[t]
  \centering
    \begin{subfigure}[b]{0.47\textwidth}
    \centering
    \includegraphics[width=\textwidth]{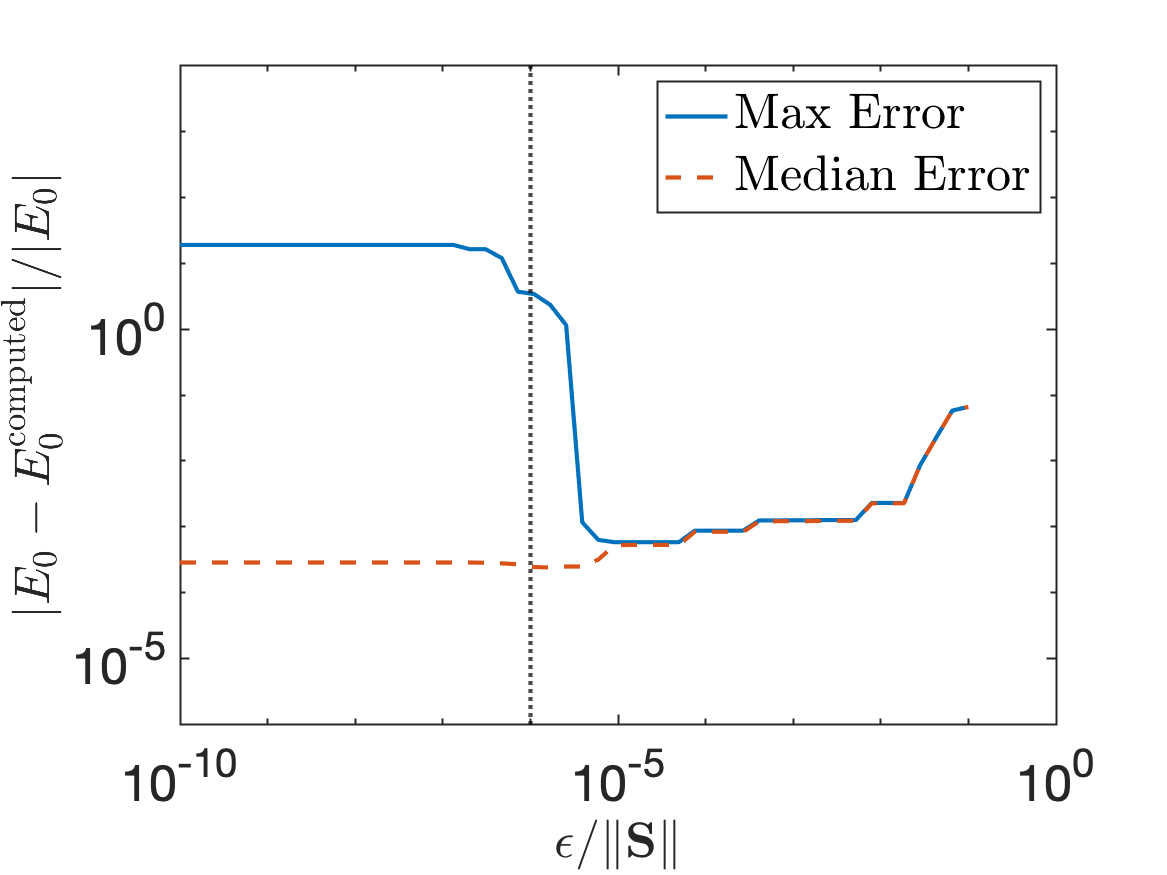}
    \caption{Hubbard, $n = 20$}\label{fig:threshold_choice_hubbard_l10_u8_q20}
  \end{subfigure}
  ~
  \begin{subfigure}[b]{0.47\textwidth}
    \centering
    \includegraphics[width=\textwidth]{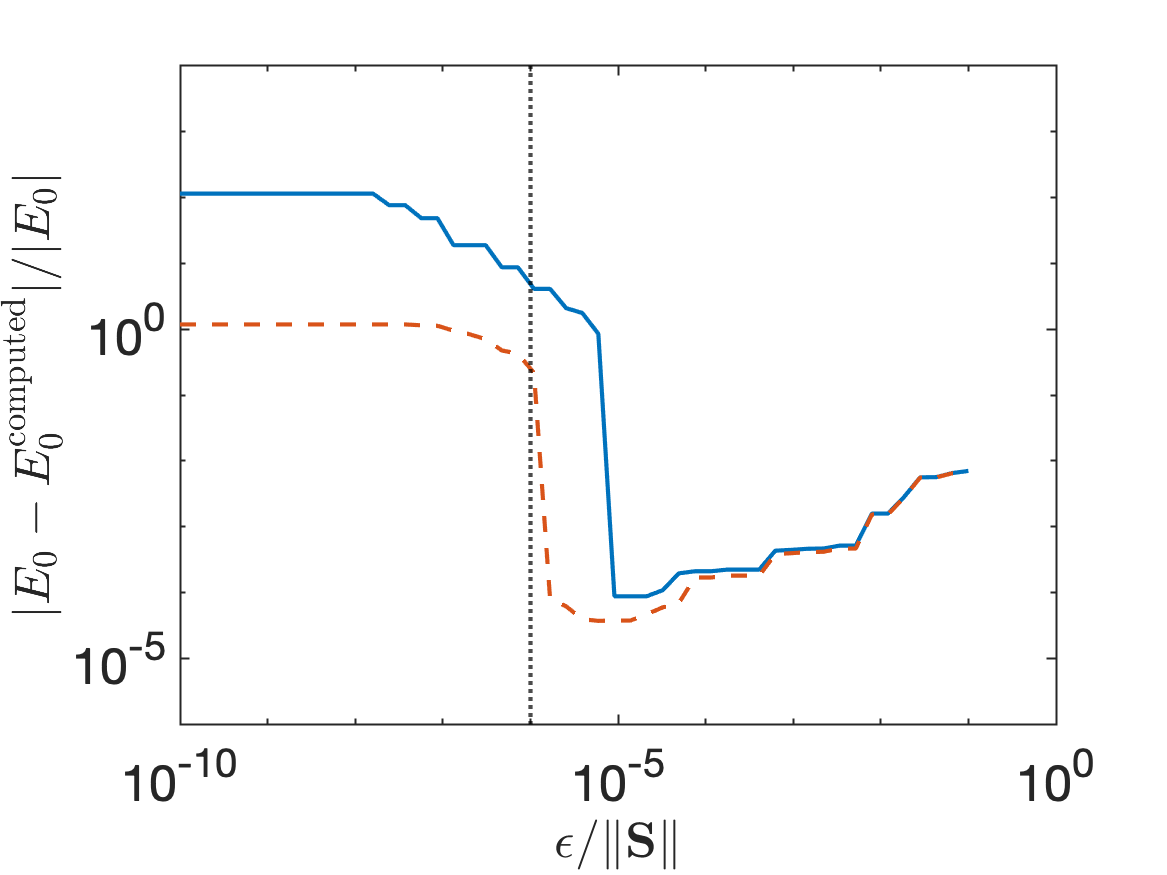}
    \caption{Hubbard, $n = 40$}\label{fig:threshold_choice_hubbard_l10_u8_q40}
  \end{subfigure}  

  \begin{subfigure}[b]{0.47\textwidth}
    \centering
    \includegraphics[width=\textwidth]{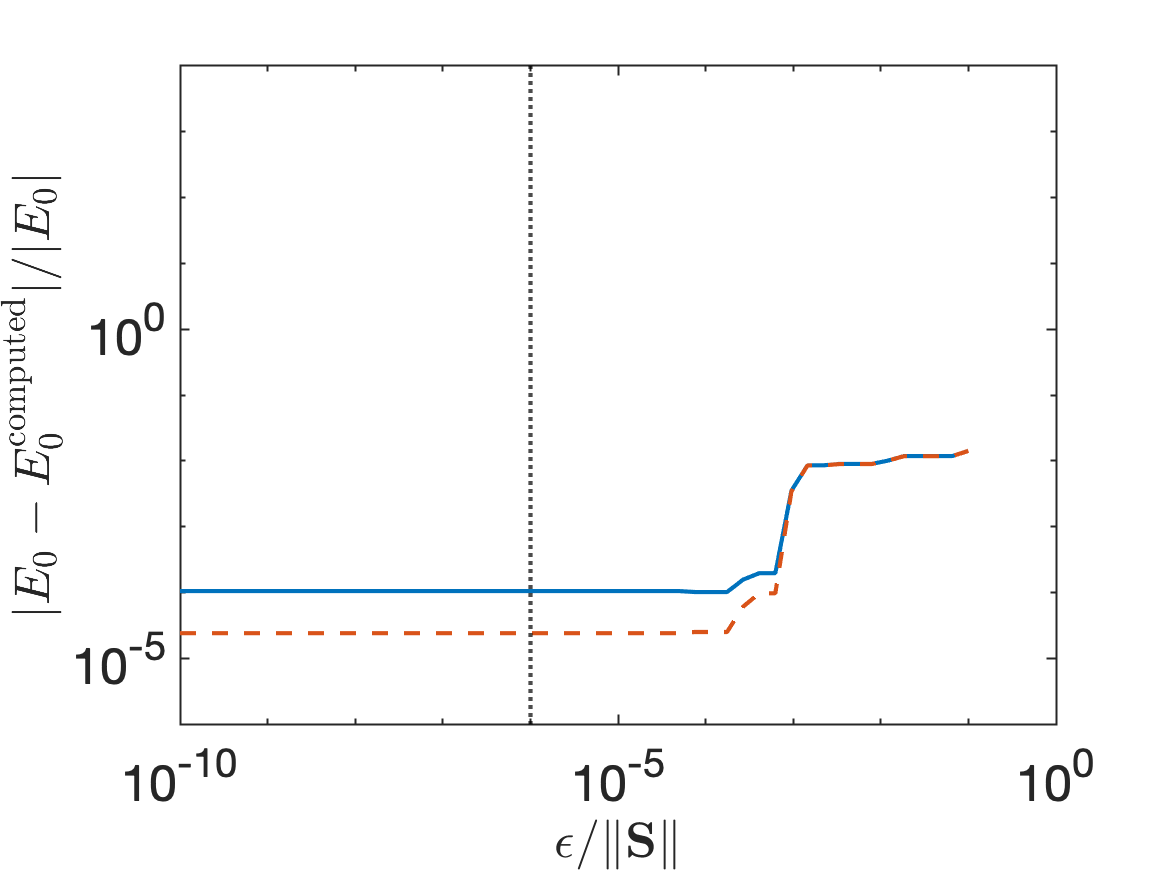}
    \caption{Ising, $n = 20$}\label{fig:threshold_choice_ising_l10_q20}
  \end{subfigure}
  ~
  \begin{subfigure}[b]{0.47\textwidth}
    \centering
    \includegraphics[width=\textwidth]{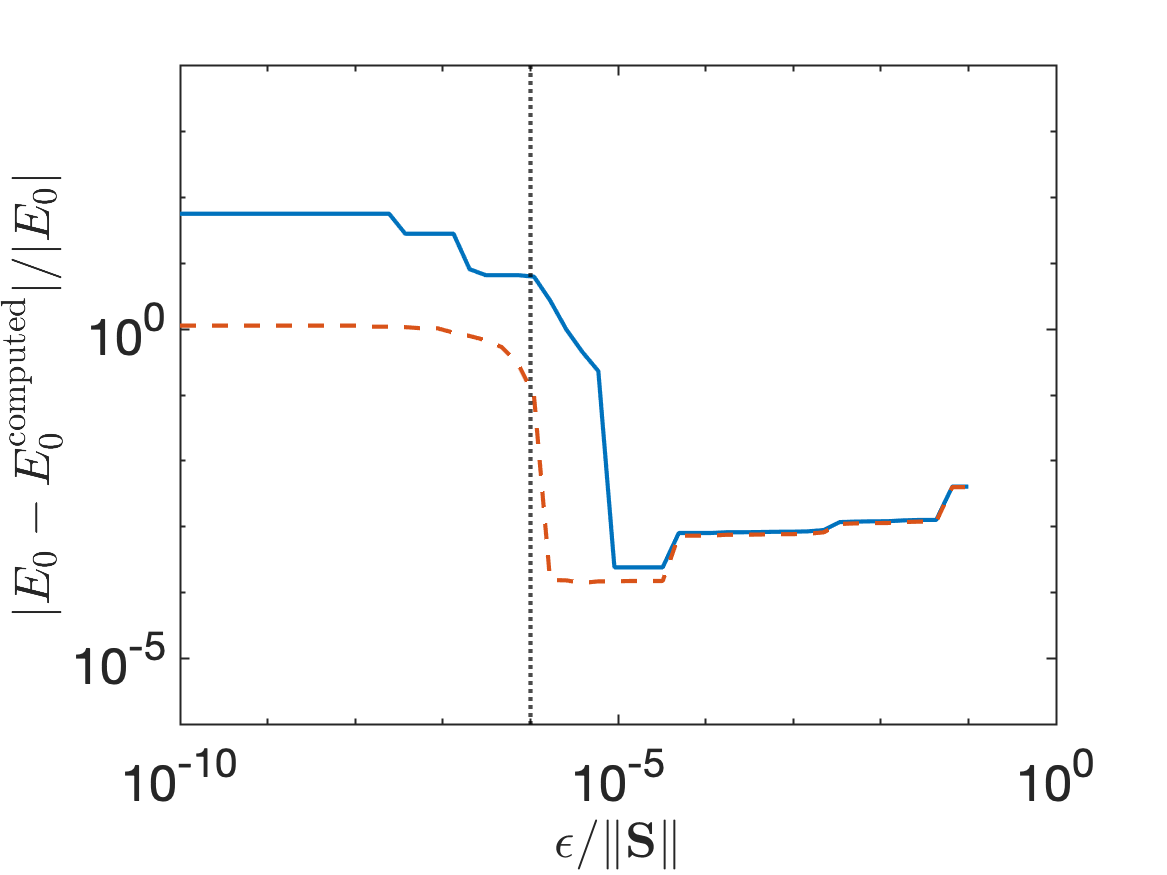}
    \caption{Ising, $n = 40$}\label{fig:threshold_choice_ising_l10_q40}
  \end{subfigure}
  
  \caption{Maximum (blue solid) and median (red dashed) error over 100 initializations for eigenvalues computed from the noise-perturbed pair $(\mat{\tilde{H}},\mat{\tilde{S}})$ using thresholding for various values of the threshold $\epsilon$ for a fixed noise level $\sigma = 10^{-6}$ (dotted black line) for Hubbard model (top) and Ising model (bottom) with $n = 20$ (left) and $40$ (right).}
  \label{fig:threshold_choice}
\end{figure}

The choice of the thresholding parameter is critical to the success of the method, as is demonstrated in Figure~\ref{fig:threshold_choice}.
In these plots, we show the median and maximum error for the computed eigenvalue over 100 random initializations of the noise (fixed to $\sigma = 10^{-6}$) for different thresholding levels.
In the normative case, the error decreases as the threshold parameter is decreased up until the threshold parameter reaches a modest multiple of the noise level, after which the error sharply rises.
In the Supplementary Materials (section~\ref{sec:auto_thresh}), we discuss an automatically tuned variant of the thresholding procedure which can help in picking a good threshold level $\epsilon$.

\section{Conclusions}
\label{sec:conclusions}

In this article, we have presented the first theoretical analysis of the accuracy of the quantum subspace diagonalization method with a thresholding procedure.
Our explanation has two parts:

\begin{enumerate}
    \item With an appropriate choice of time sequence $\{t_j\}$, QSD with thresholding (Algorithm~\ref{alg:thresholding}) \rev{can compute} an accurate approximation to the ground-state energy. (Theorem~\ref{thm:a_priori_bound})
    \item \rev{Under appropriate conditions,} the thresholding procedure \rev{can} robustly \rev{determine} the smallest eigenvalue in the presence of perturbations to the pair $(\mat{H},\mat{S})$. (Theorem~\ref{thm:main_theorem_formal})
\end{enumerate}

These two pieces combine to give a bound on the total error of the QSD procedure (comprising Rayleigh--Ritz, thresholding, and perturbation errors) in Informal Theorem~\ref{thm:main_theorem_big}. 
The conditions of our theory are natural, and many of the parameters in our bounds can be estimated in the presence of noise, allowing our bounds to be able to give approximate bounds on the error \emph{a posteriori}.
Our numerical experiments (including additional experiments in the Supplementary Material) support the conclusion that QSD is accurate when implemented with thresholding (and not accurate when implemented without).

Our theoretical estimate\rev{s} can still be significantly improved, such as the bound 
$\chi_{\mat{H}} \le \order(\eta_{\mat{S}}/\epsilon^\alpha + \eta_{\mat{H}})$ (see Theorem~\ref{thm:H_projection_error}) for the discrepancy between the thresholded $\mat{H}$ and $\mat{\tilde{H}}$ matrices, where $0\le \alpha \le 1/2$.
This suggests that we need to take $\epsilon = \Omega(\eta_{\mat{S}}^{1/(1+\alpha)})$ for accurate recovery to be guaranteed by Corollary~\ref{cor:mathias_li}, which has hypothesis $\chi \le \epsilon / q$.
This is in contradiction to our numerical experiments, where $\epsilon$ can be chosen to be a small multiple of $\eta$.
Synthetically generated worst-case examples (see section~\ref{sec:evid-tightn-theor}) suggests the bound $\chi_{\mat{H}} \le \order(\eta_{\mat{S}}/\epsilon^\alpha + \eta_{\mat{H}})$ is tight, but it remains possible a better bound can be derived for pairs $(\mat{H},\mat{S})$ generated by QSD.
An interesting open question is to give a convincing explanation for why we appear to have $\alpha = 1/4$ and $\mu \approx \max |\Lambda(\mat{H},\mat{S})|$ in \eqref{eq:geometric_mean} for many QSD instances.
Despite the modest size of $n$ in practice, the polynomial dependence on $n$ in Corollary~\ref{cor:mathias_li} and Theorems~\ref{thm:H_projection_error}, \ref{thm:S_projection_error}, and \ref{thm:low_rank_approximation} can lead to significant overestimates of the error.
Therefore, another natural question is whether these dimensional factors can be improved.  

Natural extensions of this work are to generalized our analysis to excited states (interior eigenvalues) and to develop bounds on the accuracy of the computed eigenvectors.
\rev{Although the QSD algorithm cannot be used to coherently prepare an eigenstate on a quantum computer, we may still compute other physical observables from the approximate eigenstate.}
Mathias and Li's theory \cite[\S6]{ML04} suggests eigenvectors might be more sensitive to the noise than the eigenvalues, 
and we plan to study the accuracy of the computed eigenvectors in future work.

\section*{Acknowledgments}
This work is supported by the U.S.\ Department of Energy, Office of Science, Office of Advanced Scientific Computing Research, Department of Energy Computational Science Graduate Fellowship under Award Number DE-SC0021110 (ENE), by the Department of Energy under Grant No. DE-SC0017867, and by the NSF Quantum Leap Challenge Institute (QLCI) program through grant number OMA-2016245 (LL).  LL is a Simons Investigator.
ENE thanks the Lawrence Berkeley Laboratory summer student program for providing a welcoming environment to perform this work.
We thank Yulong Dong, Yu Tong, Norman Tubman\rev{, and Robert Webber} for helpful discussions.

\section*{Disclaimer}
\label{sec:disclaimer}

This report was prepared as an account of work sponsored by an agency of the United States Government. Neither the United States Government nor any agency thereof, nor any of their employees, makes any warranty, express or implied, or assumes any legal liability or responsibility for the accuracy, completeness, or usefulness of any information, apparatus, product, or process disclosed, or represents that its use would not infringe privately owned rights. Reference herein to any specific commercial product, process, or service by trade name, trademark, manufacturer, or otherwise does not necessarily constitute or imply its endorsement, recommendation, or favoring by the United States Government or any agency thereof. The views and opinions of authors expressed herein do not necessarily state or reflect those of the United States Government or any agency thereof.

\bibliographystyle{siamplain}
\bibliography{references}

\ifarxiv
\newpage

\begin{center}
  \textbf{SUPPLEMENTARY MATERIAL}
\end{center}

\setcounter{section}{0}
\renewcommand{\thesection}{SM\arabic{section}}
\setcounter{figure}{0}
\renewcommand{\thefigure}{SM\arabic{figure}}
\input{supplement_core.tex}
\fi

\end{document}

%% file: ex_shared.tex
\usepackage{lipsum}
\usepackage{amsfonts}
\usepackage{graphicx}
\usepackage{epstopdf}
\DeclareGraphicsExtensions{.eps,.pdf,.png,.jpg}

\input{our_macros.tex}

\newsiamremark{remark}{Remark}
\newsiamremark{hypothesis}{Hypothesis}
\crefname{hypothesis}{Hypothesis}{Hypotheses}
\newsiamthm{claim}{Claim}
\newsiamthm{fact}{Fact}
\newsiamthm{inftheorem}{Informal Theorem}

\headers{A Theory of Quantum Subspace Diagonalization}{E. N. Epperly, L. Lin, and Y. Nakatsukasa}

\title{A Theory of Quantum Subspace Diagonalization
  \thanks{This is a preprint version of \emph{A Theory of Quantum Subspace Diagonalization} (\url{https://doi.org/10.1137/21M145954X}), which appeared in the SIAM Journal on Matrix Analysis and Applications on August 1, 2022.  
  }
}

\author{Ethan N. Epperly\thanks{Department of Computing and Mathematical Sciences, California Institute of Technology, Pasadena, CA, USA 
  (\email{eepperly@caltech.edu}).}
\and Lin Lin\thanks{Department of Mathematics, and Challenge Institute of Quantum Computation, University of California Berkeley, Berkeley, CA, USA and Computational Research Division, Lawrence Berkeley National Laboratory, Berkeley, CA, USA 
  (\email{linlin@math.berkeley.edu}).}
\and Yuji Nakatsukasa\thanks{Mathematical Institute, Oxford University, Oxford, UK (\email{nakatsukasa@maths.ox.ac.uk}).}}

\usepackage{amsopn}


%% file: our_macros.tex
\usepackage{amssymb}
\usepackage{subcaption}
\usepackage{enumitem}
\usepackage{mathtools}
\usepackage{mleftright}
\usepackage{color}
\usepackage{booktabs}

\usepackage{algorithm, algpseudocode, algorithmicx}

\usepackage{soul}

\newcommand{\complex}{\mathbb{C}}

\DeclareMathOperator{\tr}{tr}
\DeclareMathOperator{\diag}{diag}
\newcommand{\mat}[1]{\boldsymbol{#1}}
\renewcommand{\vec}[1]{\boldsymbol{#1}}
\usepackage{stmaryrd}
\newcommand{\lowrank}[1]{\left\llbracket #1 \right\rrbracket}
\DeclarePairedDelimiter\norm{\|}{\|}
\DeclareMathOperator{\spn}{span}

\newcommand{\expmat}[1]{\begin{bmatrix} #1 \end{bmatrix}}
\newcommand{\twobytwo}[4]{\expmat{#1 & #2 \\ #3 & #4}}

\newcommand{\order}{\mathcal{O}}

\newcommand{\set}[1]{\mathsf{#1}}
\newcommand{\e}{\mathrm{e}}
\newcommand{\iu}{\mathrm{i}}

\renewcommand{\hat}[1]{\widehat{#1}}
\renewcommand{\tilde}[1]{\widetilde{#1}}
\newcommand{\rev}[1]{{#1}}

\DeclarePairedDelimiter\ket{\lvert}{\rangle}
\DeclarePairedDelimiterX\braket[2]{\langle}{\rangle}{#1 \,\delimsize\vert\, #2}
\DeclarePairedDelimiterX\braoperket[3]{\langle}{\rangle}{#1 \,\delimsize\vert\, #2 \,\delimsize\vert\, #3}
\newcommand{\oper}[1]{\rev{\mat{\hat{#1}}}}
\newcommand{\trigpoly}{\mathcal{T}}

%% file: supplement_core.tex
\section{Proof of Theorem~\ref{thm:low_rank_approximation}}
\label{sec:proof_theorem_4.1}

For reference, we provide a complete proof of Theorem~\ref{thm:low_rank_approximation}.

\begin{proof}[Proof of Theorem~\ref{thm:low_rank_approximation}]
  Let $\mat{\Pi}$ and $\mat{\tilde{\Pi}}$ be the spectral projectors onto the dominant $m$-dimensional invariant subspaces of $\mat{A}$ and $\mat{A}+\mat{\Delta}$ respectively.
  First, we bound
  \begin{align*}
    \norm*{\lowrank{\mat{A}+\mat{\Delta}}_m - \lowrank{\mat{A}}_m}_{\rm QUI}
    &= \norm{\mat{\tilde{\Pi}}(\mat{A}+\mat{\Delta})\mat{\tilde{\Pi}} - \mat{\Pi}\mat{A}\mat{\Pi}}_{\rm QUI} \\
    &\le \norm{\mat{\tilde{\Pi}}\mat{\Delta}\mat{\tilde{\Pi}}}_{\rm QUI} + \norm{\mat{\tilde{\Pi}}\mat{A}\mat{\tilde{\Pi}} - \mat{\Pi}\mat{A}\mat{\Pi}}_{\rm QUI}.
  \end{align*}
  For the first term, we bound $\norm{\mat{\tilde{\Pi}}\mat{\Delta}\mat{\tilde{\Pi}}}_{\rm QUI} \le \norm{\mat{\Delta}}_{\rm QUI}$ using the fact $\norm{\cdot}_{\rm QUI}$ is \emph{symmetric} in the sense that $\norm{\mat{B}_1\mat{B}_2\mat{B}_3}_{\rm QUI} \le \norm{\mat{B}_1}\norm{\mat{B}_2}_{\rm QUI}\norm{\mat{B}_3}$ \cite[Thm.~3.9]{SS90}.
  Every quadratic unitarily invariant norm is bounded by the Frobenius norm, which follows from the definition of quadratic unitarily invariant norm and the fact that the nuclear norm bounds every unitarily invariant norm \cite[Eq.~(IV.38)]{Bha97}.
  Thus, the second term is  bounded as
  \begin{equation*}
    \norm{\mat{\tilde{\Pi}}\mat{A}\mat{\tilde{\Pi}} - \mat{\Pi}\mat{A}\mat{\Pi}}_{\rm QUI} \le \norm{\mat{\tilde{\Pi}}\mat{A}\mat{\tilde{\Pi}} - \mat{\Pi}\mat{A}\mat{\Pi}}_{\rm F},
  \end{equation*}
  which is then bounded by Theorem~\ref{thm:S_projection_error} with $\epsilon = \lambda_{m+1} + \norm{\mat{\Delta}}$ and $\rho = (\lambda_m - \lambda_{m+1} - \norm{\mat{\Delta}})/(\lambda_{m+1} - \norm{\mat{\Delta}})$.
\end{proof}

\section{The Failure of Heuristics}
\label{sec:failure-heuristics}

In the main text, we show how accurate recovery for the QSD algorithm usually fails, if one solves the noisy generalized eigenvalue problem \eqref{eq:perturbed_gep} with no special treatment. 
A forthcoming example (Figure~\ref{fig:fixed_threshold}) shows that using a fixed threshold independent of the noise level may not perform much better.
An alternate strategy, which we initially believed to be more promising than thresholding, is to compute the eigenvalues (either with no thresholding at all or with a small threshold independent of the noise level) and attempt to determine real from spurious eigenvalues by means of some property of the computed eigenvector.
In this section, we shall consider a couple variants of such an approach and ultimately conclude the performance of these heuristics can still be unsatisfactory when compared to thresholding.

Two natural heuristics for whether $\tilde{E}$ is a plausible candidate for the ground state energy suggest themselves.
Let $\vec{\tilde{c}}$ be the unit-norm eigenvector associated with a computed eigenvalue $\tilde{E}$.
The Ritz vector $\rev{\vec{\tilde{\psi}}_0} := \sum_{j=0}^{n-1} \vec{\tilde{c}}_j \rev{\vec{\varphi}_j}$ is supposed to be close to the true ground-state eigenvector $\rev{\vec{\psi}_0}$ of $\oper{H}$.
Our heuristics are as follows:

\begin{enumerate}
\item \textbf{Require $h_1 := \vec{\tilde{c}}^{\rev{*}}\mat{\tilde{S}}\vec{\tilde{c}}$ to be large.}
  The squared norm of the Ritz vector is precisely $\vec{\tilde{c}}^{\rev{*}}\mat{S}\vec{\tilde{c}} \approx h_1$.
  If $h_1$ is small, then the norm of the Ritz vector is very small due to cancellations in the sum $\sum_{j=0}^{n-1} \vec{\tilde{c}}_j \rev{\vec{\varphi}_j}$ and should thus be treated as suspect because of the noise.
  Thus, it is natural to insist on a large value of $h_1$.\footnote{$h_1$ is also related to the conditioning of the eigenvalue \cite[Eq.~(VI.2.2)]{SS90}.}
\item \textbf{Require the estimated overlap $h_2 := |\vec{e}_0^{\rev{*}}\mat{\tilde{S}}\vec{\tilde{c}}| \approx |\rev{\vec{\varphi}_0^*\vec{\tilde{\psi}}_0}|$ to be large.}
  It is important that the initial vector $\rev{\vec{\varphi}_0}$ has a relatively large initial overlap $|\rev{\vec{\varphi}_0^*\vec{\psi}_0}|$ with the eigenvector of interest---indeed, our analysis suggests accurate recovery of the ground-state energy requires this (see Theorem~\ref{thm:a_priori_bound}).
  As such, it is natural to use the overlap (or its surrogate $h_1$ computable from the noise-corrupted $\mat{\tilde{S}}$ matrix) as a measure of whether an eigenvalue is a genuine candidate for the ground-state energy.
  Note that by unit-norm scaling $\vec{\tilde{c}}$ (rather than adopting the normalization $\vec{\tilde{c}}^{\rev{*}}\mat{\tilde{S}}\vec{\tilde{c}} = 1$), we are implicitly also incorporating the condition for $\rev{\vec{\tilde{\psi}}_0}$ to be a stable linear combination of the basis states which motivated our interest in $h_1$.
\end{enumerate}

There are several ways of using a heuristic $h \in \{h_1,h_2\}$ as an algorithm for computing the ground-state eigenvalue: (a) pick $E$ with the highest $h$, (b) pick the smallest $E$ of the eigenvalues with the top $k$ values of $h$, and (c) pick the smallest $E$ with $h$ above some thresholding $h_0$ (or simply the largest $h$ if none exceeds $h_0$).

Unfortunately, unlike thresholding where there is a natural choice of the threshold parameter (related to the noise level $\eta$ \eqref{eq:eta} which can usually be reliably estimated), we are unaware of any good systematic ways to pick the parameters $k$ and $h_0$ for strategies (b) and (c).
These heuristics thus usually require some tuning to make them accurate for a given problem instance, with the parameters needing to be readjusted when a new problem is encountered.
This reduces the reliability of these heuristics, when the ground truth is unavailable to compare against.
The robustness of heuristics such as (a), (b), and (c) can be improved by medians of repeated trials or by comparing the results of different heuristics against each other.
However, even with such improvements, without rigorous guarantees, the validity of these heuristics remains conjectural when the genuine ground-state energy is unavailable to be validated against.

\begin{figure}[t]
  \centering
  \begin{subfigure}[b]{0.97\textwidth}
    \centering
    \includegraphics[width=\textwidth]{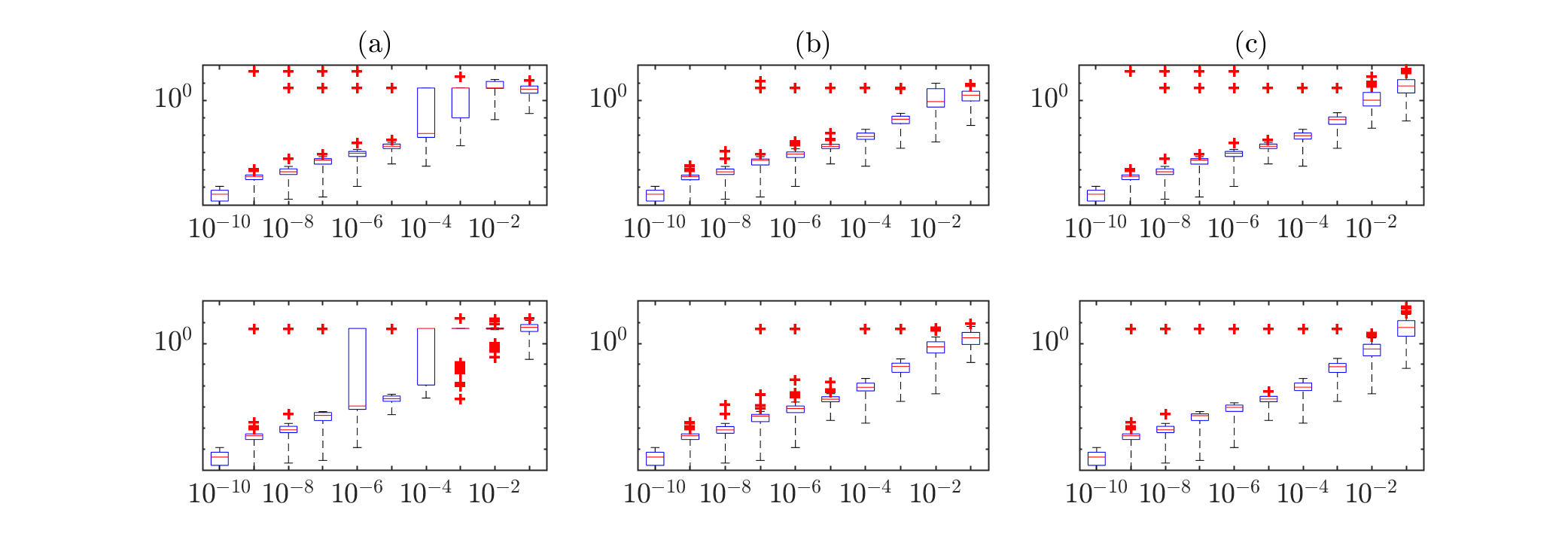}
    \caption{Hubbard, $n = 20$}\label{fig:heuristic_hubbard_q20}
  \end{subfigure}  
  
  \begin{subfigure}[b]{0.97\textwidth}
    \centering
    \includegraphics[width=\textwidth]{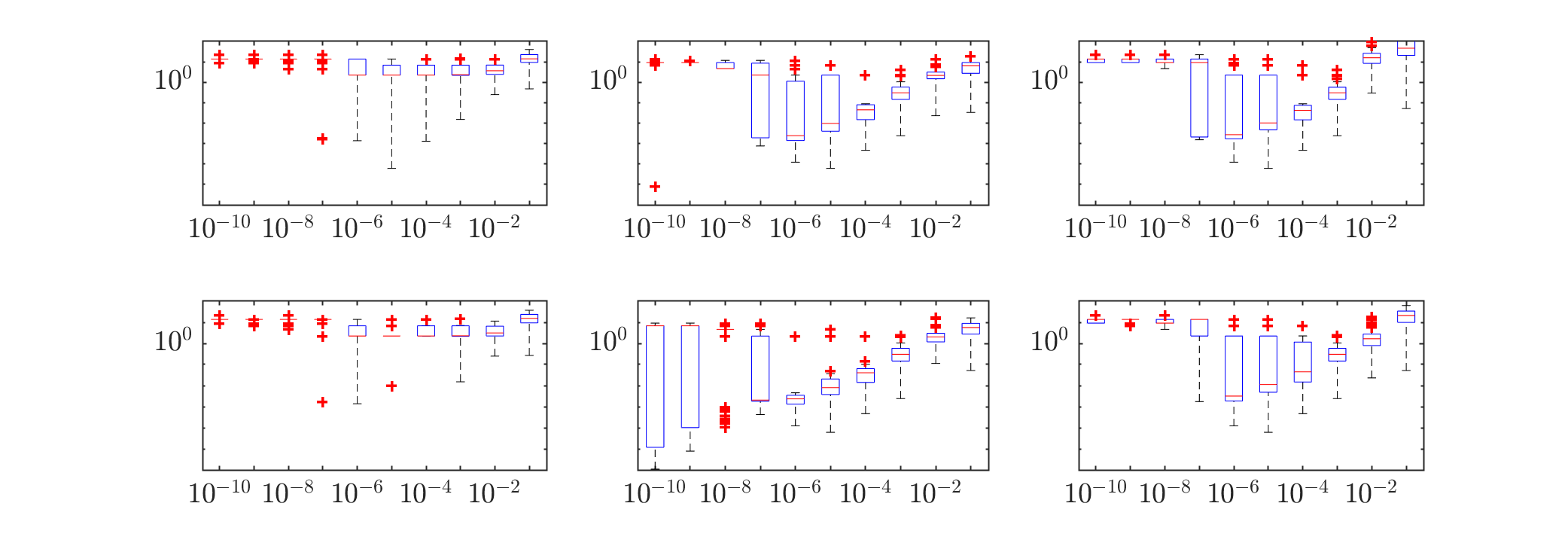}
    \caption{Ising, $n = 40$}\label{fig:heuristic_ising_q40}
  \end{subfigure}

  \caption{Errors (vertical axis) for eigenvalues computed from the perturbed pair $(\mat{\tilde{H}},\mat{\tilde{S}})$ with heuristics (a), (b), and (c) described in the text for quality metrics $h_1$ (first row) and $h_2$ (second row).
    The noisy generalized eigenvalue problem \eqref{eq:perturbed_gep} is solved using thresholding with a fixed threshold parameter $\epsilon = 10^{-12}\norm{\mat{\tilde{S}}}$.
    Shown are 100 random initializations of the noise for several random noise levels $\sigma$ (horizontal axis) for the Hubbard example with $n = 20$ (top) and Ising example with $n = 40$ (bottom).}
  \label{fig:heuristics}
\end{figure}

Figure~\ref{fig:heuristics} shows the suggested heuristics (a), (b), and (c) with the figures of merit $h_1$ and $h_2$ with $k = 5$ and $h_0 = 10^{-2}\norm{\mat{\tilde{S}}}$.
The first subfigure, Figure~\ref{fig:heuristic_hubbard_q20}, shows a relatively optimistic case for the heuristics.
For low levels of noise, the eigenvalue is generally recovered with low error with the exception of a few outliers which could be ameliorated by the median trick.
Figure~\ref{fig:heuristic_ising_q40} shows the potential danger of applying these heuristics; despite working well for the Hubbard example with $n=20$ in Figure~\ref{fig:heuristic_hubbard_q20}, the heuristics fail with the same parameter choices for the Ising example with $n=40$.
For this problem, the eigenvalues are observed to be recovered accurately only with very small probability.
Improvements to both plots are likely possible by more careful choice of the heuristic parameters or more complicated heuristics, but this is a point against such heuristics rather than for them: we ideally want a method which works well without tuning problem-dependent parameters.

\section{Automatic Thresholding} \label{sec:auto_thresh}

\begin{algorithm}[t]
  \caption{Automatically tuned thresholding procedure for finding the least eigenvalue of a noise-corrupted generalized eigenvalue problem.} \label{alg:auto_thresholding}
  \begin{algorithmic}
    \Procedure{AutoThresholding}{$\mat{H}$, $\mat{S}$, $\epsilon_0$, $r$}
    \State $E \leftarrow \Call{Thresholding}{\mat{H},\mat{S},\epsilon_0}$
    \State $\Lambda \leftarrow \{ \lambda \in \mathtt{eig}(\mat{S}) : \lambda < \epsilon_0 \}$
    \While{$\Lambda\ne \emptyset$}
    \State $\epsilon \leftarrow \max \Lambda$, $\Lambda \leftarrow \Lambda \setminus \{\epsilon\}$
    \State $E' \leftarrow \Call{Thresholding}{\mat{H},\mat{S},\epsilon}$
    \If{$|E - E'| / \rev{\min(|E|, |E'|)} > r$}
    \State \textbf{break}
    \EndIf
    \State $E \leftarrow E'$
    \EndWhile 
    \State \Return $E$
    \EndProcedure
  \end{algorithmic}
\end{algorithm}

\begin{figure}[t]
  \centering
    \begin{subfigure}[b]{0.47\textwidth}
    \centering
    \includegraphics[width=\textwidth]{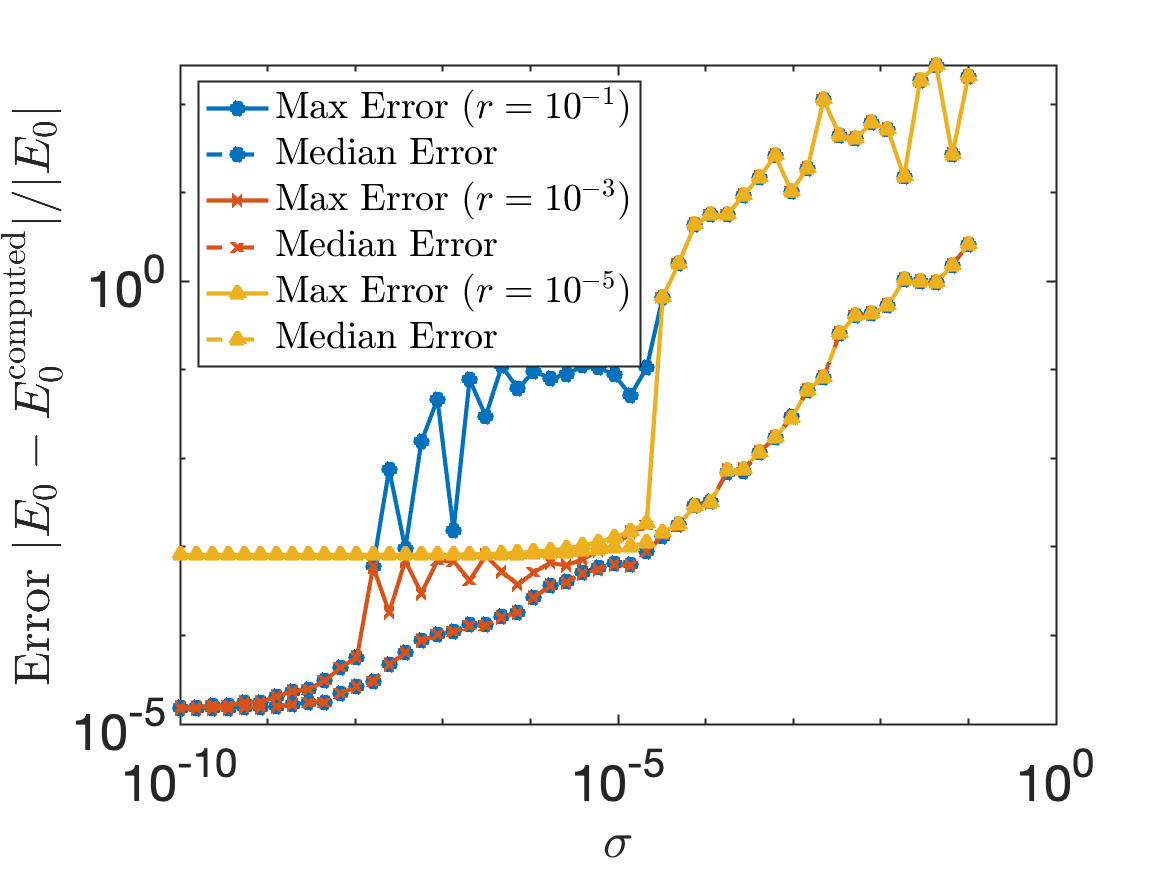}
    \caption{Hubbard, $n = 20$}\label{fig:auto_thresholding_hubbard_l10_u8_q20}
  \end{subfigure}
  ~
  \begin{subfigure}[b]{0.47\textwidth}
    \centering
    \includegraphics[width=\textwidth]{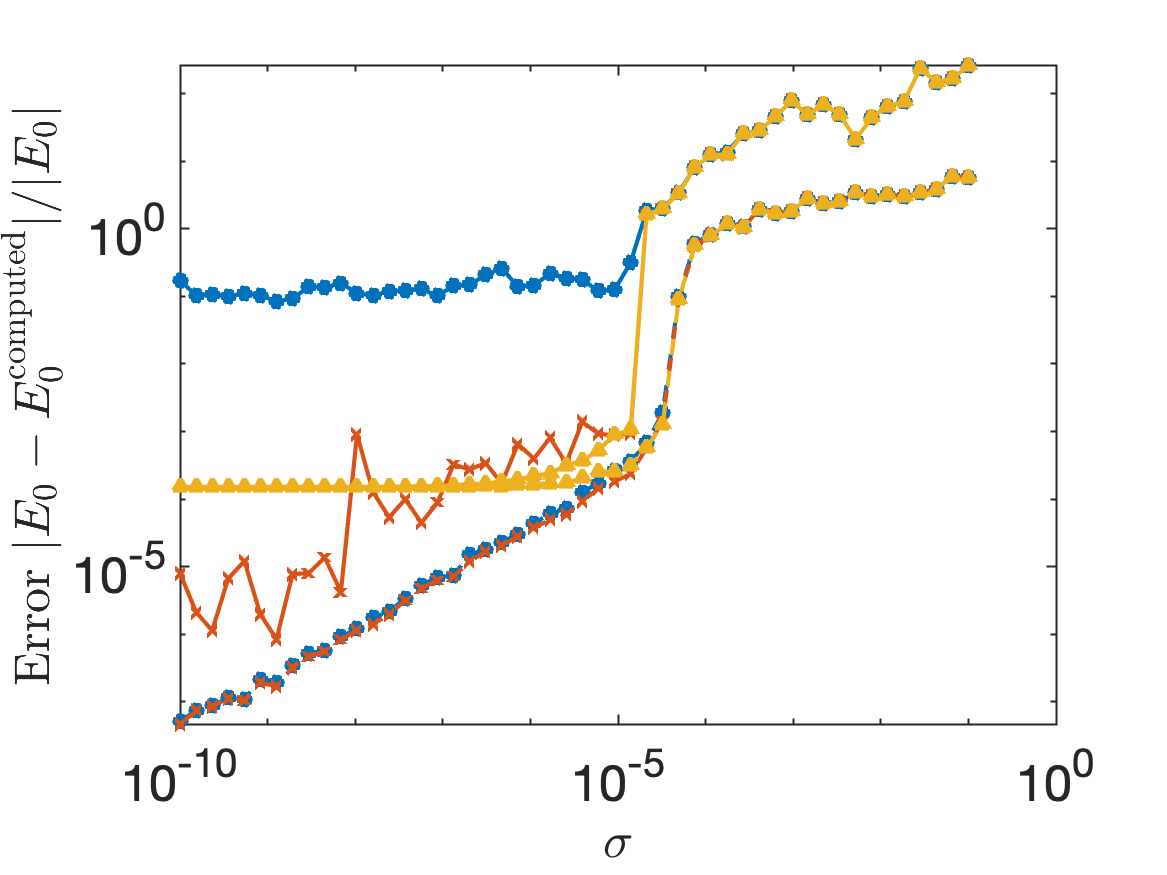}
    \caption{Hubbard, $n = 40$}\label{fig:auto_thresholding_hubbard_l10_u8_q40}
  \end{subfigure}  

  \begin{subfigure}[b]{0.47\textwidth}
    \centering
    \includegraphics[width=\textwidth]{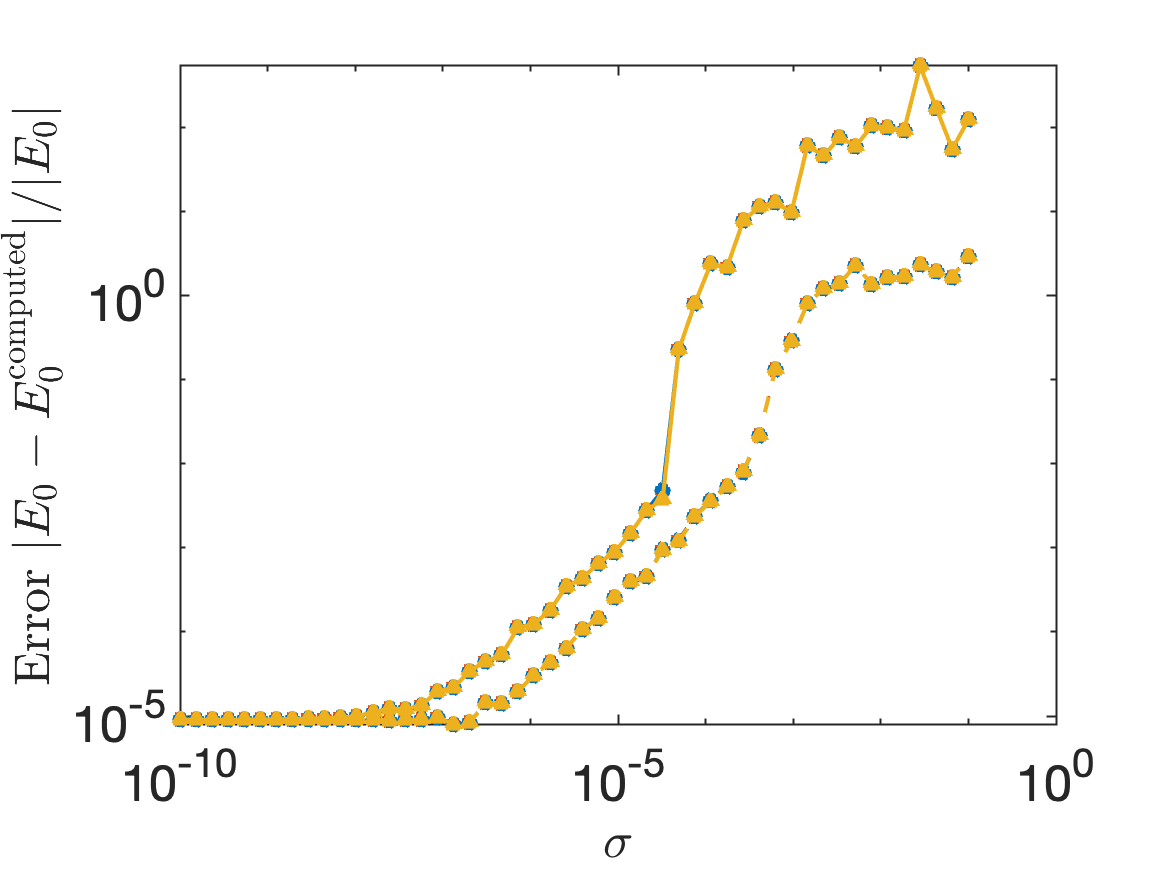}
    \caption{Ising, $n = 20$}\label{fig:auto_thresholding_ising_l10_q20}
  \end{subfigure}
  ~
  \begin{subfigure}[b]{0.47\textwidth}
    \centering
    \includegraphics[width=\textwidth]{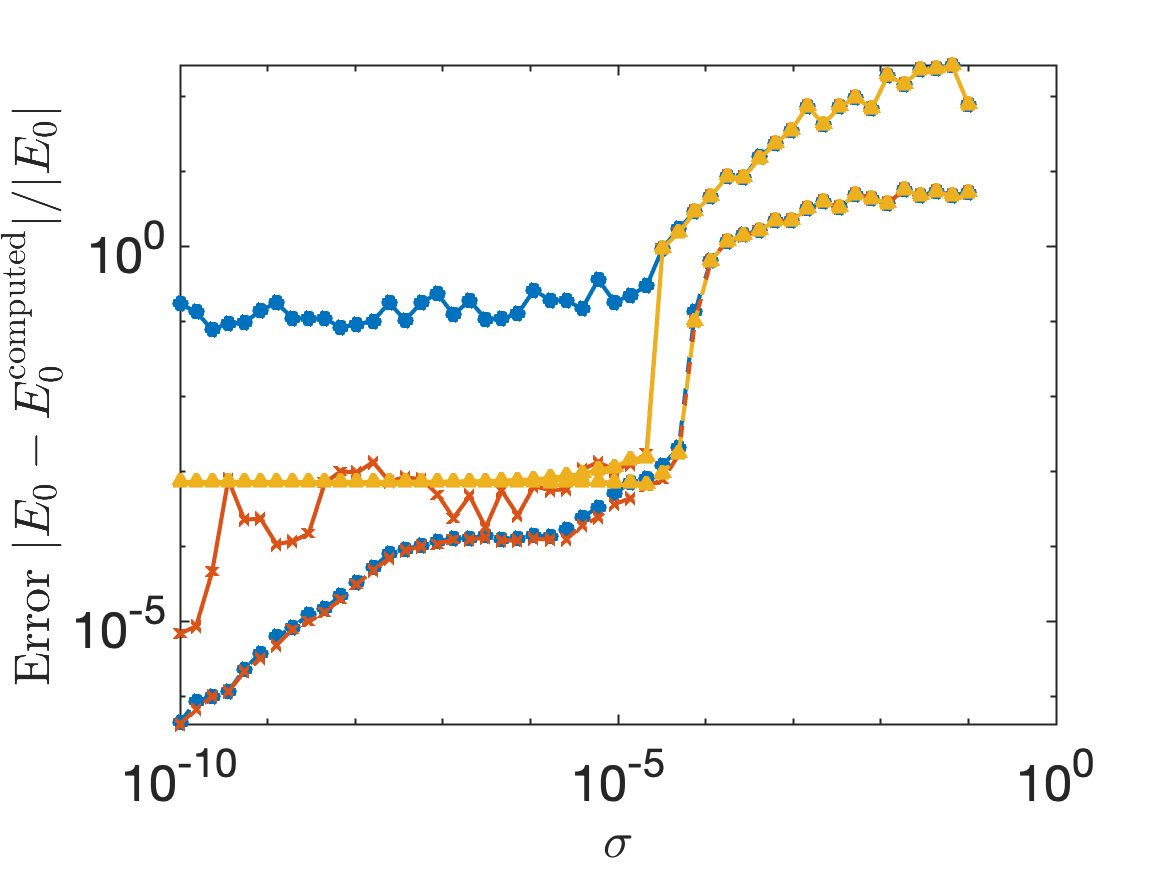}
    \caption{Ising, $n = 40$}\label{fig:auto_thresholding_ising_l10_q40}
  \end{subfigure}
  
  \caption{Maximum and median error over 100 initializations for eigenvalues computed from the noise-perturbed pair $(\mat{\tilde{H}},\mat{\tilde{S}})$ using automatically tuned thresholding (Algorithm~\ref{alg:auto_thresholding}) for three cutoffs $r \in \{ 10^{-1}, 10^{-3}, 10^{-5} \}$ for various values of the noise level $\sigma$ for Hubbard model (top) and Ising model (bottom) with $n = 20$ (left) and $40$ (right).}
  \label{fig:auto_thresholding}
\end{figure}

As we saw in the main text, choosing a good maximum thresholding level $\epsilon$ is critical to the success of the thresholding procedure Algorithm~\ref{alg:thresholding}.
A useful ``sanity check'' is thus to solve the problem using a handful of plausible thresholding parameters to make sure the computed eigenvalues are close to each other.
(A variant of this strategy is proposed by Parlett for the Fix--Heiberger procedure \cite[\S15.5]{Par98}.)
A more ambitious strategy is to solve with a range of thresholding parameters beginning with a conservative (but not comically large) threshold parameter $\epsilon_0$ and then tuning it down until the eigenvalue ``jumps'' to a presumably spurious value.
The best approximation to the ground-state energy suggested by this procedure is the last value before this jump.
If one wishes to automate this procedure, one needs to have a mechanistic way of deciding whether a jump has occurred: For this purpose, we shall test whether the relative difference exceeds a cutoff $r$.
This procedure is demonstrated in Algorithm~\ref{alg:auto_thresholding}.
The success of this procedure relies on the choice of $\epsilon_0$ not being too large as the method uses the eigenvalue recovered with parameter $\epsilon_0$ as a baseline, large deviations from which are characterized as erroneous.
Usually, one will have some good estimate of the amount of noise so picking a sensible $\epsilon_0$ should be possible.

The performance of the automatic thresholding procedure Algorithm~\ref{alg:auto_thresholding} with three choices of the parameter $r$ are shown in Figure~\ref{fig:auto_thresholding}.
These plots represent the worst-case situation where the noise level is completely unknown and the choice one has available for $\epsilon_0$ is a constant multiple of $\norm{\mat{\tilde{S}}}$.
The best case scenario is shown in the $r = 10^{-3}$ lines in Figures~\ref{fig:auto_thresholding_hubbard_l10_u8_q40} and \ref{fig:auto_thresholding_ising_l10_q20}; in these cases, the error decays nicely as the noise does with the procedure being relatively robust (as shown by the error over a maximum over 100 trials being similar to the median).
This automatic thresholding procedure can still be somewhat delicate, with the maximum error over 100 runs being near the cutoff $r$ for the $r = 10^{-1}$ in Figures~\ref{fig:auto_thresholding_hubbard_l10_u8_q20}, \ref{fig:auto_thresholding_hubbard_l10_u8_q40}, and \ref{fig:auto_thresholding_ising_l10_q20}; this suggests, in the worst case, one must be willing to accept an error level on the order $r$ due to overly aggressive automatic tuning of the thresholding parameter.
However, these same plots illustrate the importance of not being too cautious either, with the maximum error being $\approx 10^{-3}$ with $r = 10^{-5}$ due to overly conservative automatic tuning of the thresholding parameter.
In totality, Figure~\ref{fig:auto_thresholding} shows that the automatic thresholding procedure cannot determine a near-optimal choice for the thresholding parameter in all cases, but it can be useful in ``upgrading'' an overly cautious threshold parameter $\epsilon_0$ to a better choice for $\epsilon$, obtaining a couple more decimal digits of accuracy in the best case.
As a final comment, observe that thresholding is an inherently discrete process since each eigenvalue must either be discarded or not.
Therefore, even with automatically tuned thresholding, there can be plateaus in the noise-level vs accuracy curve, owing to the importance of a single eigenvalue to the overall error; this is shown in Figure~\ref{fig:auto_thresholding_ising_l10_q20}. 

\section{Validation of Theorems~\ref{thm:a_priori_bound} and \ref{thm:thresholding}}
\label{sec:noise-less-case}

\begin{figure}[t]
  \centering
    \begin{subfigure}[b]{0.47\textwidth}
    \centering
    \includegraphics[width=\textwidth]{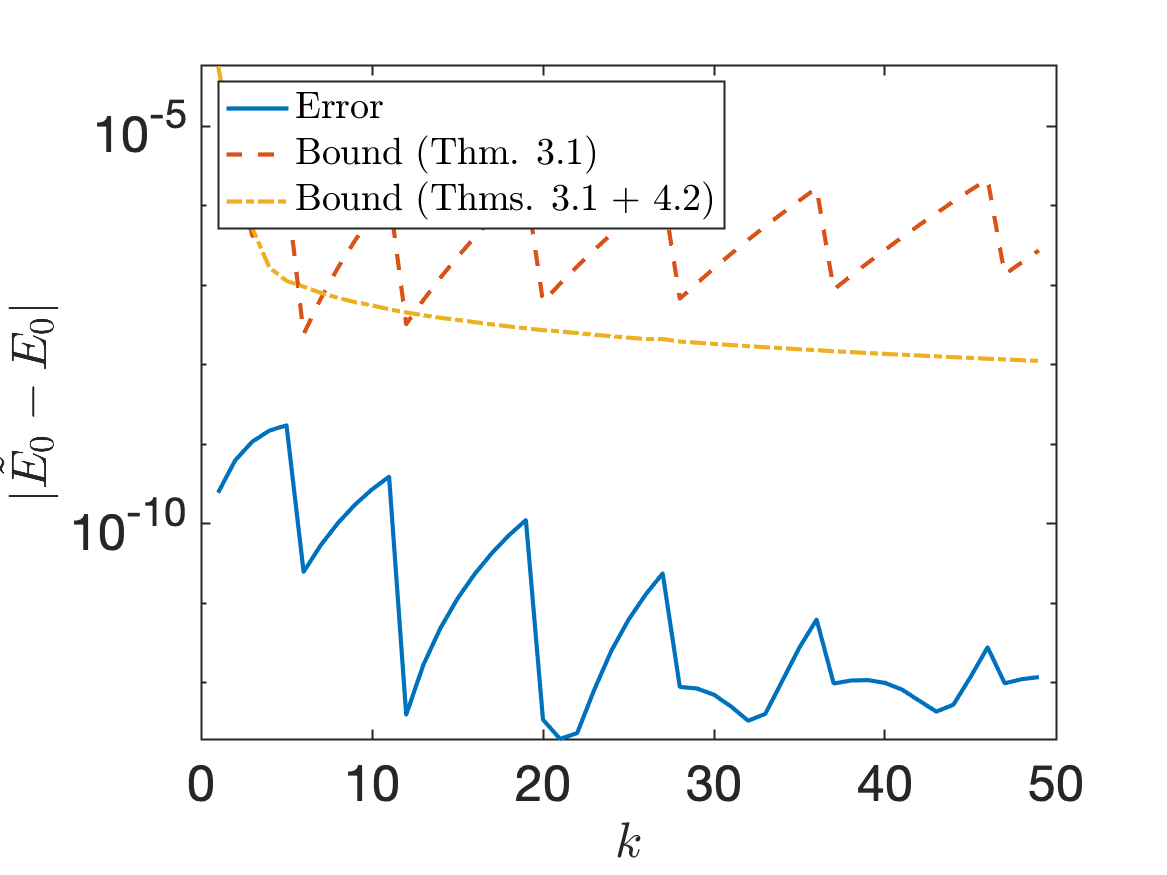}
    \caption{$\oper{H} = \oper{H}_1$, $\rev{\vec{\varphi}_0} = \rev{\vec{\xi}_{\rm I}}$}\label{fig:noiseless_k_1}
  \end{subfigure}
  ~
  \begin{subfigure}[b]{0.47\textwidth}
    \centering
    \includegraphics[width=\textwidth]{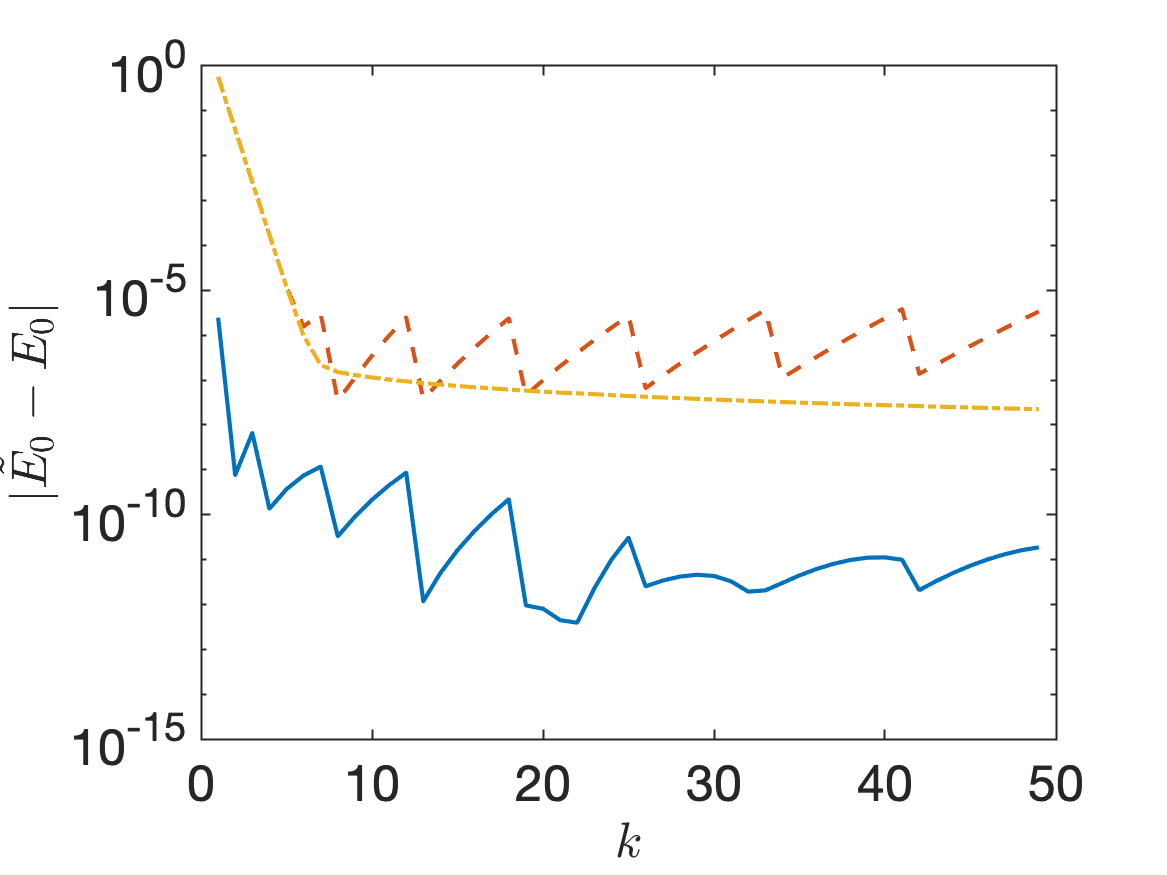}
    \caption{$\oper{H} = \oper{H}_1$, $\rev{\vec{\varphi}_0} = \rev{\vec{\xi}_{\rm II}}$}\label{fig:noiseless_k_2}
  \end{subfigure}  

  \begin{subfigure}[b]{0.47\textwidth}
    \centering
    \includegraphics[width=\textwidth]{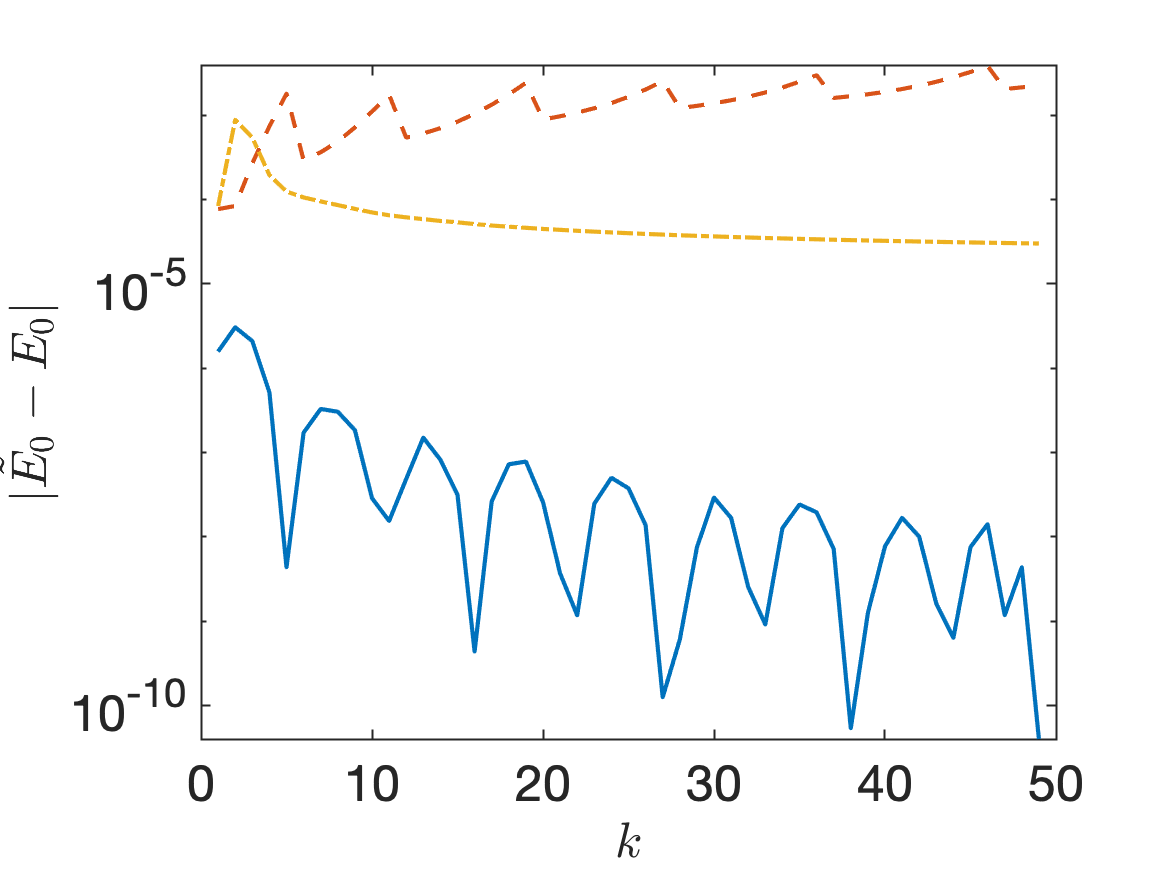}
    \caption{$\oper{H} = \oper{H}_2$, $\rev{\vec{\varphi}_0} = \rev{\vec{\xi}_{\rm III}}$}\label{fig:noiseless_k_3}
  \end{subfigure}
  ~
  \begin{subfigure}[b]{0.47\textwidth}
    \centering
    \includegraphics[width=\textwidth]{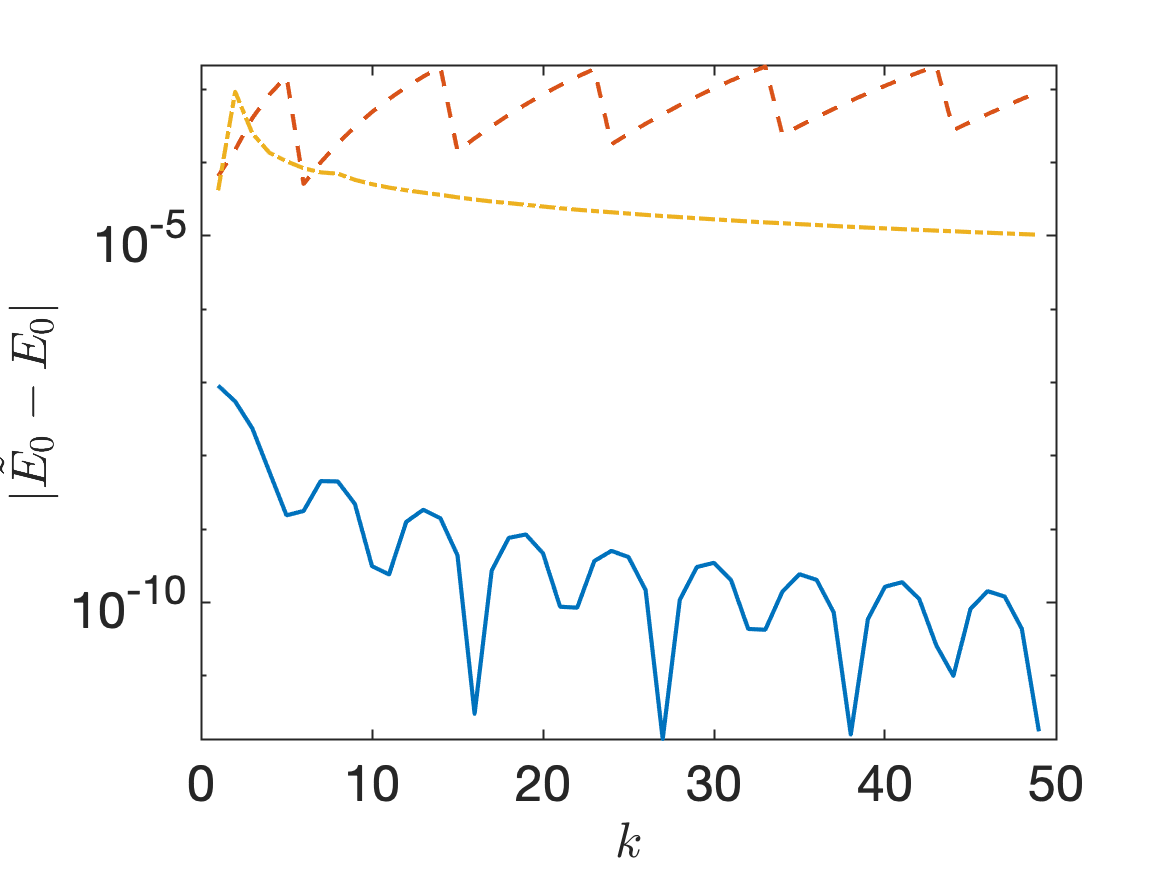}
    \caption{$\oper{H} = \oper{H}_2$, $\rev{\vec{\varphi}_0} = \rev{\vec{\xi}_{\rm IV}}$}\label{fig:noiseless_k_4}
  \end{subfigure}  
  
  \caption{Error for QSD with the time sequence from the hypotheses of Theorem~\ref{thm:a_priori_bound} with threshold parameter $\epsilon = 10^{-6}$ for various values $k$ and four different sets of input data.}
  \label{fig:noiseless_k}
\end{figure}

Despite its desirable theoretical implications, Theorem~\ref{thm:a_priori_bound} may still significantly overestimate the error incurred by thresholding in practice.
Consider the following examples:
\begin{enumerate}[label=(\Roman*)]
\item $\oper{H} = \oper{H}_1$ and $\rev{\vec{\varphi}_0} = \rev{\vec{\xi}_{\rm I}}$,
\item $\oper{H} = \oper{H}_1$ and $\rev{\vec{\varphi}_0} = \rev{\vec{\xi}_{\rm II}}$,
\item $\oper{H} = \oper{H}_2$ and $\rev{\vec{\varphi}_0} = \rev{\vec{\xi}_{\rm III}}$,
\item $\oper{H} = \oper{H}_2$ and $\rev{\vec{\varphi}_0} = \rev{\vec{\xi}_{\rm IV}}$.
\end{enumerate}
where
\begin{align*}
  \rev{\vec{\xi}_{\rm I}} &= \mleft(\sqrt{1-10^{-4}},\sqrt{\frac{10^{-4}}{998}},\sqrt{\frac{10^{-4}}{998}},\ldots,\sqrt{\frac{10^{-4}}{998}}\mright) \in \complex^{999}, \\
  \rev{\vec{\xi}_{\rm II}} &= \mleft(\sqrt{0.5},\sqrt{\frac{0.5}{998}},\sqrt{\frac{0.5}{998}},\ldots,\sqrt{\frac{0.5}{998}}\mright) \in \complex^{999}, \\
  \rev{\vec{\xi}_{\rm III}} &= \mleft(\sqrt{1-10^{-4}-10^{-8}},\sqrt{\frac{10^{-4}}{998}},\sqrt{\frac{10^{-4}}{998}},\ldots,\sqrt{\frac{10^{-4}}{998}},10^{-4}\mright) \in \complex^{1000} \\
  \rev{\vec{\xi}_{\rm IV}} &= \mleft(1,\frac{0.01}{2},\frac{0.01}{3},\ldots,\frac{0.01}{1000}\mright) / \norm*{\mleft(1,\frac{0.01}{2},\frac{0.01}{3},\ldots,\frac{0.01}{1000}\mright)} \in \complex^{1000}, \\
\end{align*}
and
\begin{align*}
    \oper{H}_1 &= \diag\mleft(1,2+0\cdot\frac{0.1}{997},2+1\cdot \frac{0.1}{997},\cdots,2+997\cdot\frac{0.1}{997}\mright)\in \complex^{999\times 999}, \\
  \oper{H}_2 &= \diag\mleft(1,2+0\cdot\frac{0.1}{997},2+1\cdot \frac{0.1}{997},\cdots,2+997\cdot\frac{0.1}{997},1000\mright)\in \complex^{1000\times 1000}.
\end{align*}
These examples are  artificial: They are engineered to have a large spectral gap $\Delta E_1$ but a small spectral range $\Delta E_{999}$.
Even with these artificial examples, Theorem~\ref{thm:a_priori_bound} (as well as Theorems~\ref{thm:a_priori_bound} and \ref{thm:thresholding} together) still overestimates the error by several orders of magnitude.
See Figure~\ref{fig:noiseless_k}.

\begin{figure}[t]
  \centering
  \includegraphics[width=0.7\textwidth]{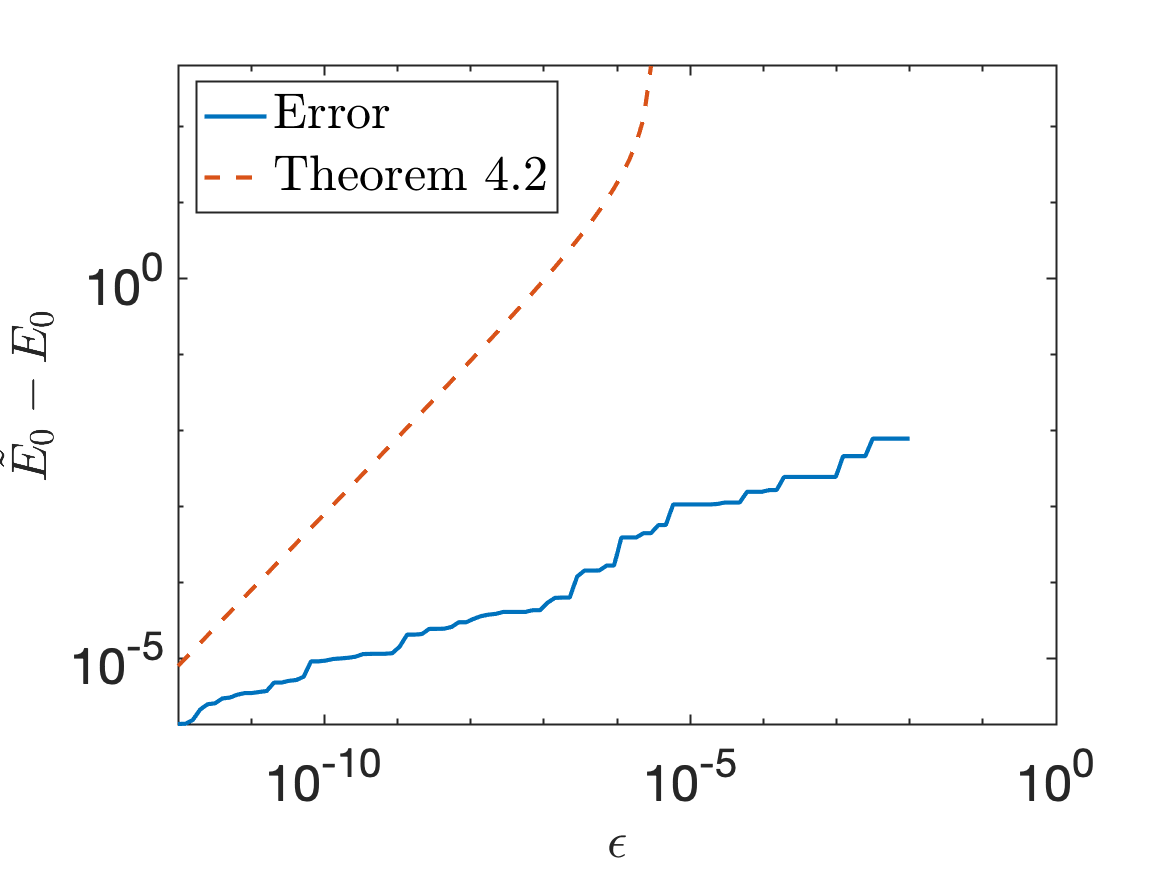}
  \caption{Error due to thresholding and error bound from Theorem~\ref{thm:thresholding} for least eigenvalue computed from a synthetically generated pair $(\mat{H},\mat{S})$ for various threshold parameters $\epsilon$.}
  \label{fig:thresholding_alone}
\end{figure}

We consider the bound Theorem~\ref{thm:thresholding} by itself in Figure~\ref{fig:thresholding_alone}.
Shown is the error $\tilde{E}_0 - E_0$ between the recovered least eigenvalue $\tilde{E}_0$ and the true least eigenvalue $E_0$ for different choices of threshold parameter $\epsilon$.
For this example, we used $\mat{H} = \mat{K}^{\rev{*}}\diag(1,2,\ldots,100)\mat{K}$ and $\mat{S} = \mat{K}^{\rev{*}}\mat{K}$, where $\mat{K}$ is a product of diagonal matrix with $(j,j)$th entry $j^{-2}$ and a ``\texttt{randsvd}'' matrix from MATLAB's \texttt{gallery} with approximate condition number $10^3$.
This was chosen to give a fairly ill-conditioned ``$\mat{S}$'' matrix ($\kappa(\mat{S}) \approx 10^{12}$) in which the $\mat{S}$-normalized ground state eigenvector of fairly small norm ($\norm{\vec{c}_0}\approx 10^2$).
We find the bound from Theorem~\ref{thm:thresholding} becomes increasingly conservative as $\epsilon$ increases, diverging to $+\infty$ at $\epsilon \approx 10^{-6}$ which the true error $\tilde{E}_0 - E_0$ remains bounded $\le 10^{-2}$ for $\epsilon \le 10^{-2}$.

\section{Evidence for Tightness of Theorem~\ref{thm:H_projection_error}}
\label{sec:evid-tightn-theor}

First, we present a synthetically generated numerical example which suggests that the $\eta/\epsilon^\alpha$ behavior in Theorem~\ref{thm:H_projection_error} is necessary, at least without further assumptions.
As our example, we set $\mat{A} = (\mat{G}+\mat{G}^{\rev{*}})/2$ to be the Hermitian part of a $5\times 5$ real standard Gaussian matrix $\mat{G}$ and pick $\mat{S} = \diag(1,0.1,3\times 10^{-10},2\times 10^{-10},10^{-10})$ and $\mat{H} = \mat{S}^{1/2}\mat{A}\mat{S}^{1/2}$.
By construction, this example obeys the geometric mean bound \eqref{eq:geometric_mean} with $\alpha = 1/2$ and $\mu = 0.5\lambda_{\rm max}(\mat{G}+\mat{G}^{\rev{*}})$ which is $\lessapprox 10$ with high probability.
We choose a threshold level of $\epsilon = 1.5\times 10^{-10}$, so that the thresholded problem has dimension four.
As perturbation, we take $\mat{\Delta}_{\mat{S}} = 10^{-12} \cdot (\mat{\Gamma}+\mat{\Gamma}^{\rev{*}})/2$ (for a $5\times 5$ real standard Gaussian matrix $\mat{\Gamma}$).

Let $\mat{\Pi}$ and $\mat{\tilde{\Pi}}$ denote the spectral projectors onto the eigenvectors $>\epsilon$ for $\mat{S}$ and $\mat{\tilde{S}} = \mat{S} + \mat{\Delta}_{\mat{S}}$ respectively.
For one random initialization of the Gaussian test matrices (which we find is broadly representative of repeat trials), we computed
\begin{equation} \label{eq:projection_error_example}
  \norm{\mat{\tilde{\Pi}}\mat{H}\mat{\tilde{\Pi}} - \mat{\Pi}\mat{H}\mat{\Pi}} = 6.3\times 10^{-8} \approx 10^{-7}\approx \norm{\mat{\Delta}_{\mat{S}}}/\epsilon^{1/2}.
\end{equation}
Were the $\epsilon^{-\alpha}$ dependence in Theorem~\ref{thm:H_projection_error} unnecessary, we would expect that the projection error $\norm{\mat{\tilde{\Pi}}\mat{H}\mat{\tilde{\Pi}} - \mat{\Pi}\mat{H}\mat{\Pi}}$ would be bounded by $\mu (1+\rho^{-1})5^3\norm{\mat{\Delta}_{\mat{S}}} \approx 10^{-9}$.
We take this as evidence that the $\epsilon^{-\alpha}$ factor in the bound in Theorem~\ref{thm:H_projection_error} is necessary, at least without additional assumptions.

\section{The Value of \texorpdfstring{$\alpha$}{alpha} in Eq.~\eqref{eq:geometric_mean}}
\label{sec:value-alpha}

As we argued in the main text, any pair $(\mat{H},\mat{S})$ obeys the geometric mean bound \eqref{eq:geometric_mean} with $\alpha = 1/2$ and $\gamma = \max |\Lambda(\mat{H},\mat{S})|$.
In this section, we present numerical evidence that \eqref{eq:geometric_mean} often holds with $\alpha = 1/4$ and $\gamma \approx \max |\Lambda(\mat{H},\mat{S})|$ for QSD problem instances, a substantial improvement on the provable bound.
This $\alpha = 1/4$ behavior remains somewhat mysterious to us, and we have yet to discover a convincing explanation for why this behavior emerges.

\begin{figure}[t]
  \centering
    \begin{subfigure}[b]{0.47\textwidth}
    \centering
    \includegraphics[width=\textwidth]{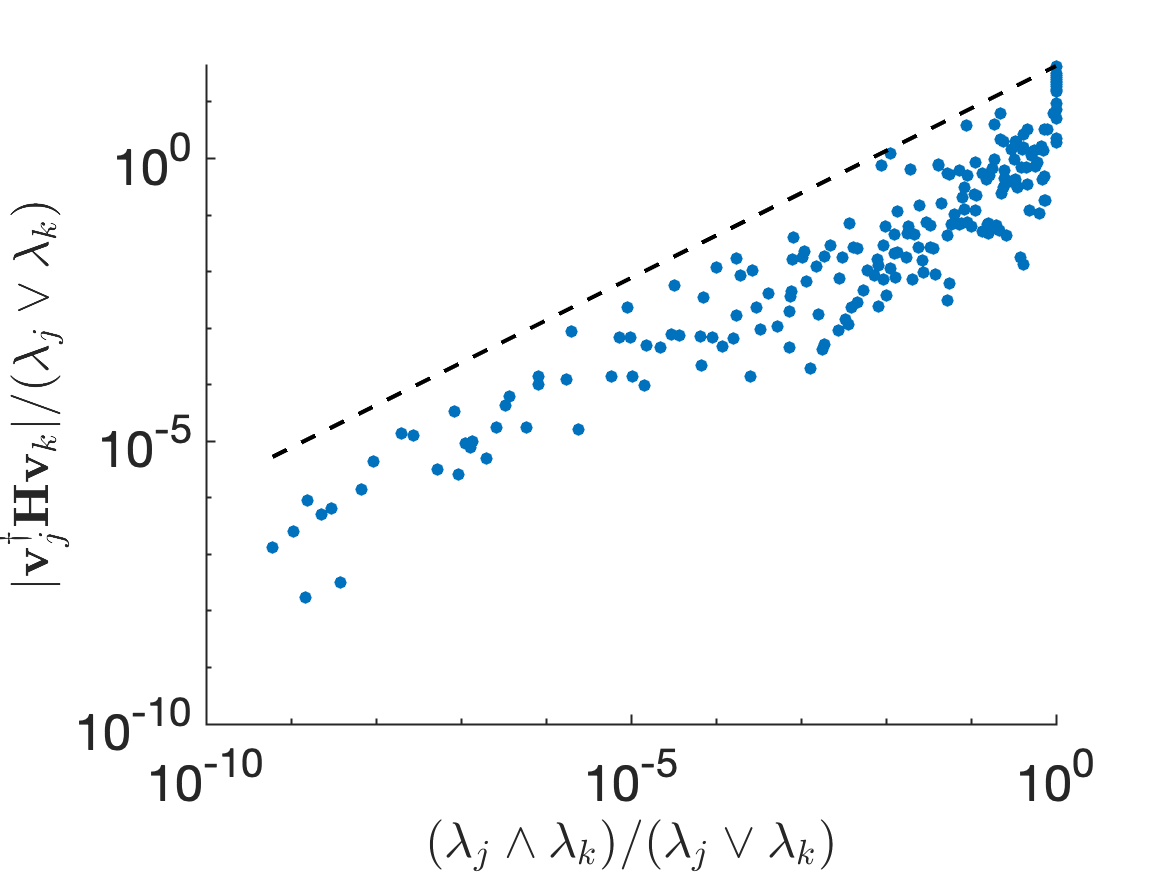}
    \caption{Hubbard, $n = 20$}\label{fig:alpha_hubbard_l10_u8_q20}
  \end{subfigure}
  ~
  \begin{subfigure}[b]{0.47\textwidth}
    \centering
    \includegraphics[width=\textwidth]{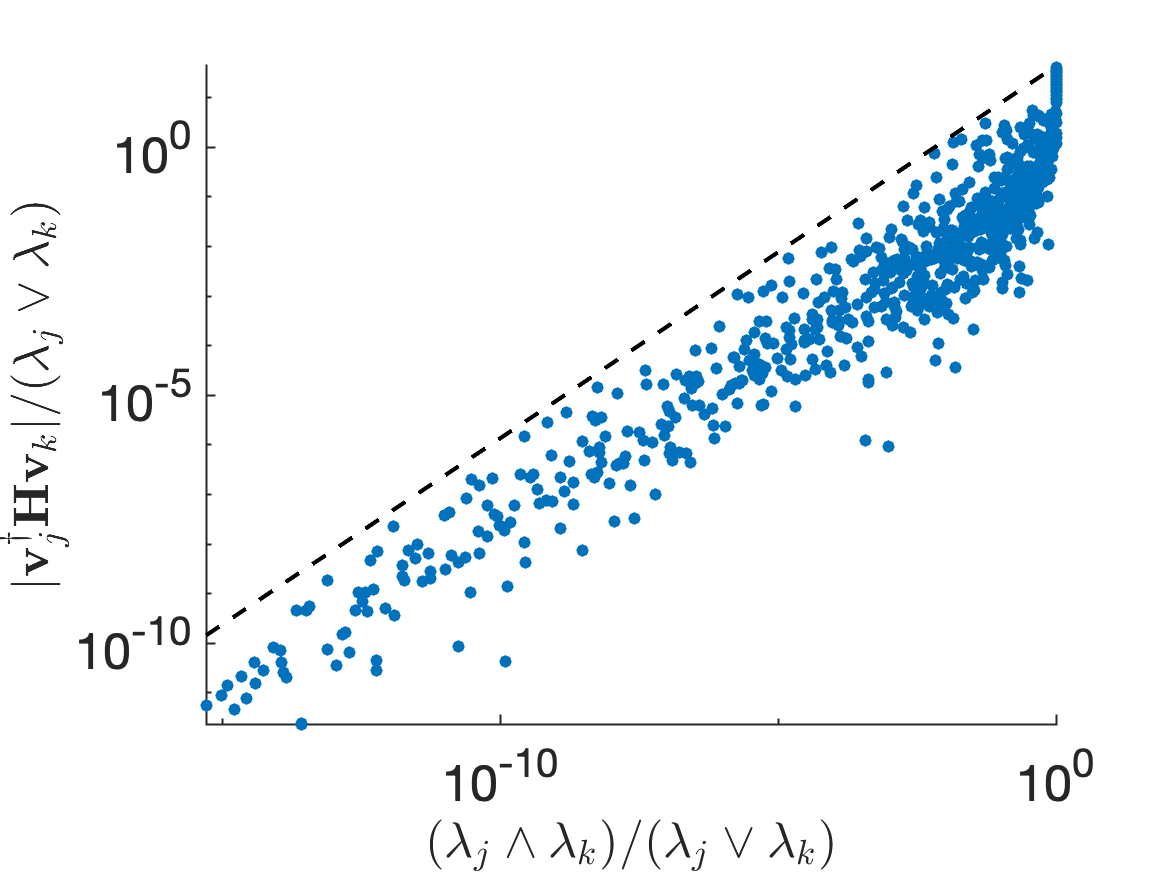}
    \caption{Hubbard, $n = 40$}\label{fig:alpha_hubbard_l10_u8_q40}
  \end{subfigure}  

  \begin{subfigure}[b]{0.47\textwidth}
    \centering
    \includegraphics[width=\textwidth]{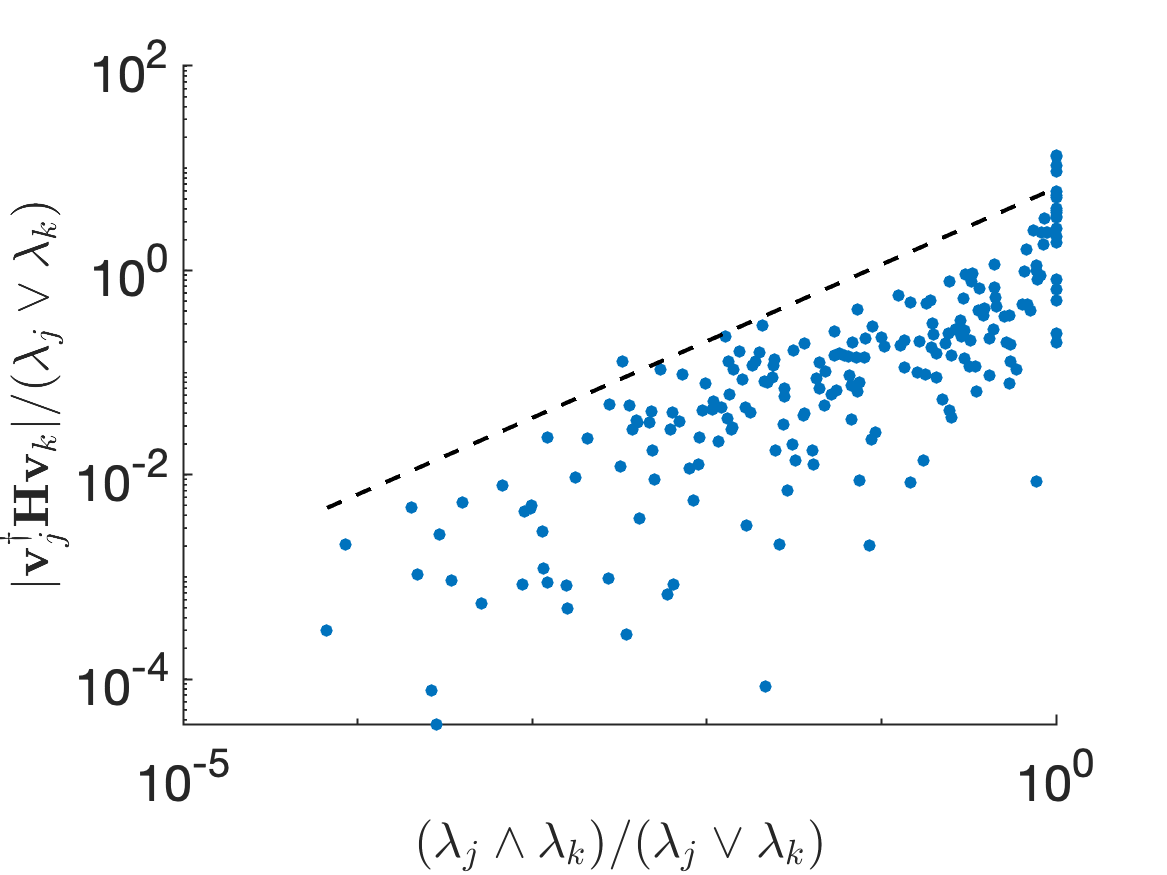}
    \caption{Ising, $n = 20$}\label{fig:alpha_ising_l10_q20}
  \end{subfigure}
  ~
  \begin{subfigure}[b]{0.47\textwidth}
    \centering
    \includegraphics[width=\textwidth]{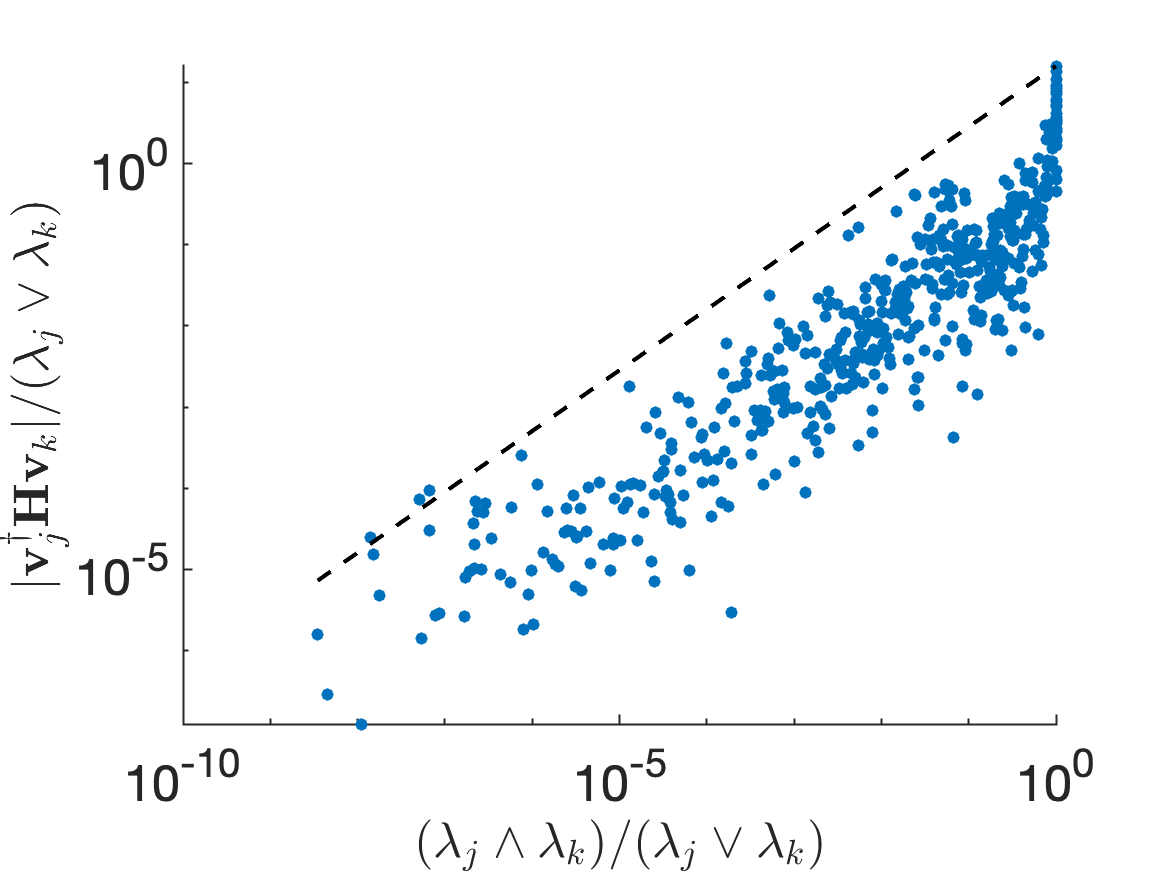}
    \caption{Ising, $n = 40$}\label{fig:alpha_ising_l10_q40}
  \end{subfigure}
  
  \caption{Scatter plot of values $y = \mleft|\vec{v}_j^{\rev{*}} \mat{H}\vec{v}_k\mright|/(\lambda_j \vee \lambda_k)$ versus $x = \rev{\min(\lambda_j,\lambda_k)/\max(\lambda_j,\lambda_k)}$ over all indices $j,k = 1,\ldots,n$ for which $\rev{\min(\lambda_j, \lambda_k)} \ge 10^{-16}\lambda_1$ for Hubbard model (top) and Ising model (bottom) with $n = 20$ (left) and $40$ (right).
    Shown as a dashed black line is $\max |\Lambda(\mat{H},\mat{S})| \cdot x^{3/4}$, demonstrating that \eqref{eq:geometric_mean} holds numerically with $\alpha = 1/4$ and $\gamma \approx \max |\Lambda(\mat{H},\mat{S})|$.}  
  \label{fig:alpha}
\end{figure}

The numerical validity of \eqref{eq:geometric_mean} with $\alpha = 1/4$ and $\gamma \approx \max |\Lambda(\mat{H},\mat{S})|$ is demonstrated in Figure~\ref{fig:alpha}.
In these plots, we plot $y = \mleft|\vec{v}_j^{\rev{*}} \mat{H}\vec{v}_k\mright|/\rev{\max(\lambda_j, \lambda_k)}$ against $x = \rev{\min(\lambda_j,\lambda_k)/\max(\lambda_j,\lambda_k)}$ over all indices $j$ and $k$ for several different QSD instances, where we use the notation from section~\ref{sec:how-do-perturbations} that $(\lambda_j,\vec{v}_j)$ represents the $j$th largest eigenpair of $\mat{S}$.
Since the accurately computable eigenvalues span a range roughly on the order of the inverse machine precision ($\approx 10^{16}$ in double precision), we only plot pairs $(x,y)$ corresponding to indices $j$ and $k$ for which $\rev{\min(\lambda_j, \lambda_k)} \ge 10^{-16}\lambda_1$.
The bound \eqref{eq:geometric_mean} holds only if all points $(x,y)$ (as well as those numerically incomputable) lie below a power law curve $y \le \gamma x^{1-\alpha}$.
The curve $\max |\Lambda(\mat{H},\mat{S})| \cdot x^{0.75}$ is shown on each of the subplots in Figure~\ref{fig:alpha}, and it lies above almost all of the pairs $(x,y)$.
We consider this  convincing evidence of the validity of \eqref{eq:geometric_mean} with $\alpha = 1/4$ and $\gamma \approx \max |\Lambda(\mat{H},\mat{S})|$ for the QSD examples we tested.
We suspect this relation will continue to hold for ``reasonable'' QSD instances, though we lack a precise definition of ``reasonable'' and a formal argument justifying this suspicion.

\section{Extra Figures}
\label{sec:extra-figures}

\begin{figure}[t]
  \centering
    \begin{subfigure}[b]{0.47\textwidth}
    \centering
    \includegraphics[width=\textwidth]{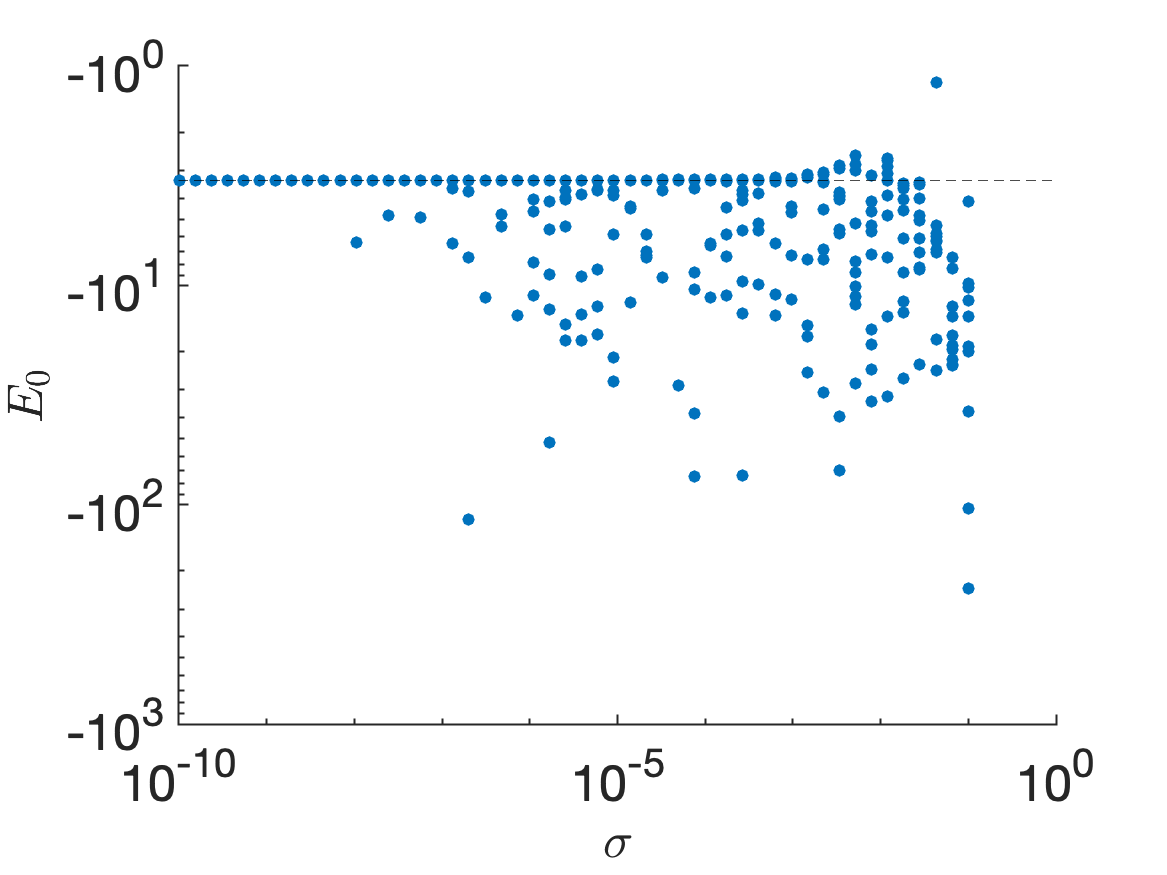}
    \caption{$n = 20$}\label{fig:fixed_threshold_q20}
  \end{subfigure}
  ~
  \begin{subfigure}[b]{0.47\textwidth}
    \centering
    \includegraphics[width=\textwidth]{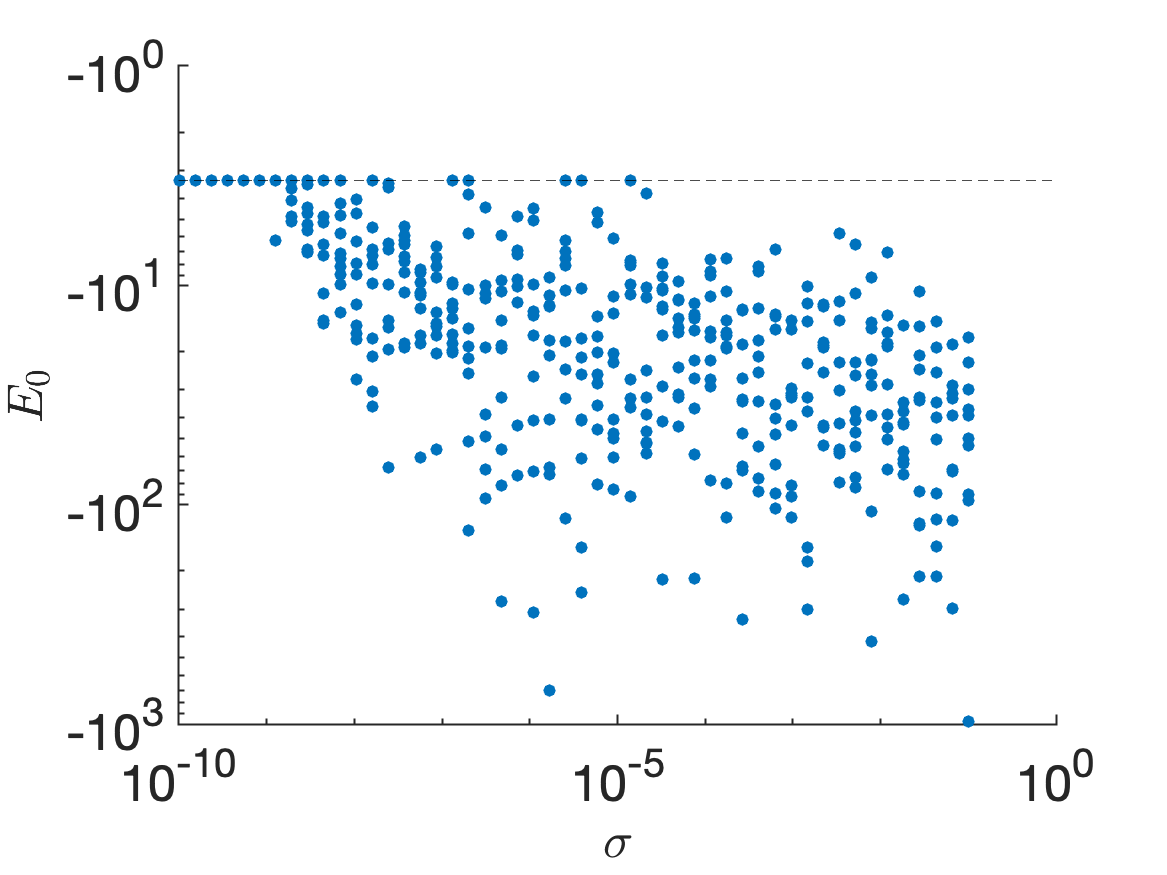}
    \caption{$n = 80$}\label{fig:fixed_threshold_q80}
  \end{subfigure}  

  \caption{Least eigenvalues computed from the perturbed pair $(\mat{\tilde{H}},\mat{\tilde{S}})$ with a fixed threshold $10^{-8}\norm*{\mat{S}}$. Shown are 10 random initializations of the noise for several random noise levels $\sigma$ for the Hubbard example with $n = 20$ (left) and $n = 80$ (right).
  The true eigenvalue is shown for reference as a horizontal dashed line.}
  \label{fig:fixed_threshold}
\end{figure}

\begin{figure}[t]
  \centering
    \begin{subfigure}[b]{0.47\textwidth}
    \centering
    \includegraphics[width=\textwidth]{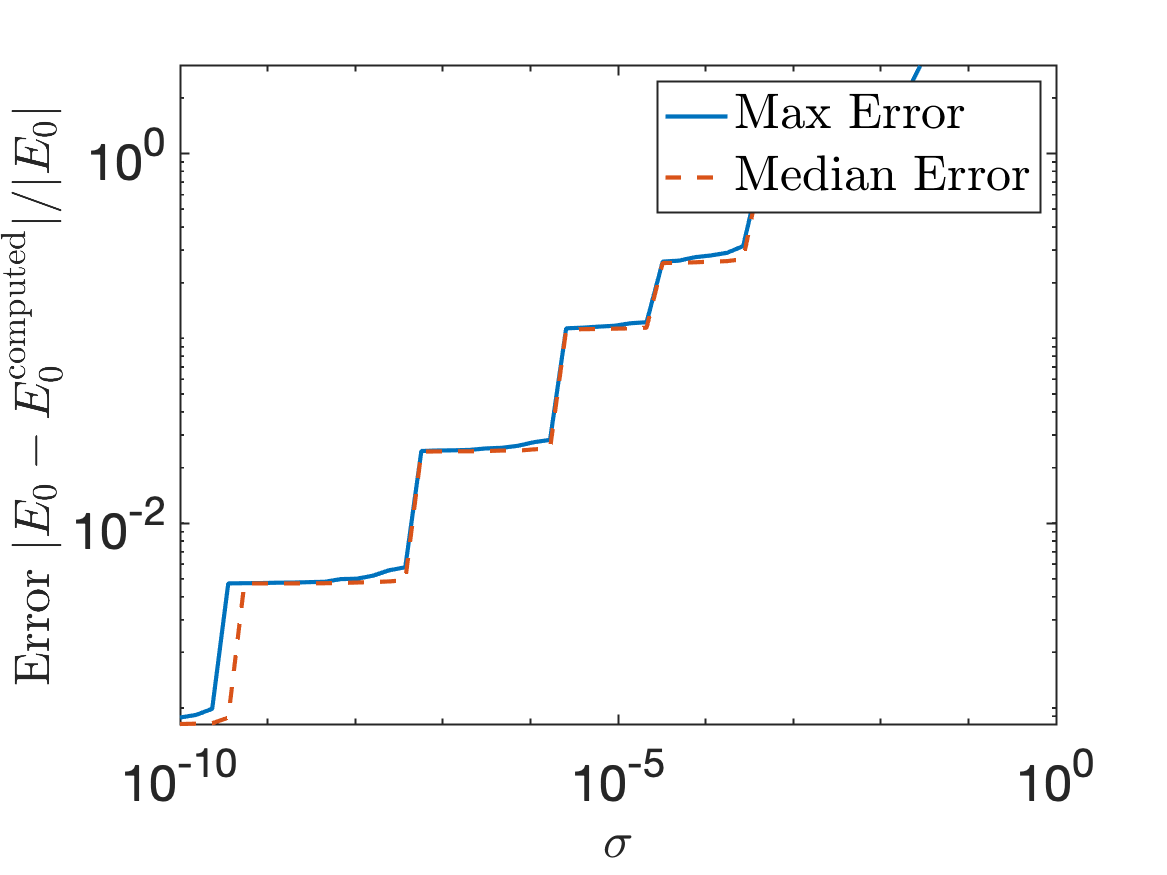}
    \caption{Hubbard, $L = 10$, $U = 8$, $n = 10$}\label{fig:thresholding_hubbard_l10_u8_q10}
  \end{subfigure}
  ~
  \begin{subfigure}[b]{0.47\textwidth}
    \centering
    \includegraphics[width=\textwidth]{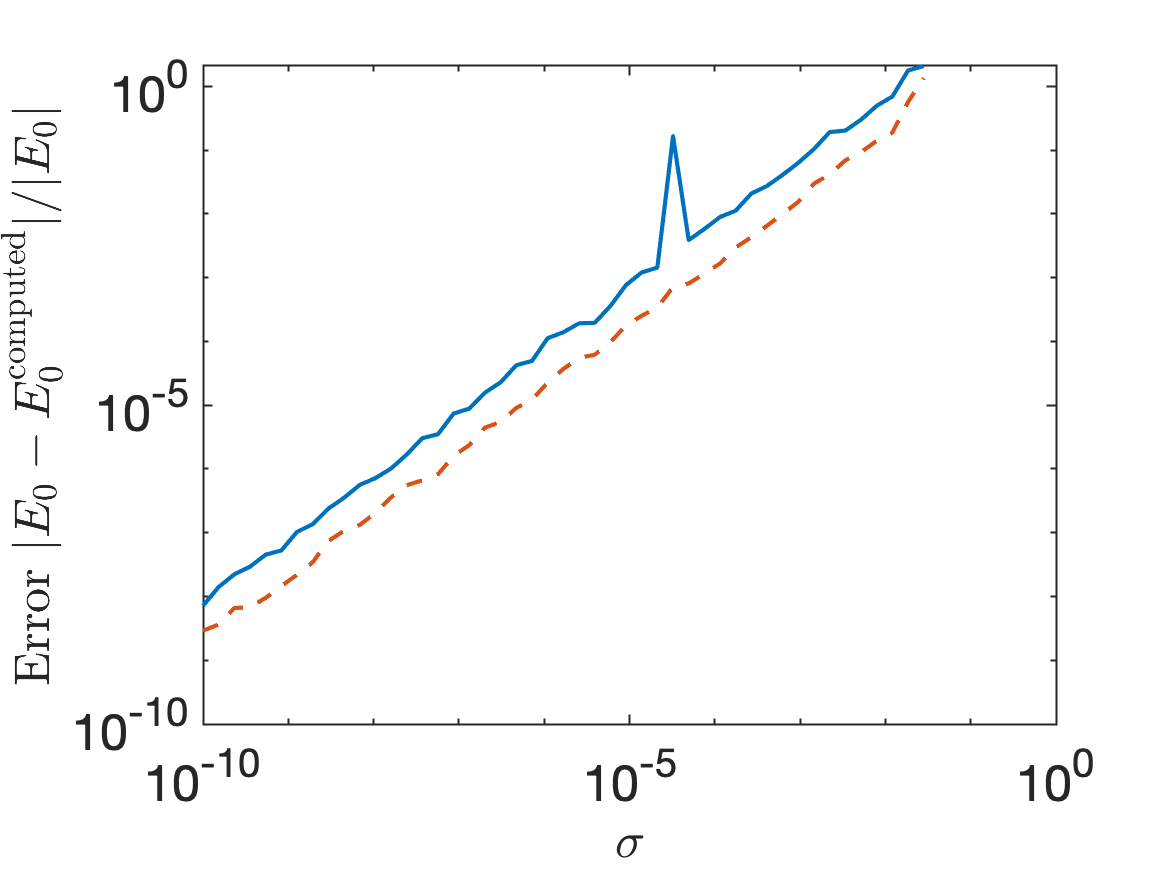}
    \caption{Hubbard, $L = 10$, $U = 8$, $n = 80$}\label{fig:thresholding_hubbard_l10_u8_q80}
  \end{subfigure}  

  \begin{subfigure}[b]{0.47\textwidth}
    \centering
    \includegraphics[width=\textwidth]{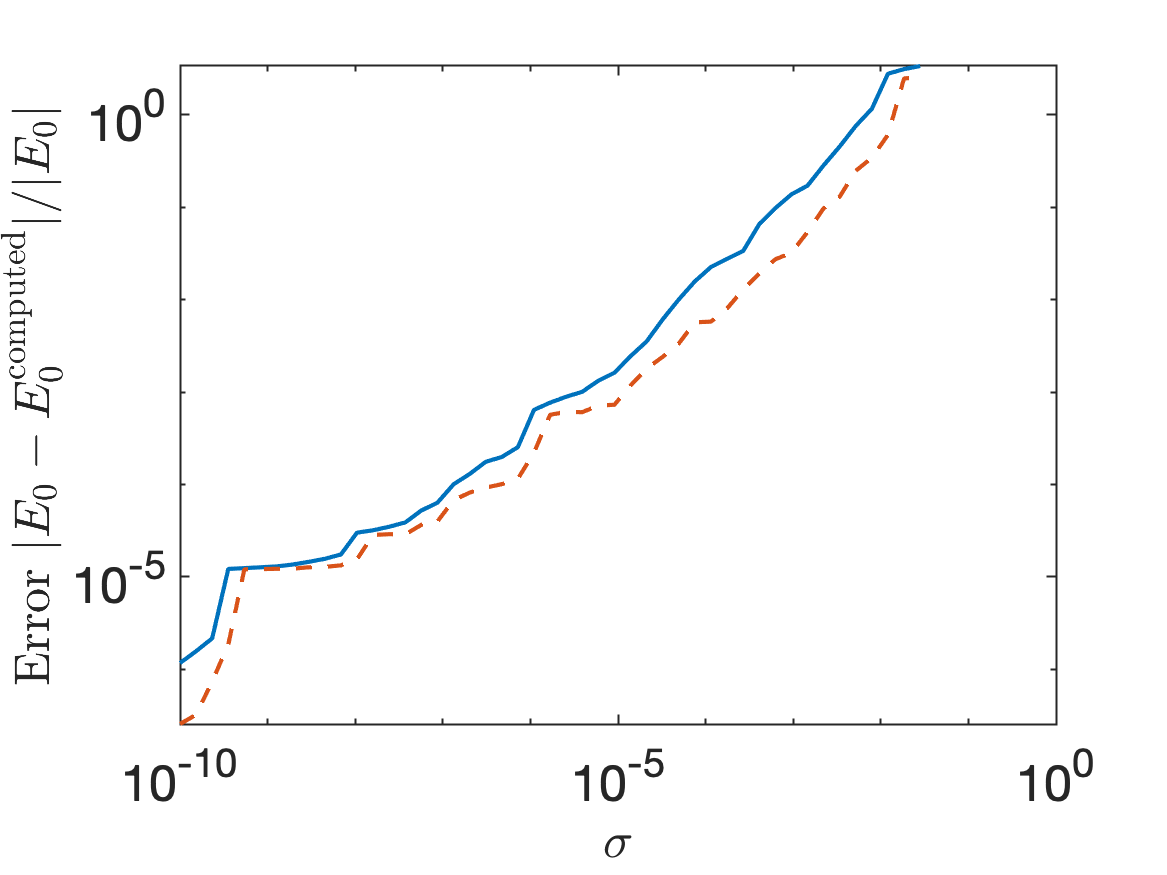}
    \caption{Hubbard, $L = 10$, $U = 10$, $n = 30$}\label{fig:thresholding_hubbard_l10_u10_q30}
  \end{subfigure}
  ~
  \begin{subfigure}[b]{0.47\textwidth}
    \centering
    \includegraphics[width=\textwidth]{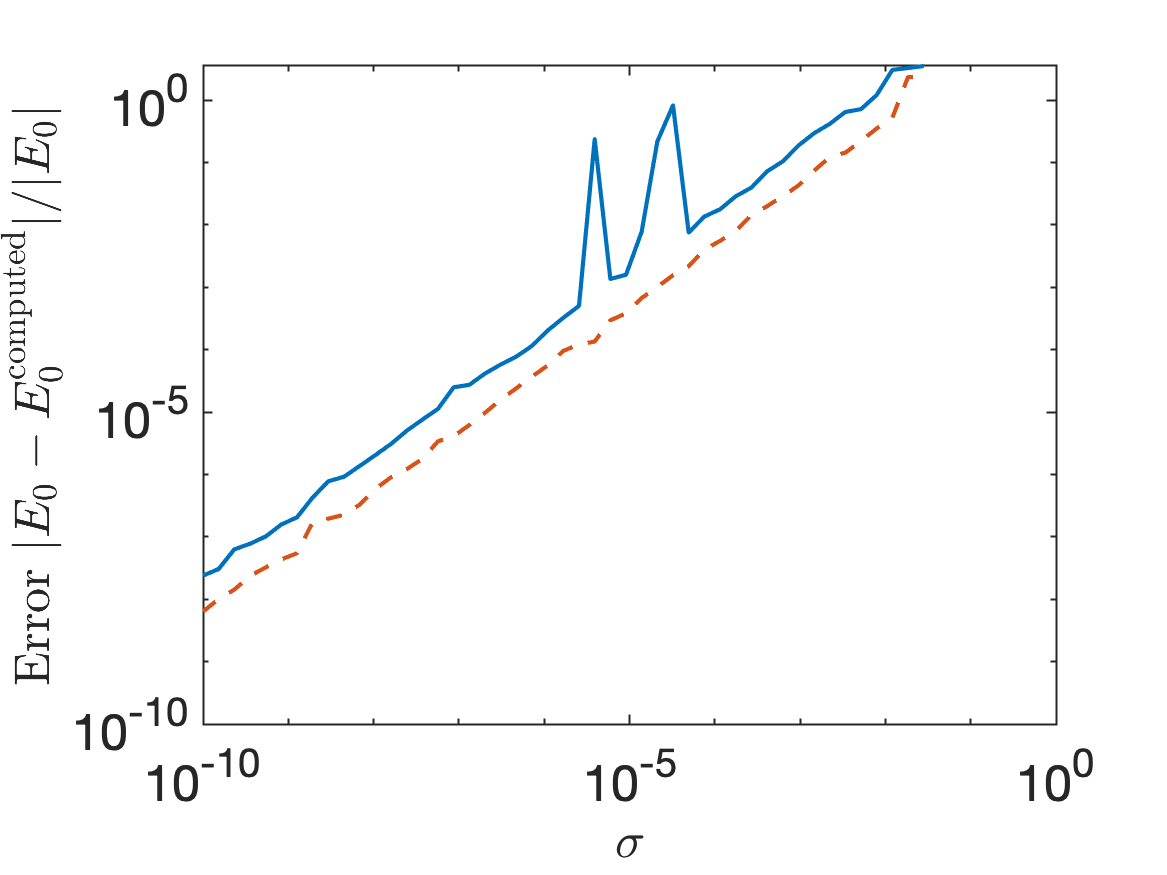}
    \caption{Hubbard, $L = 10$, $U = 10$, $n = 80$}\label{fig:thresholding_hubbard_l10_u10_q80}
  \end{subfigure}

\begin{subfigure}[b]{0.47\textwidth}
    \centering
    \includegraphics[width=\textwidth]{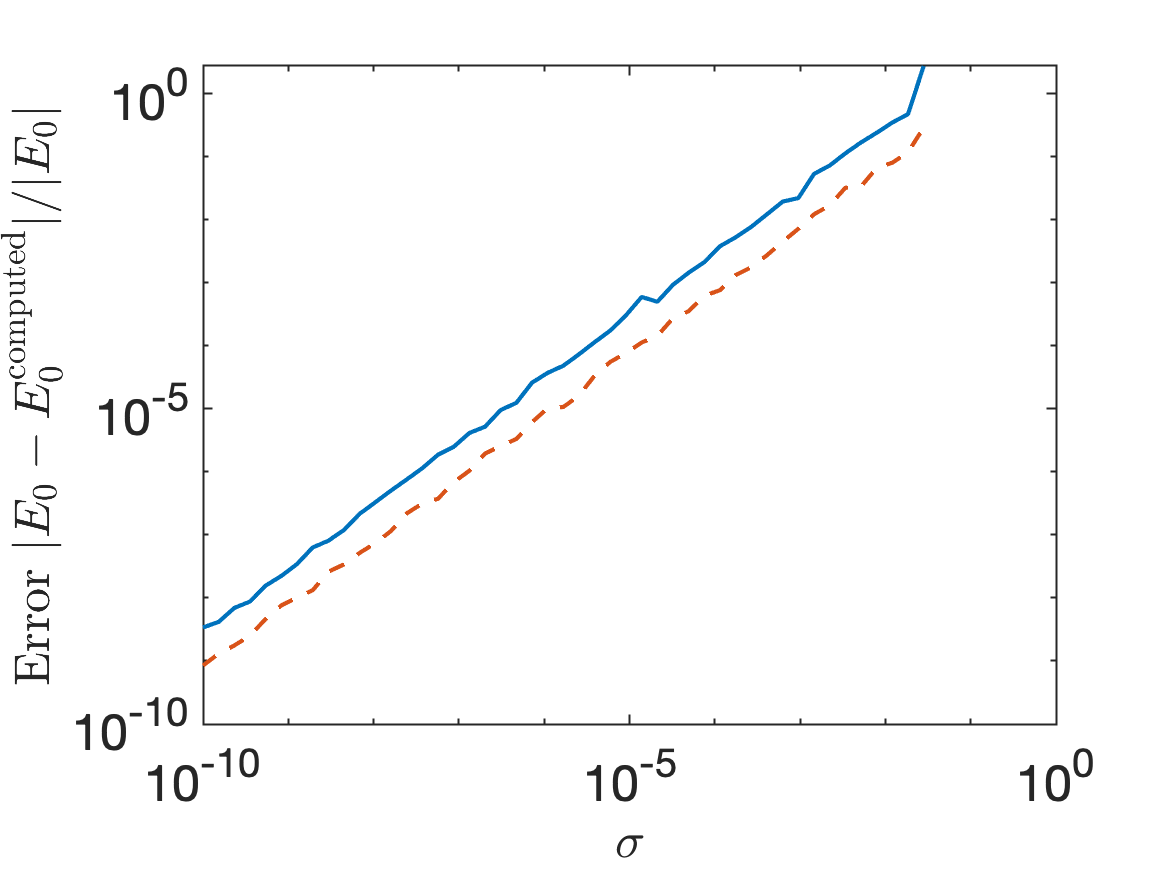}
    \caption{Hubbard, $L = 6$, $U = 8$, $n = 40$}\label{fig:thresholding_hubbard_l6_u8_q40}
  \end{subfigure}
  ~
  \begin{subfigure}[b]{0.47\textwidth}
    \centering
    \includegraphics[width=\textwidth]{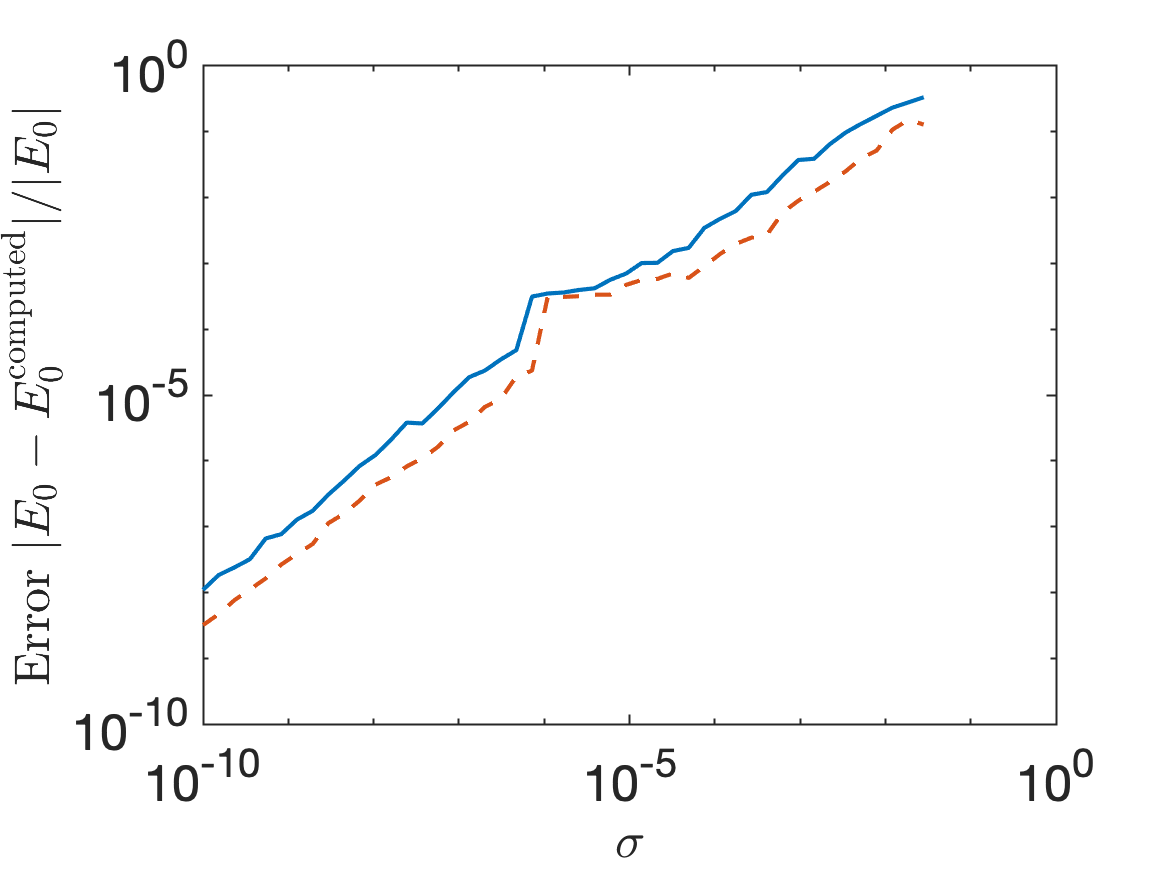}
    \caption{Ising, $L = 8$, $n = 20$}\label{fig:thresholding_ising_l8_q20}
  \end{subfigure}
  
  \caption{Maximum (blue solid) and median (red dashed) error over 100 initializations for eigenvalues computed from the noise-perturbed pair $(\mat{\tilde{H}},\mat{\tilde{S}})$ using thresholding with threshold parameter $25\sigma\norm{\mat{\tilde{S}}}$ for Hubbard and Ising models for various parameters not considered in Figure~\ref{fig:thresholding}.}
  \label{fig:thresholding_more}
\end{figure}

\begin{figure}[t]
  \centering
    \begin{subfigure}[b]{0.47\textwidth}
    \centering
    \includegraphics[width=\textwidth]{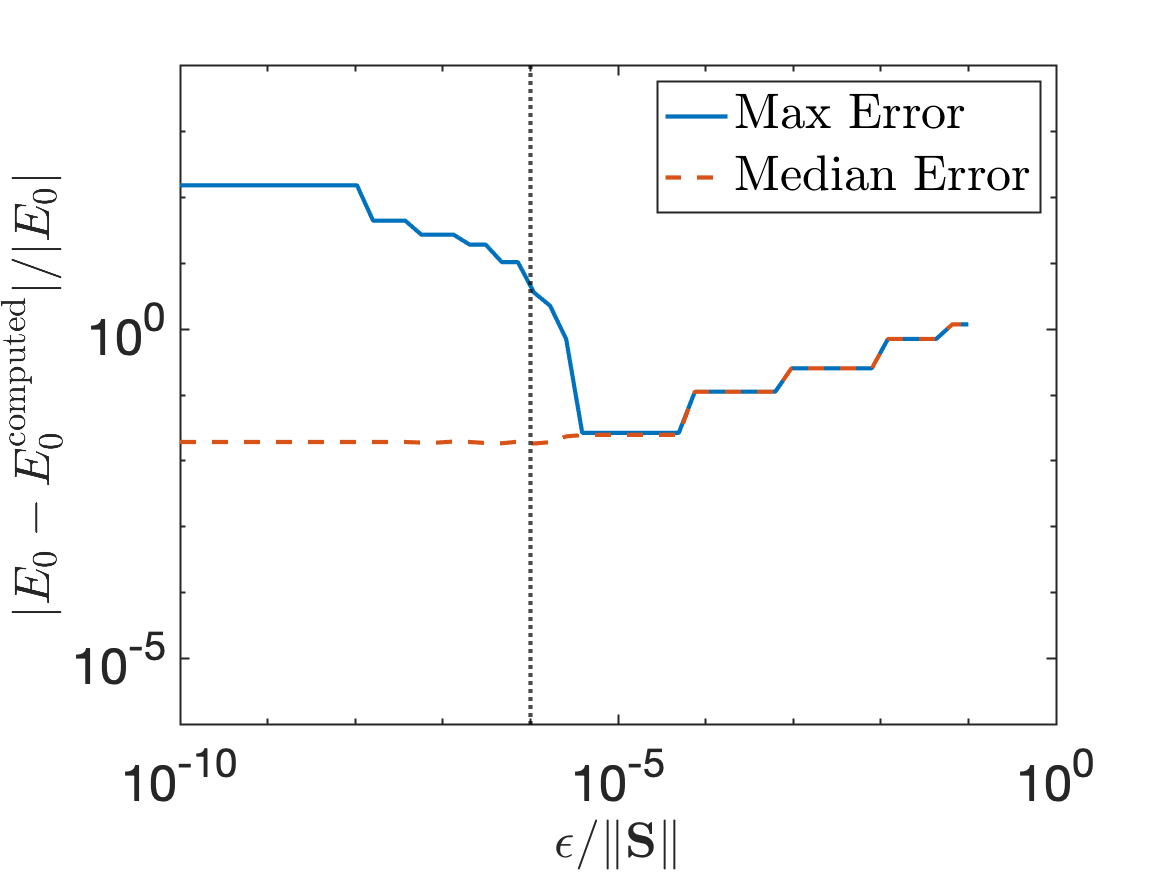}
    \caption{Hubbard, $L = 10$, $U = 8$, $n = 10$}\label{fig:threshold_choice_hubbard_l10_u8_q10}
  \end{subfigure}
  ~
  \begin{subfigure}[b]{0.47\textwidth}
    \centering
    \includegraphics[width=\textwidth]{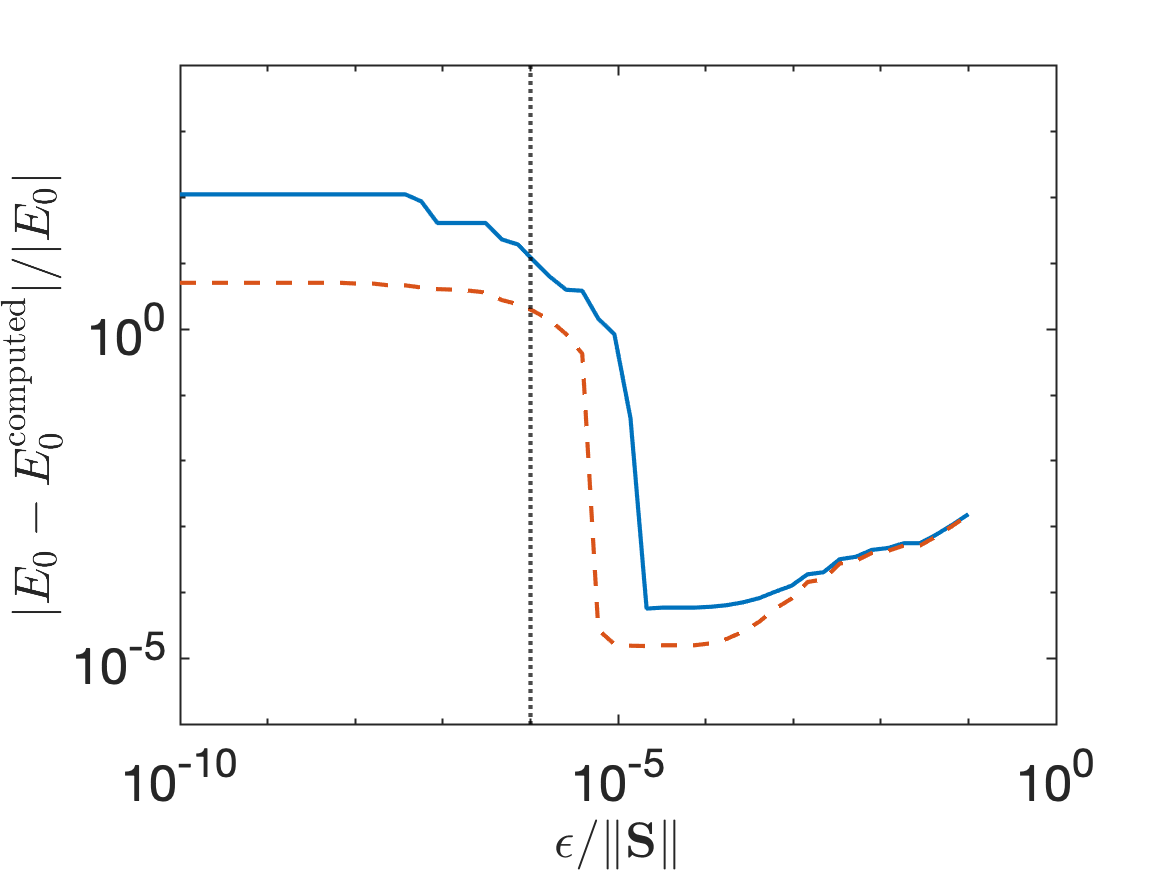}
    \caption{Hubbard, $L = 10$, $U = 8$, $n = 80$}\label{fig:threshold_choice_hubbard_l10_u8_q80}
  \end{subfigure}  

  \begin{subfigure}[b]{0.47\textwidth}
    \centering
    \includegraphics[width=\textwidth]{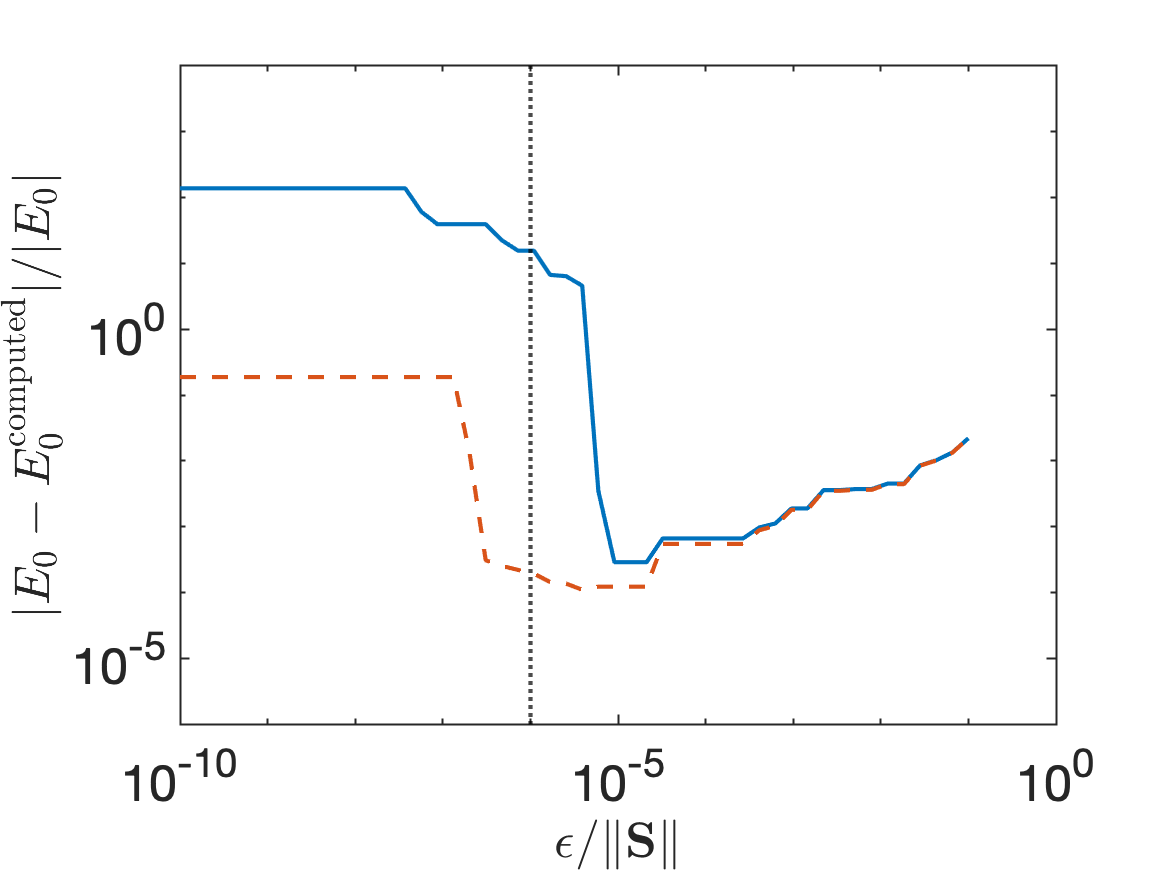}
    \caption{Hubbard, $L = 10$, $U = 10$, $n = 30$}\label{fig:threshold_choice_hubbard_l10_u10_q30}
  \end{subfigure}
  ~
  \begin{subfigure}[b]{0.47\textwidth}
    \centering
    \includegraphics[width=\textwidth]{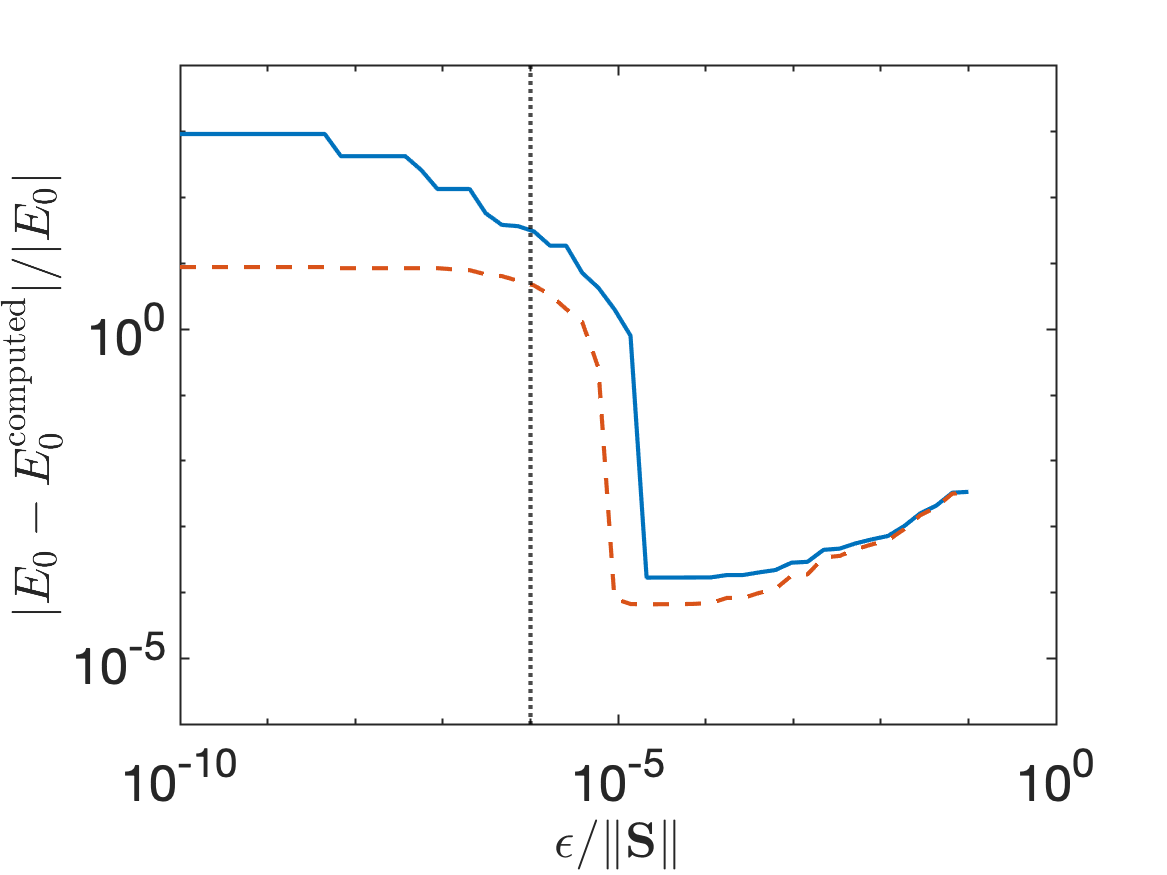}
    \caption{Hubbard, $L = 10$, $U = 10$, $n = 80$}\label{fig:threshold_choice_hubbard_l10_u10_q80}
  \end{subfigure}

\begin{subfigure}[b]{0.47\textwidth}
    \centering
    \includegraphics[width=\textwidth]{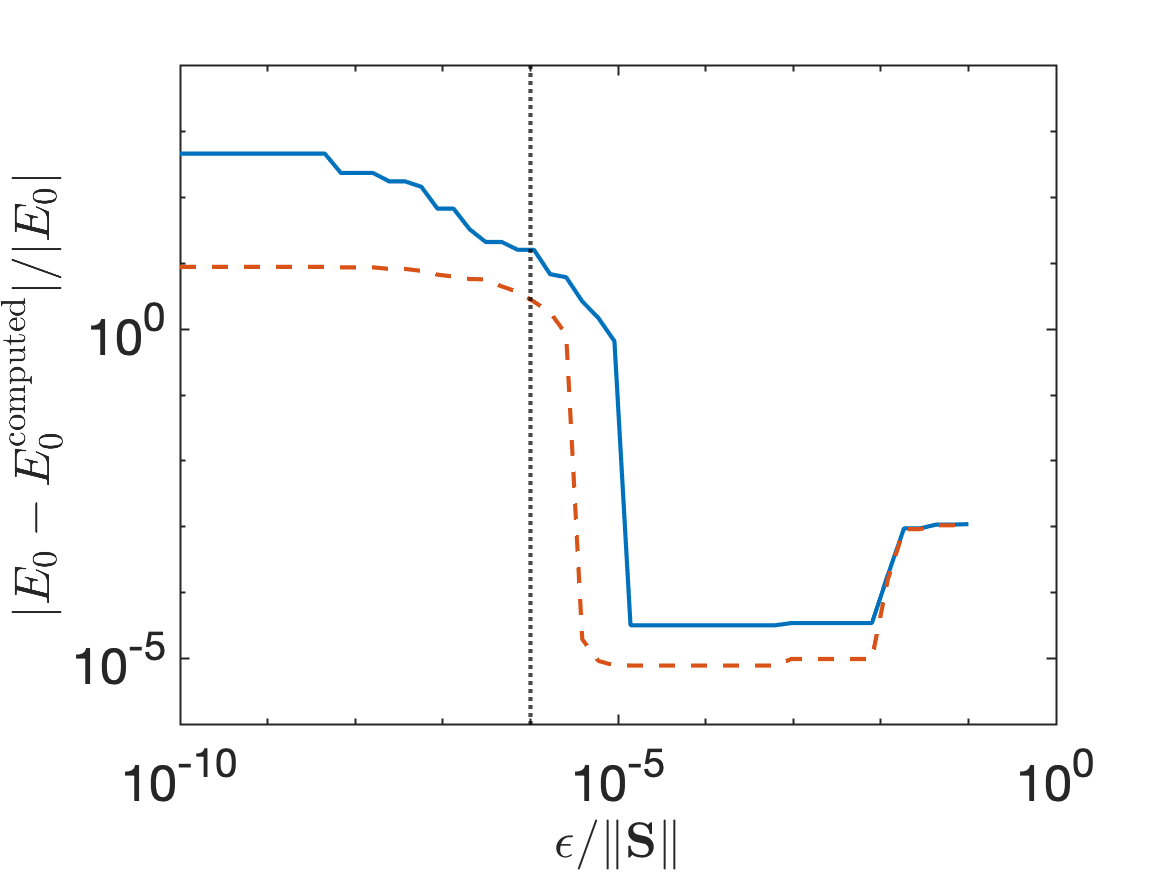}
    \caption{Hubbard, $L = 6$, $U = 8$, $n = 40$}\label{fig:threshold_choice_hubbard_l6_u8_q40}
  \end{subfigure}
  ~
  \begin{subfigure}[b]{0.47\textwidth}
    \centering
    \includegraphics[width=\textwidth]{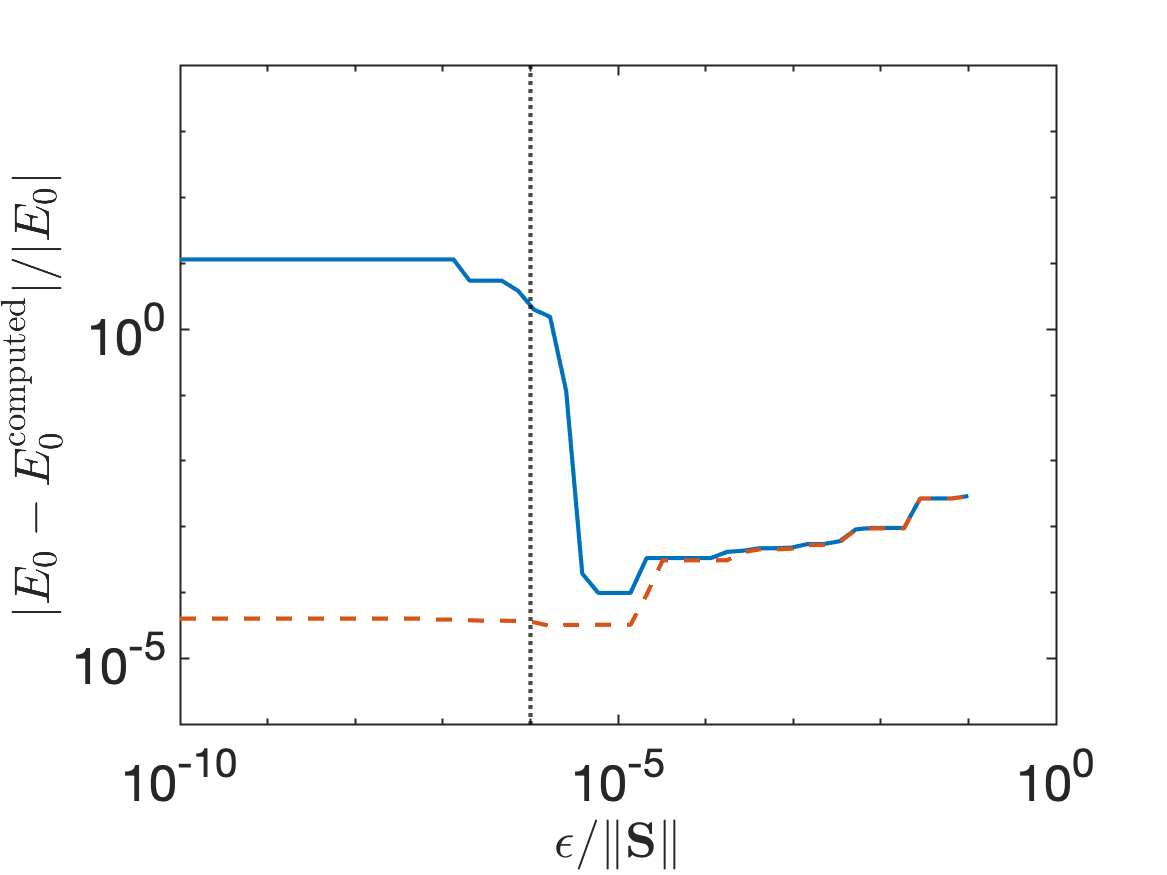}
    \caption{Ising, $L = 8$, $n = 20$}\label{fig:threshold_choice_ising_l8_q20}
  \end{subfigure}
  
  \caption{Maximum (blue solid) and median (red dashed) error over 100 initializations for eigenvalues computed from the noise-perturbed pair $(\mat{\tilde{H}},\mat{\tilde{S}})$ using thresholding for various values of the threshold $\epsilon$ for a fixed noise level $\sigma = 10^{-6}$ (dotted black line) for Hubbard and Ising models for various parameters not considered in Figure~\ref{fig:threshold_choice}.}
  \label{fig:threshold_choice_more}
\end{figure}

\begin{figure}[t]
  \centering
    \begin{subfigure}[b]{0.47\textwidth}
    \centering
    \includegraphics[width=\textwidth]{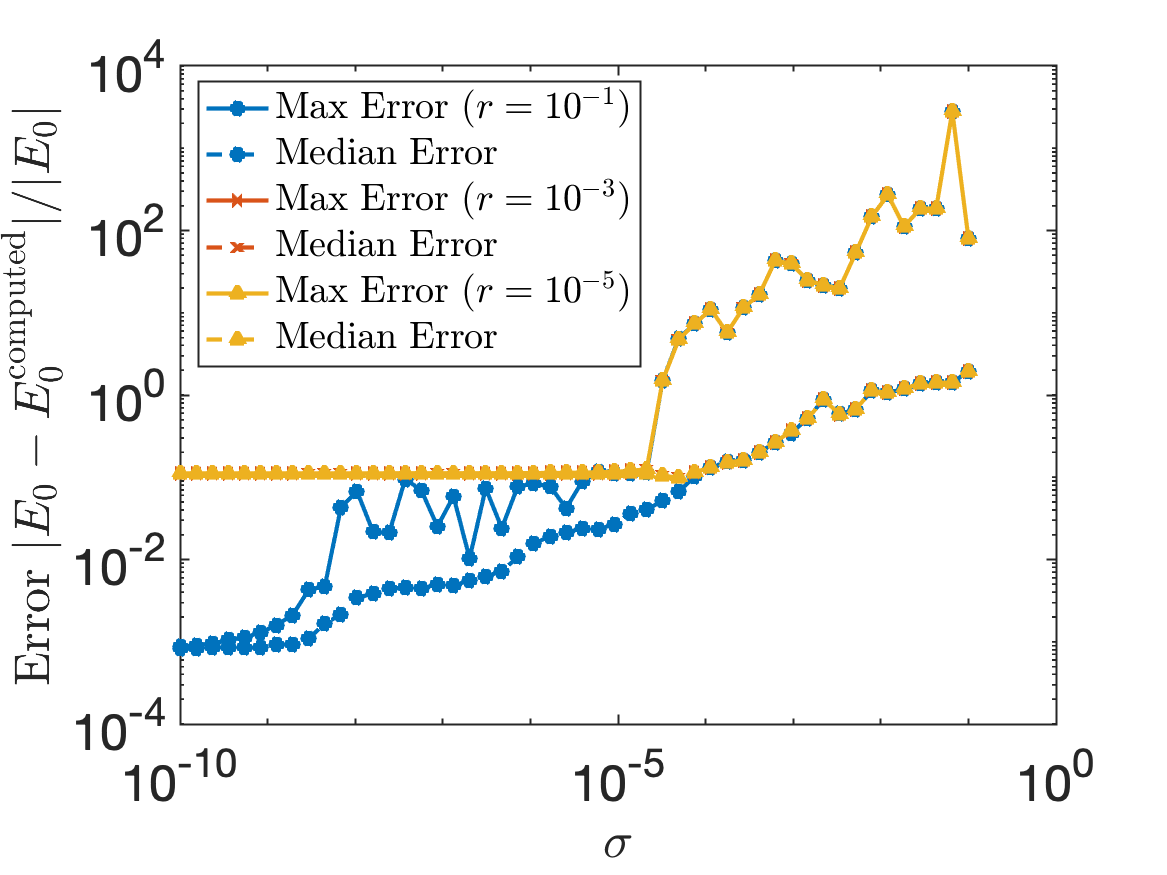}
    \caption{Hubbard, $L = 10$, $U = 8$, $n = 10$}\label{fig:auto_thresholding_hubbard_l10_u8_q10}
  \end{subfigure}
  ~
  \begin{subfigure}[b]{0.47\textwidth}
    \centering
    \includegraphics[width=\textwidth]{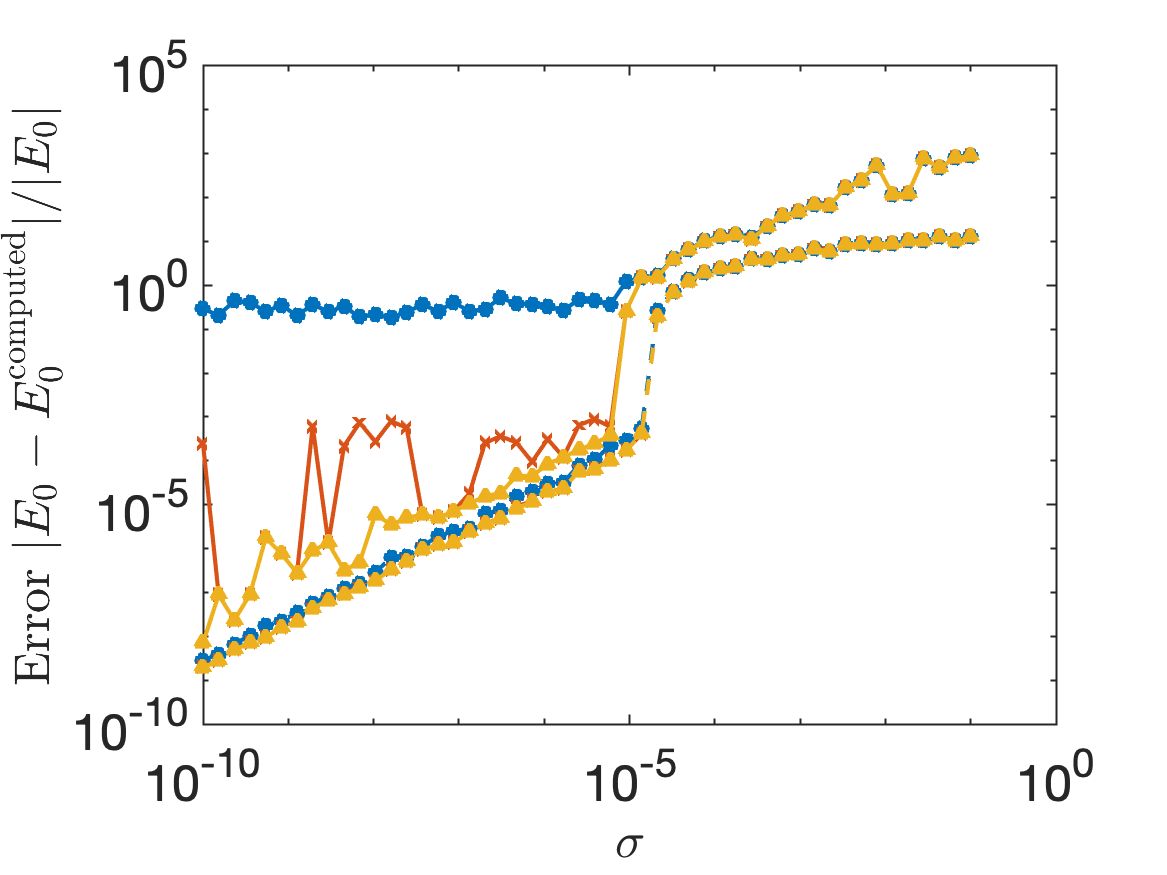}
    \caption{Hubbard, $L = 10$, $U = 8$, $n = 80$}\label{fig:auto_thresholding_hubbard_l10_u8_q80}
  \end{subfigure}  

  \begin{subfigure}[b]{0.47\textwidth}
    \centering
    \includegraphics[width=\textwidth]{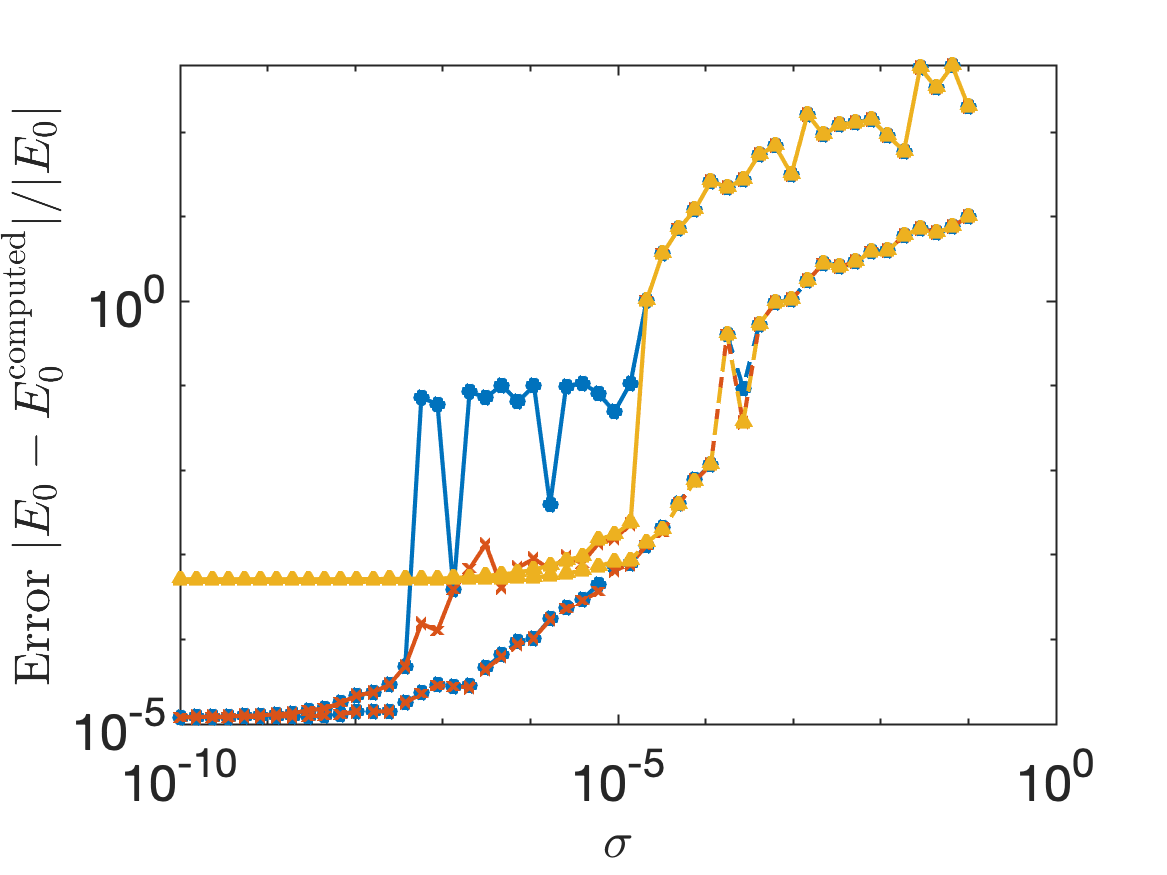}
    \caption{Hubbard, $L = 10$, $U = 10$, $n = 30$}\label{fig:auto_thresholding_hubbard_l10_u10_q30}
  \end{subfigure}
  ~
  \begin{subfigure}[b]{0.47\textwidth}
    \centering
    \includegraphics[width=\textwidth]{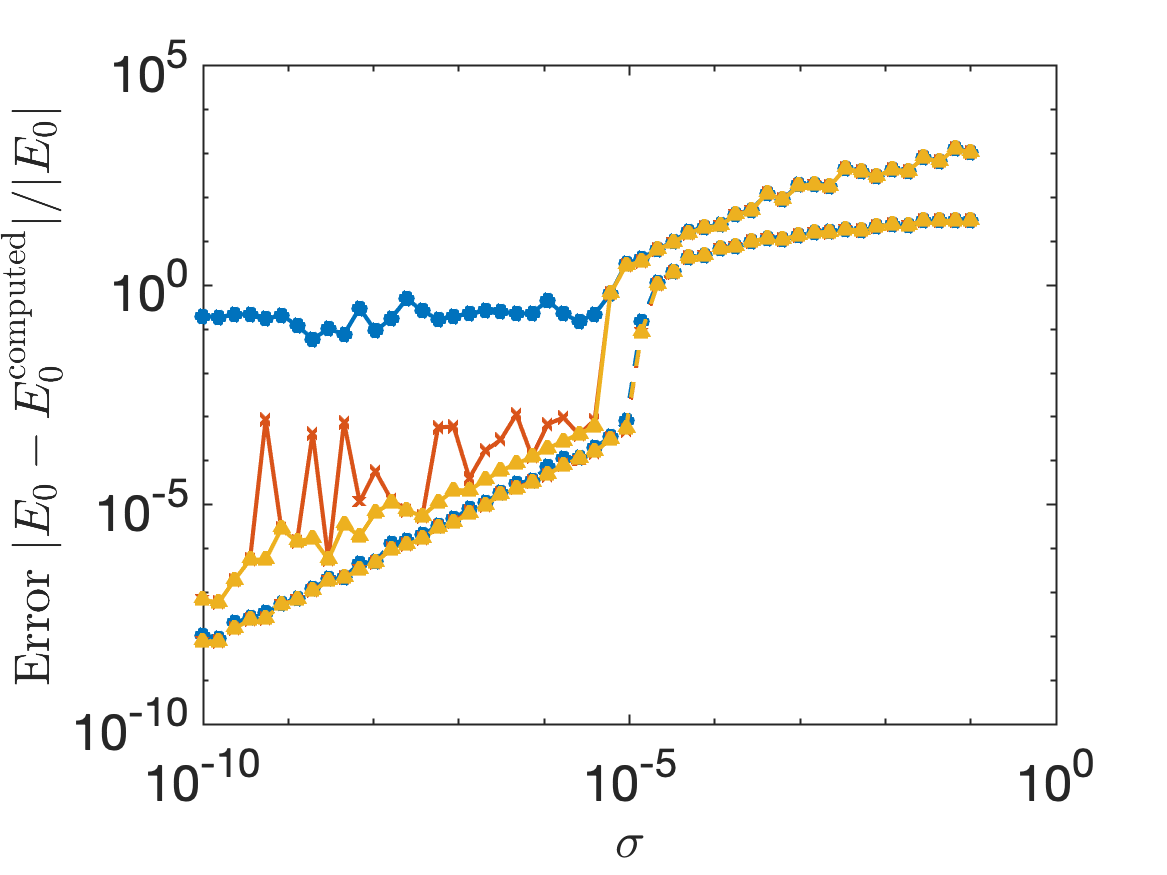}
    \caption{Hubbard, $L = 10$, $U = 10$, $n = 80$}\label{fig:auto_thresholding_hubbard_l10_u10_q80}
  \end{subfigure}

\begin{subfigure}[b]{0.47\textwidth}
    \centering
    \includegraphics[width=\textwidth]{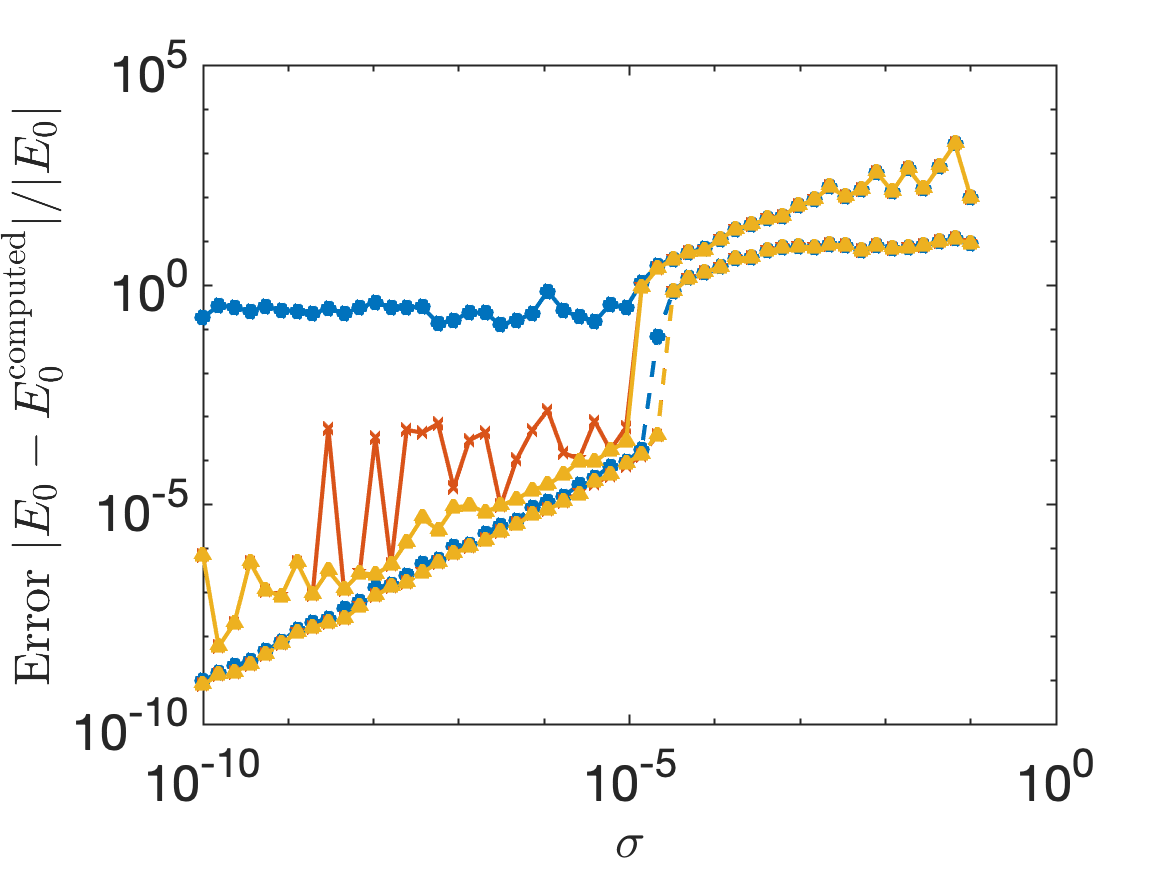}
    \caption{Hubbard, $L = 6$, $U = 8$, $n = 40$}\label{fig:auto_thresholding_hubbard_l6_u8_q40}
  \end{subfigure}
  ~
  \begin{subfigure}[b]{0.47\textwidth}
    \centering
    \includegraphics[width=\textwidth]{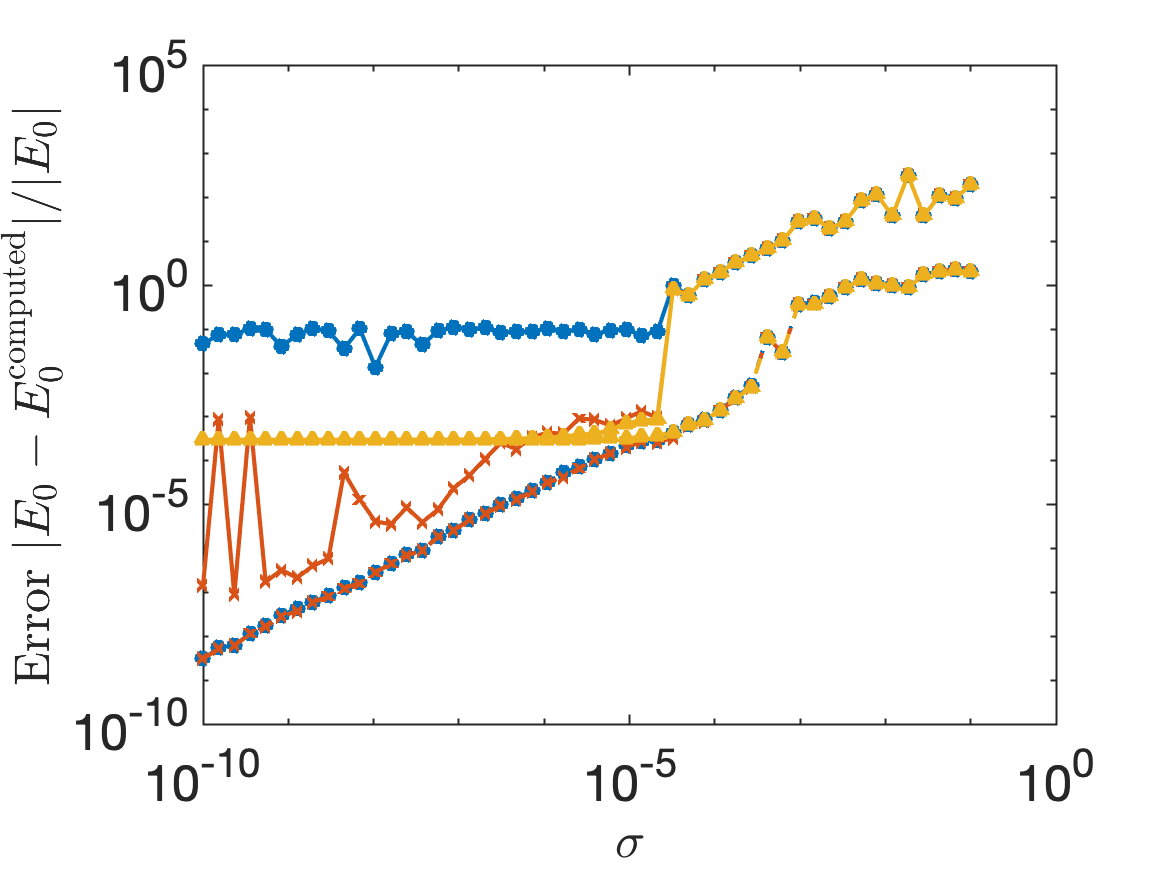}
    \caption{Ising, $L = 8$, $n = 20$}\label{fig:auto_thresholding_ising_l8_q20}
  \end{subfigure}
  
  \caption{Maximum and median error over 100 initializations for eigenvalues computed from the noise-perturbed pair $(\mat{\tilde{H}},\mat{\tilde{S}})$ using automatically tuned thresholding (Algorithm~\ref{alg:auto_thresholding}) for three cutoffs $r \in \{ 10^{-1}, 10^{-3}, 10^{-5} \}$ for various values of the noise level $\sigma$ for Hubbard and Ising models for various parameters not considered in Figure~\ref{fig:auto_thresholding}.}
  \label{fig:auto_thresholding_more}
\end{figure}

Finally, we conclude with some additional figures concerning additional numerical experiments.
Figure~\ref{fig:fixed_threshold} shows the smallest eigenvalue computed when a fixed threshold parameter is used, independent of the noise level.
Figures~\ref{fig:thresholding_more}, \ref{fig:threshold_choice_more}, and \ref{fig:auto_thresholding_more} provide more parameter settings for the Hubbard and Ising models for the Figures~\ref{fig:thresholding}, \ref{fig:threshold_choice}, and \ref{fig:auto_thresholding}.

%% file: revision1.bbl
\def\noopsort#1{}\def\l{\char32l}\def\v#1{{\accent20 #1}}
  \let\^^_=\v\def\hbk{hardback}\def\pbk{paperback}
\begin{thebibliography}{10}

\bibitem{Bha97}
{\sc R.~Bhatia}, {\em Matrix Analysis}, vol.~169 of Graduate {{Texts}} in
  {{Mathematics}}, {Springer-Verlag, New York}, 1997,
  \url{https://doi.org/10.1007/978-1-4612-0653-8}.

\bibitem{Bjo15}
{\sc {\AA}.~Bj{\"o}rck}, {\em Numerical {{Methods}} in {{Matrix
  Computations}}}, vol.~59 of Texts in {{Applied Mathematics}}, {Springer
  International Publishing}, 2015,
  \url{https://doi.org/10.1007/978-3-319-05089-8}.

\bibitem{CaiBaiPaskEtAl2013}
{\sc Y.~Cai, Z.~Bai, J.~E. Pask, and N.~Sukumar}, {\em Hybrid preconditioning
  for iterative diagonalization of ill-conditioned generalized eigenvalue
  problems in electronic structure calculations}, J. Comput. Phys., 255 (2013),
  pp.~16--30.

\bibitem{CST+21}
{\sc A.~M. Childs, Y.~Su, M.~C. Tran, N.~Wiebe, and S.~Zhu}, {\em Theory of
  {{Trotter Error}} with {{Commutator Scaling}}}, Phys. Rev. X, 11 (2021),
  p.~011020, \url{https://doi.org/10.1103/PhysRevX.11.011020}.

\bibitem{CRD+18}
{\sc J.~I. Colless, V.~V. Ramasesh, D.~Dahlen, M.~S. Blok, M.~E.
  {Kimchi-Schwartz}, J.~R. McClean, J.~Carter, W.~A. {de Jong}, and
  I.~Siddiqi}, {\em Computation of {{Molecular Spectra}} on a {{Quantum
  Processor}} with an {{Error}}-{{Resilient Algorithm}}}, Phys. Rev. X, 8
  (2018), p.~011021, \url{https://doi.org/10.1103/PhysRevX.8.011021}.

\bibitem{CG21}
{\sc C.~L. Cortes and S.~K. Gray}, {\em Quantum {{Krylov}} subspace algorithms
  for ground and excited state energy estimation}, arXiv:2109.06868,  (2021),
  \url{https://arxiv.org/abs/2109.06868}.

\bibitem{DK70}
{\sc C.~Davis and W.~M. Kahan}, {\em The {{Rotation}} of {{Eigenvectors}} by a
  {{Perturbation}}. {{III}}}, SIAM J. Numer. Anal., 7 (1970), pp.~1--46,
  \url{https://doi.org/10.1137/0707001}.

\bibitem{DI19}
{\sc P.~Drineas and I.~C.~F. Ipsen}, {\em Low-{{Rank Matrix Approximations Do
  Not Need}} a {{Singular Value Gap}}}, SIAM J. Matrix Anal. Appl., 40 (2019),
  pp.~299--319, \url{https://doi.org/10.1137/18M1163658}.

\bibitem{EV82}
{\sc A.~{Emami-Naeini} and P.~Van~Dooren}, {\em Computation of zeros of linear
  multivariable systems}, Automatica, 18 (1982), pp.~415--430,
  \url{https://doi.org/10.1016/0005-1098(82)90070-X}.

\bibitem{FH72}
{\sc G.~Fix and R.~Heiberger}, {\em An {{Algorithm}} for the
  {{Ill}}-{{Conditioned Generalized Eigenvalue Problem}}}, SIAM J. Numer.
  Anal., 9 (1972), pp.~78--88, \url{https://doi.org/10.1137/0709009}.

\bibitem{GolubVanLoan2013}
{\sc G.~H. Golub and C.~F. Van~Loan}, {\em Matrix computations}, Johns Hopkins
  Univ. Press, Baltimore, fourth~ed., 2013.

\bibitem{GN86}
{\sc L.~A. Gribov and B.~K. Novosadov}, {\em Use of overcomplete basis sets in
  quantum-chemical calculations}, Journal of Molecular Structure: THEOCHEM, 136
  (1986), pp.~387--389, \url{https://doi.org/10.1016/0166-1280(86)80152-X}.

\bibitem{HLB+20}
{\sc W.~J. Huggins, J.~Lee, U.~Baek, B.~O'Gorman, and K.~B. Whaley}, {\em A
  non-orthogonal variational quantum eigensolver}, New Journal of Physics, 22
  (2020), p.~073009, \url{https://doi.org/10.1088/1367-2630/ab867b}.

\bibitem{JungenKaufmann1992}
{\sc M.~Jungen and K.~Kaufmann}, {\em The {F}ix--{H}eiberger procedure for
  solving the generalized ill-conditioned symmetric eigenvalue problem}, Int.
  J. Quantum Chem., 41 (1992), pp.~387--397.

\bibitem{KMC+21}
{\sc K.~Klymko, C.~{Mejuto-Zaera}, S.~J. Cotton, F.~Wudarski, M.~Urbanek,
  D.~Hait, M.~{Head-Gordon}, K.~B. Whaley, J.~Moussa, N.~Wiebe, W.~A. {de
  Jong}, and N.~M. Tubman}, {\em Real time evolution for ultracompact
  {{Hamiltonian}} eigenstates on quantum hardware}, arXiv:2103.08563,  (2021),
  \url{https://arxiv.org/abs/2103.08563}.

\bibitem{LN20}
{\sc M.~Lotz and V.~Noferini}, {\em Wilkinson's {{Bus}}: {{Weak Condition
  Numbers}}, with an {{Application}} to {{Singular Polynomial Eigenproblems}}},
  Found. Comput. Math., 20 (2020), pp.~1439--1473,
  \url{https://doi.org/10.1007/s10208-020-09455-y}.

\bibitem{Low67}
{\sc P.-O. L{\"o}wdin}, {\em Group {{Algebra}}, {{Convolution Algebra}}, and
  {{Applications}} to {{Quantum Mechanics}}}, Reviews of Modern Physics, 39
  (1967), pp.~259--287, \url{https://doi.org/10.1103/RevModPhys.39.259}.

\bibitem{MT97}
{\sc V.~A. Mandelshtam and H.~S. Taylor}, {\em Harmonic inversion of time
  signals and its applications}, J. Chem. Phys., 107 (1997), pp.~6756--6769,
  \url{https://doi.org/10.1063/1.475324}.

\bibitem{ML04}
{\sc R.~Mathias and C.-K. Li}, {\em The definite generalized eigenvalue
  problem: {{A}} new perturbation theory}, T-{{NAREP}} No. 457, {inst-MCCM},
  {inst-MCCM:adr}, Oct. 2004.

\bibitem{MKCd17}
{\sc J.~R. McClean, M.~E. {Kimchi-Schwartz}, J.~Carter, and W.~A. {de Jong}},
  {\em Hybrid quantum-classical hierarchy for mitigation of decoherence and
  determination of excited states}, Phys. Rev. A, 95 (2017), p.~042308,
  \url{https://doi.org/10.1103/PhysRevA.95.042308}.

\bibitem{MST+20}
{\sc M.~Motta, C.~Sun, A.~T.~K. Tan, M.~J. O'Rourke, E.~Ye, A.~J. Minnich,
  F.~G. S.~L. Brand{\~a}o, and G.~K.-L. Chan}, {\em Determining eigenstates and
  thermal states on a quantum computer using quantum imaginary time evolution},
  Nature Physics, 16 (2020), pp.~205--210,
  \url{https://doi.org/10.1038/s41567-019-0704-4}.

\bibitem{Nannicini2020}
{\sc G.~Nannicini}, {\em An introduction to quantum computing, without the
  physics}, SIAM Rev., 62 (2020), pp.~936--981.

\bibitem{NegeleOrland1988}
{\sc J.~W. Negele and H.~Orland}, {\em Quantum many-particle systems},
  Westview, 1988.

\bibitem{NielsenChuang2000}
{\sc M.~A. Nielsen and I.~Chuang}, {\em Quantum computation and quantum
  information}, 2000.

\bibitem{Par98}
{\sc B.~N. Parlett}, {\em The {{Symmetric Eigenvalue Problem}}}, Classics in
  {{Applied Mathematics}}, SIAM, 1998,
  \url{https://doi.org/10.1137/1.9781611971163}.

\bibitem{PM19}
{\sc R.~M. Parrish and P.~L. McMahon}, {\em Quantum {{Filter Diagonalization}}:
  {{Quantum Eigendecomposition}} without {{Full Quantum Phase Estimation}}},
  arXiv:1909.08925,  (2019), \url{https://arxiv.org/abs/1909.08925}.

\bibitem{Preskill2018}
{\sc J.~Preskill}, {\em Quantum computing in the {NISQ} era and beyond},
  Quantum, 2 (2018), p.~79.

\bibitem{Preskill2021}
{\sc J.~Preskill}, {\em Quantum computing 40 years later}, arXiv:2106.10522,
  (2021).

\bibitem{Saa80}
{\sc Y.~Saad}, {\em On the {{Rates}} of {{Convergence}} of the {{Lanczos}} and
  the {{Block}}-{{Lanczos Methods}}}, SIAM J. Numer. Anal., 17 (1980),
  pp.~687--706, \url{https://doi.org/10.1137/0717059}.

\bibitem{SMS20}
{\sc I.~Sabzevari, A.~Mahajan, and S.~Sharma}, {\em An accelerated linear
  method for optimizing non-linear wavefunctions in variational {{Monte
  Carlo}}}, J. Chem. Phys., 152 (2020), p.~024111,
  \url{https://doi.org/10.1063/1.5125803}.

\bibitem{SekiYunoki2021}
{\sc K.~Seki and S.~Yunoki}, {\em Quantum power method by a superposition of
  time-evolved states}, Phys. Rev. X Quantum, 2 (2021), p.~010333.

\bibitem{SHE20}
{\sc N.~H. Stair, R.~Huang, and F.~A. Evangelista}, {\em A {{Multireference
  Quantum Krylov Algorithm}} for {{Strongly Correlated Electrons}}}, J. Chem.
  Theory Comput., 16 (2020), pp.~2236--2245,
  \url{https://doi.org/10.1021/acs.jctc.9b01125}.

\bibitem{Ste79}
{\sc G.~W. Stewart}, {\em Perturbation bounds for the definite generalized
  eigenvalue problem}, Linear Algebra Appl., 23 (1979), pp.~69--85,
  \url{https://doi.org/10.1016/0024-3795(79)90094-6}.

\bibitem{SS90}
{\sc G.~W. Stewart and J.-G. Sun}, {\em Matrix Perturbation Theory}, Computer
  Science and Scientific Computing, {Academic Press}, 1st edition~ed., 1990.

\bibitem{Tro15}
{\sc J.~A. Tropp}, {\em An introduction to matrix concentration inequalities},
  Foundations and Trends in Machine Learning, 8 (2015), pp.~1--230.

\bibitem{WN95}
{\sc M.~R. Wall and D.~Neuhauser}, {\em Extraction, through
  filter-diagonalization, of general quantum eigenvalues or classical normal
  mode frequencies from a small number of residues or a short-time segment of a
  signal. {{I}}. {{Theory}} and application to a quantum-dynamics model}, The
  Journal of Chemical Physics, 102 (1995), pp.~8011--8022,
  \url{https://doi.org/10.1063/1.468999}.

\bibitem{WeinbergBukov2017}
{\sc P.~Weinberg and M.~Bukov}, {\em {QuSpin: a Python Package for Dynamics and
  Exact Diagonalisation of Quantum Many Body Systems part I: spin chains}},
  SciPost Phys., 2 (2017), p.~003.

\bibitem{WeinbergBukov2019}
{\sc P.~Weinberg and M.~Bukov}, {\em {QuSpin: a Python Package for Dynamics and
  Exact Diagonalisation of Quantum Many Body Systems. Part II: bosons, fermions
  and higher spins}}, SciPost Phys., 7 (2019), p.~20.

\bibitem{Wil79}
{\sc J.~H. Wilkinson}, {\em Kronecker's canonical form and the {{QZ}}
  algorithm}, Linear Algebra Appl., 28 (1979), pp.~285--303,
  \url{https://doi.org/10.1016/0024-3795(79)90140-X}.

\end{thebibliography}
